\documentclass{aa}  

\usepackage{graphicx}
\usepackage{txfonts}
\usepackage{float} 
\usepackage{xcolor}

\begin{document}

   \title{Neutron-capture elements in dwarf galaxies III:\\ A homogenized analysis of 13 dwarf spheroidal and ultra-faint galaxies.\thanks{Based on data obtained with VLT and the W.M. Keck Observatory. Information of the used program IDs is given in the acknowledgments. Abundances and stellar parameters are only available in electronic form at the CDS via anonymous ftp to cdsarc.u-strasbg.fr (130.79.128.5) or via http://cdsweb.u-strasbg.fr/cgi-bin/qcat?J/A+A/}}

   \author{  M. Reichert\inst{\ref{TUD}}   \and   C. J. Hansen\inst{\ref{MPI},\ref{Darkcosm}} \and M. Hanke\inst{\ref{ARI}} \and \'A. Sk{\'u}lad{\'o}ttir\inst{\ref{MPI},\ref{firenze1},\ref{firenze2}} \and A. Arcones\inst{\ref{TUD},\ref{GSI}}   \and E. K. Grebel\inst{\ref{ARI}}   }
         
 \institute{Technische Universit\"at Darmstadt, Institut f\"ur Kernphysik, Schlossgartenstr. 2, 64289 Darmstadt, Germany \label{TUD}\\
 \email{mreichert@theorie.ikp.physik.tu-darmstadt.de}
              \and
             Max-Planck-Institut f\"ur Astronomie, K\"onigstuhl 17, D-69117 Heidelberg, Germany \label{MPI}\\
            \email{hansen@mpia-hd.mpg.de}
             \and
             Copenhagen University, Dark Cosmology Centre, The Niels Bohr Institute, 
Vibenshuset, Lyngbyvej 2, DK-2100 Copenhagen, Denmark \label{Darkcosm}
\and
             Astronomisches Rechen-Institut, Zentrum f\"ur Astronomie der Universit\"at Heidelberg, M\"onchhofstr.12-14, 69120 Heidelberg, Germany\label{ARI}
\and
Dipartimento di Fisica e Astronomia, Universit\'a degli Studi di Firenze, Via G. Sansone 1, I-50019 Sesto Fiorentino, Italy \label{firenze1}
\and      
INAF/Osservatorio Astrofisico di Arcetri, Largo E. Fermi 5, I-50125 Firenze, Italy \label{firenze2}
\and
GSI Helmholtzzentrum für Schwerionenforschung GmbH, Planckstr. 1, D-64291 Darmstadt, Germany\label{GSI}
              }

   \date{Received ; accepted }

  \abstract{
  We present a large homogeneous set of stellar parameters and abundances across a broad range of metallicities, involving $13$ classical dwarf spheroidal (dSph) and ultra-faint dSph (UFD) galaxies. In total, this study includes $380$ stars in Fornax, Sagittarius, Sculptor, Sextans, Carina, Ursa Minor, Draco, Reticulum II, Bootes I, Ursa Major II, Leo I, Segue I, and Triangulum II. This sample represents the largest, homogeneous, high-resolution study of dSph galaxies to date.
  }
   {With our homogeneously derived catalog, we are able to search for similar and deviating trends across different galaxies. We investigate the mass dependence of the individual systems on the production of $\alpha$-elements, but also try to shed light on the long-standing puzzle of the dominant production site of r-process elements.}
   {We used data from the Keck observatory archive and the ESO reduced archive to reanalyze stars from these $13$ classical dSph and UFD galaxies. We automatized the step of obtaining stellar parameters, but ran a full spectrum synthesis (1D, local thermal equilibrium) to derive all abundances except for iron to which we applied nonlocal thermodynamic equilibrium corrections where possible.}
   {The homogenized set of abundances yielded the unique possibility of deriving a relation between the onset of type Ia supernovae and the stellar mass of the galaxy. Furthermore, we derived a formula to estimate the evolution of $\alpha$-elements. This reveals a universal relation of these systems across a large range in mass. Finally, we show that between stellar masses of $2.1\cdot 10^7 \mathrm{M}_\odot$ and  $2.9\cdot 10^5 \mathrm{M}_\odot$, there is no dependence of the production of heavy r-process elements on the stellar mass of the galaxy.}
   {Placing all abundances consistently on the same scale is crucial to answering questions about the chemical history of galaxies. 
   {By homogeneously analyzing Ba and Eu in the 13 systems, we have traced the onset of the s-process and found it to increase with metallicity as a function of the galaxy's stellar mass. Moreover, the r-process material correlates with the $\alpha$-elements indicating some coproduction of these, which in turn would point toward rare core-collapse supernovae rather than binary neutron star mergers as a host for the r-process at low [Fe/H] in the investigated dSph systems.} }
   {}


\titlerunning{A homogenized analysis of 13 dwarf spheroidal and ultra-faint galaxies}
\authorrunning{M. Reichert et al.}

   \maketitle

\section{Introduction}
Dwarf spheroidal (dSph) galaxies are arguably the most frequent type of galaxy in the present-day Universe. They include the least luminous, least massive, and most dark-matter-dominated galaxies known. For summaries of their properties, see, for example,
\citet{Grebel2003}, \citet{Tolstoy2009}, and \citet{McConnachie2012}. These gas-deficient, early-type galaxies are usually found in the outskirts of massive galaxies and show a pronounced morphology-density relation, suggesting that they were subject to environmental effects during their evolution \citep[e.g.,][]{Mayer2001,Sawala2012,Ocvirk2014}.

Because of their low stellar densities and low surface brightnesses, dSph galaxies are challenging to detect. During the last fifteen years, the number of known dSphs in the Local Group (and beyond) have vastly increased thanks to deep homogeneous imaging surveys. Here, in particular, the Sloan Digital Sky Survey \citep[SDSS, e.g., in][]{Zucker2006a,Zucker2006b,Zucker2007}, the Panoramic Survey Telescope And Rapid Response System \citep[Pan-STARRS, e.g., in][]{Laevens2015a,Laevens2015b}, the Dark Energy Survey \citep[DES, e.g., reported in][]{Drlica-Wagner2015,Bechtol2015}, and the Pan-Andromeda Archaeological Survey \citep[PAndAS, reported in, e.g.,][]{Martin2009,Richardson2011} have been of key importance in increasing the dwarf galaxy census in the Local Group to about 100 objects. Most of the new discoveries are the so-called ultra-faint dSph (UFD) galaxies, which are fainter than the ``classical'' dSph limit of $M_V < -8$.

Typically exhibiting prominent, old, metal-poor stellar populations \citep[e.g.,][]{Grebel2004,Weisz2014}, dSphs are often seen as the surviving building blocks of the halos of more massive galaxies, which are believed to have formed, in part, through accretion \citep[e.g.,][]{DeLucia2008,Pillepich2015,Rodriguez-Gomez2016}. While the bulk of the accreted component of a stellar halo's mass in disk galaxies such as the Milky Way (MW) may have come from more massive dwarf systems \citep[e.g.,][]{Font2006,DeLucia2008,Cooper2010}, the accreted dSphs are believed to have particularly contributed to the outer halo and may be important sources of very metal-poor stars \citep[e.g.,][]{Salvadori2009,Starkenburg2010,Lai2011}. Moreover, the old populations in dSphs, which have been found to be ubiquitous in all systems studied in detail so far \citep[e.g.,][]{Grebel2004}, may hold the key to exploring early star formation and chemical enrichment in very low-mass halos. These deliberations are among the main motivations for studying stellar chemical abundances in the Galactic dSphs, which are sufficiently close for spectroscopic analyses of individual stars. 

An increasing number of high-resolution spectroscopic studies focus on these very faint objects. Both, classical dSph and UFD galaxies offer a good opportunity to study fundamental properties of galaxy formation and evolution. The analysis of their chemical abundances plays a key role when investigating these types of systems, and it allows us to probe the distribution of stellar masses, the initial mass function (IMF), the star formation, and chemical evolution across different environments. 

Intriguing examples are the recent discoveries of the UFD galaxies Reticulum II \citep{Ji2016a,Roederer2016} and Tucana III \citep{Hansen2017}, which provide evidence on the production site of heavy elements. Both galaxies are enhanced in neutron-capture (n-) elements. 

Heavy elements (with atomic numbers $Z>30$) are dominantly produced in the slow (s) and rapid (r) neutron-capture process. The intermediate (i) process \citep[e.g.,][]{Dardelet2015,Hampel2016,Hampel2019,Banerjee2018,Koch2019,Skuladottir2020b} and the light element primary process \citep[LEPP, e.g.,][]{Travaglio2004,Montes2007} may also contribute to the enrichment of heavy elements. Both the s- and the r- process are hosted by distinct astrophysical production sites and carry individual chemical fingerprints \citep[for reviews of the r-process and the origin of heavy elements see, e.g.,][]{Cowan1991,Sneden2008,Thielemann2011,Thielemann2017,Frebel2018,Horowitz2019,Cowan2019}.
We are particularly interested in the production of Ba, which can be used as a tracer of the s-process and Eu as a tracer of the r-process. Both elements exhibit very clean traces for each process \citep[e.g.,][]{Bisterzo2014} and are also easy to extract from the spectrum of a star. 
While the s-process is associated with asymptotic giant branch (AGB) stars \citep[e.g.,][]{Cristallo2009,Kaeppeler2011,Lugaro2012} and possibly fast rotating spin stars \citep[e.g., ][]{Pignatari2008,Meynet2002,Frischknecht2012,Chiappini2013}, the astrophysical production site of the r-process is not yet clear.
A homogenized set of abundances at low metallicities is the only way to distinguish between several r-process production sites. Possible candidates for the r-process sites are specific types of core-collapse supernovae, such as magneto-rotationally driven SNe \citep[MR-SNe, e.g., ][]{Winteler2012,Nishimura2015,Nishimura2017,Moesta2018}, collapsars \citep[e.g., ][]{Fujimoto2007,Siegel2019}, or neutron star mergers \citep[NSMs, e.g., ][]{Korobkin2012,Wanajo2014,Rosswog1999}. Recent studies speculate in favor of one or the other dominant astrophysical production sites of the r-process. The importance of an extra r-process production site in dSph galaxies in addition to NSMs have been pointed out by several studies \citep[e.g., ][]{Beniamini2016,Beniamini2018,skuladottir2019}, while other studies favor NSMs to be the dominant source \citep[][]{Duggan2018}.

The Stellar Abundances for Galactic Archaeology Database\footnote{http://sagadatabase.jp/}\citep[SAGA;][]{SAGA} contains a large set of chemical abundances of $24$ dSph and UFD galaxies. In total, it includes more than $6000$ stars, which offers a great opportunity to study the chemical evolution of dSph galaxies with good number statistics. In this database, a large number of $\alpha$-element detections are included, but only a subset of the stars have detections of n-capture elements available (2079 Mg detections and 316 Ba detections in 2017) and when comparing chemical abundances from different studies, several problems can arise. The lack of homogenization across different studies might obscure valuable information. SAGA accounts for the adopted solar abundances, however, other important factors as the difference in methods for deriving the stellar parameters require a complete reanalysis of the involved spectra. In addition, the assumption of local thermal equilibrium (LTE) can induce nonphysical abundance trends, especially when analyzing a large range of metallicities. Therefore, non-LTE (NLTE) corrections have to be taken into account. Minimizing these systematics is one goal of this work. This is a major effort. It was previously done in a subset of $59$ stars in dSphs by \citet{Mashonkina2017,Mashonkina2017b}. They derive NLTE abundances for ten chemical elements across seven dSph galaxies and the MW halo. We extend the number of homogeneously treated stars by presenting the first, large ($380$ stars), homogeneous, high-resolution study ($R>15000$, for 98\% of our sample) that consistently places classical dSph and UFD galaxies on the same chemical abundance scale, thus allowing us to probe trends in each system and to compare them without the usual offsets, biases, or systematics as outlined above.

In Sect.~\ref{sct:sample_selection}, we discuss the sample selection followed by the data reduction procedure in Sect.~\ref{sct:data_reduction}. The determination of stellar parameters is described in Sect.~\ref{sct:stellar_pars}, while the abundance analysis is outlined in Sect.~\ref{sct:abundance_analysis}. We discuss abundance trends of lighter and heavier elements in Sect.~\ref{sct:abundance_trends} and Sect.~\ref{ssct:ncapture}, respectively. Finally our conclusions are given in Sect.~\ref{sct:summary_and_conclusion}.

\section{Sample selection}
\label{sct:sample_selection}
We initially selected $5497$ stars for the analysis, which were previously identified as members of dSph and UFD galaxies. This also includes stars of dSph galaxies from the SAGA database. The sample contains high-resolution data ($R\gtrsim 15000$) that were subject of previous studies but includes also unpublished data. 
The reduced European Southern Observatory (ESO) archive and the Keck Observatory archive were explored and we selected spectra covering the ultraviolet and visible to include elemental transitions that are key for the present study. We did not include stars that only contain the infrared part of the spectrum used for Ca triplet (CaT) surveys \citep[e.g., ][]{Pont2004,Battaglia2006,Koch2006,Hendricks2014}. Stars for which we could not determine stellar parameters (i.e., effective temperature, surface gravity, metallicity, and microturbulence; Sect.~\ref{sct:stellar_pars}) and stars with insufficient data (e.g., $\mathrm{S/N}<10$) or missing spectra were removed from the analysis. For example, we did not include stars from \citet{McWilliam2005} that are found in the Keck archive, as we could not accurately (within a 3 arcsec circle) resolve their coordinates in the SIMBAD\footnote{Set of Identifications, Measurements and Bibliography for Astronomical Data, \citealt{Simbad}}, NED\footnote{The NASA/IPAC Extragalactic Database (NED) is operated by the Jet Propulsion Laboratory, California Institute of Technology, under contract with the National Aeronautics and Space Administration.} or the {\it Gaia} archives \citep{Gaia2018}. In addition, we determined radial velocities (Sect. 3.3) of each star to confirm membership in the individual galaxies. 
Stars analyzed in \citet{Tafelmeyer2010} were added manually to include coordinate corrections from \citet{Tafelmeyer2011}.
This resulted in a total sample size of $380$ stars in $13$ dwarf galaxies (see Table~\ref{tab:sample} and Fig.~\ref{fig:loc_dsph}), with $295$ observations from FLAMES/GIRAFFE \citep[Fibre Large Array Multi Element Spectrograph,][with resolving powers ranging from $17000$ to $28800$]{FLAMES}, $56$ from UVES \citep[Ultraviolet and Visual Echelle Spectrograph,][with resolving powers of $31950$ and $42310$]{UVES}, $2$ from X-shooter \citep[][with resolving powers of $6600$ and $11000$]{XSHOOTER}, and $27$ from HIRES \citep[High Resolution Echelle Spectrometer, ][with resolving powers of $35800$, $47700$, and $71600$]{HIRES}. The final sample size is dictated by the accessibility of the spectra obtained with HIRES, UVES, FLAMES/GIRAFFE, or X-shooter in the reduced archive. Unfortunately, observatories of other high-resolution instruments do not provide archives with reduced spectra. 

\section{Data reduction and homogenization}
\label{sct:data_reduction}
We downloaded data from various instruments and archives and processed the spectra as homogeneously as possible. The data were first corrected for sky contamination (Sect. 3.1), then numerous spectra of the same star were coadded (after applying heliocentric corrections; Sect. 3.2), and finally the spectra were radial velocity corrected (Sect. 3.3). Furthermore, sample stars were cross-correlated between various instruments, and the abundances were brought to the same solar abundances scale \citep{Asplund2009}. Differences in the final data reduction steps can propagate through to the abundance analysis and cause abundance differences as often seen in the literature. This we strive to prevent. However, we are limited by the observed spectral ranges and the limitations of the used methods and codes (see Sect.~\ref{sct:stellar_pars}, Sect.~\ref{sct:abundance_analysis}, and Appendix~\ref{appendix:effective_temp}).

\subsection{Sky subtraction}
Sky contamination (both, emission and absorption) influences the continuum level of the spectra and in turn the strength of absorption lines. This affects all measured chemical abundances and has to be corrected for prior to the analysis. We obtain sky emission- and absorption-subtracted spectra for UVES, HIRES, and X-shooter from the reduced archives. FLAMES/GIRAFFE spectra were corrected for sky emission and absorption, based on the sky-fiber data also included in the archive. We calculate the sky contamination as the median of all corresponding sky fibers, performing a sigma clipping to account for cosmic rays, and subtract the resulting median flux from all science spectra individually before coadding. 

The resulting median sky was multiplied by a factor $f$ before subtraction in order to study the impact on our derived stellar parameters. A small deviation around $f=1$, has only a minor impact on the temperature and metallicity (Fig.~\ref{fig:tests}).
For the final sky subtraction, we apply a factor of $f=1$ (i.e., a normal, nonscaled median) throughout this work.

\subsection{Coaddition}
We performed several tests for coadding the spectra from multiple observations, and we tried to use spectra with a signal-to-noise ratio, (S/N)$>10$. The resulting S/N has an impact on the precision of the abundances as well as the stellar parameters. We tested the average, square weighted average, median, and weighted median to coadd spectra. The best S/N was obtained by coadding the spectra using the square weighted average:
 \begin{equation}
     F = \frac{\sum \left( \mathrm{S/N}_i^2 \cdot F_i \right)}{\sum \mathrm{S/N}_i ^2},
 \end{equation}
with $F$ being the resulting coadded flux, and  $F_i$ the flux of the individual spectra, $i$. A large amount of noise can introduce higher uncertainties in our measurements. This is illustrated in the middle panel of Fig.~\ref{fig:tests} where we injected noise into the spectrum of a reference star (HD26297) and determined the stellar parameters. 
    \begin{figure}
   
   \centering
   \includegraphics[width=\hsize]{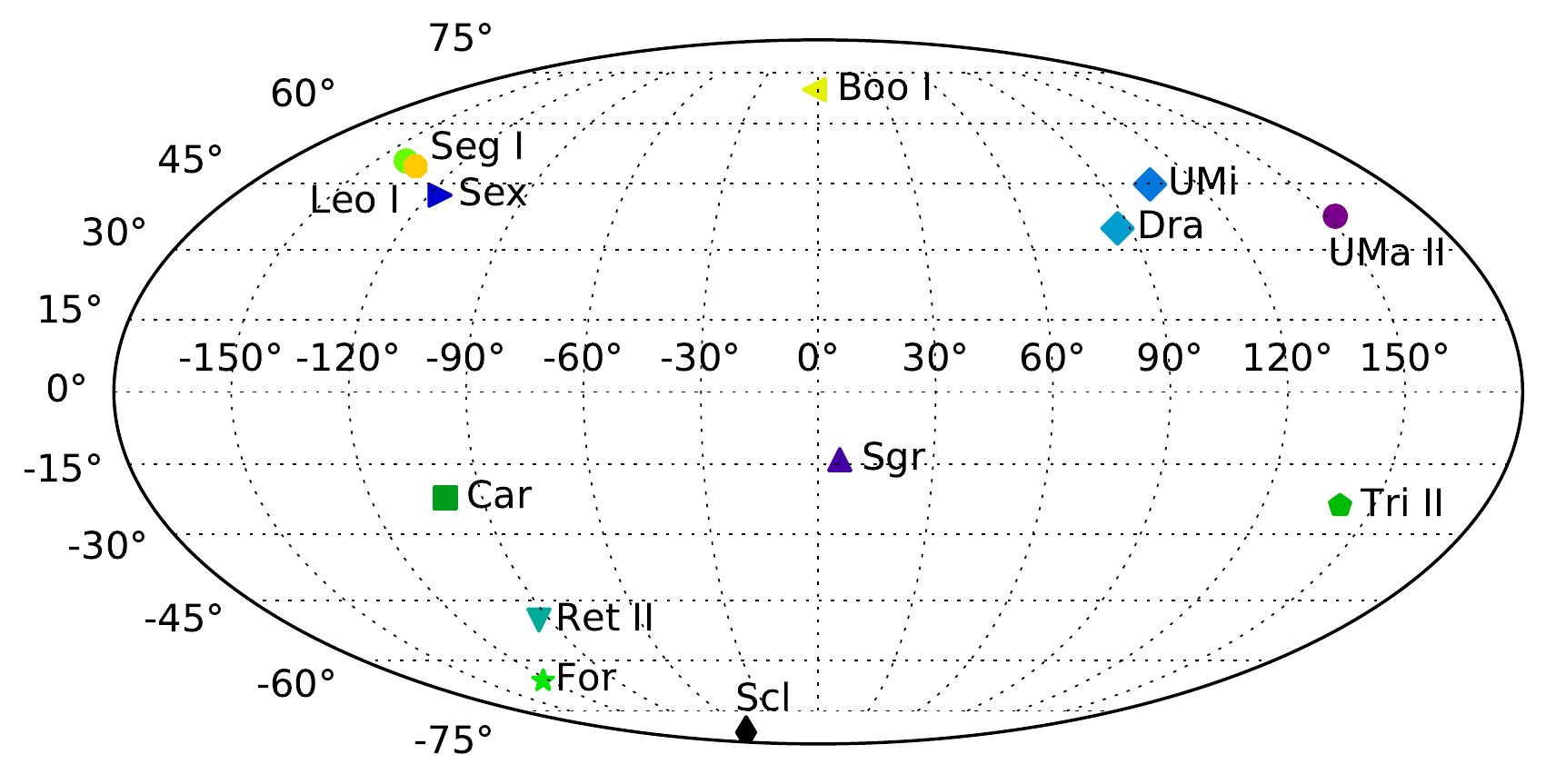}
      \caption{Locations of investigated dSph and UFD galaxies in Galactic coordinates. The different colors and symbols that are used throughout this work indicate the individual dSph galaxies.}
        \label{fig:loc_dsph}
   \end{figure}
   
 \begin{figure*}
   \centering
    \includegraphics[width=\hsize/3]{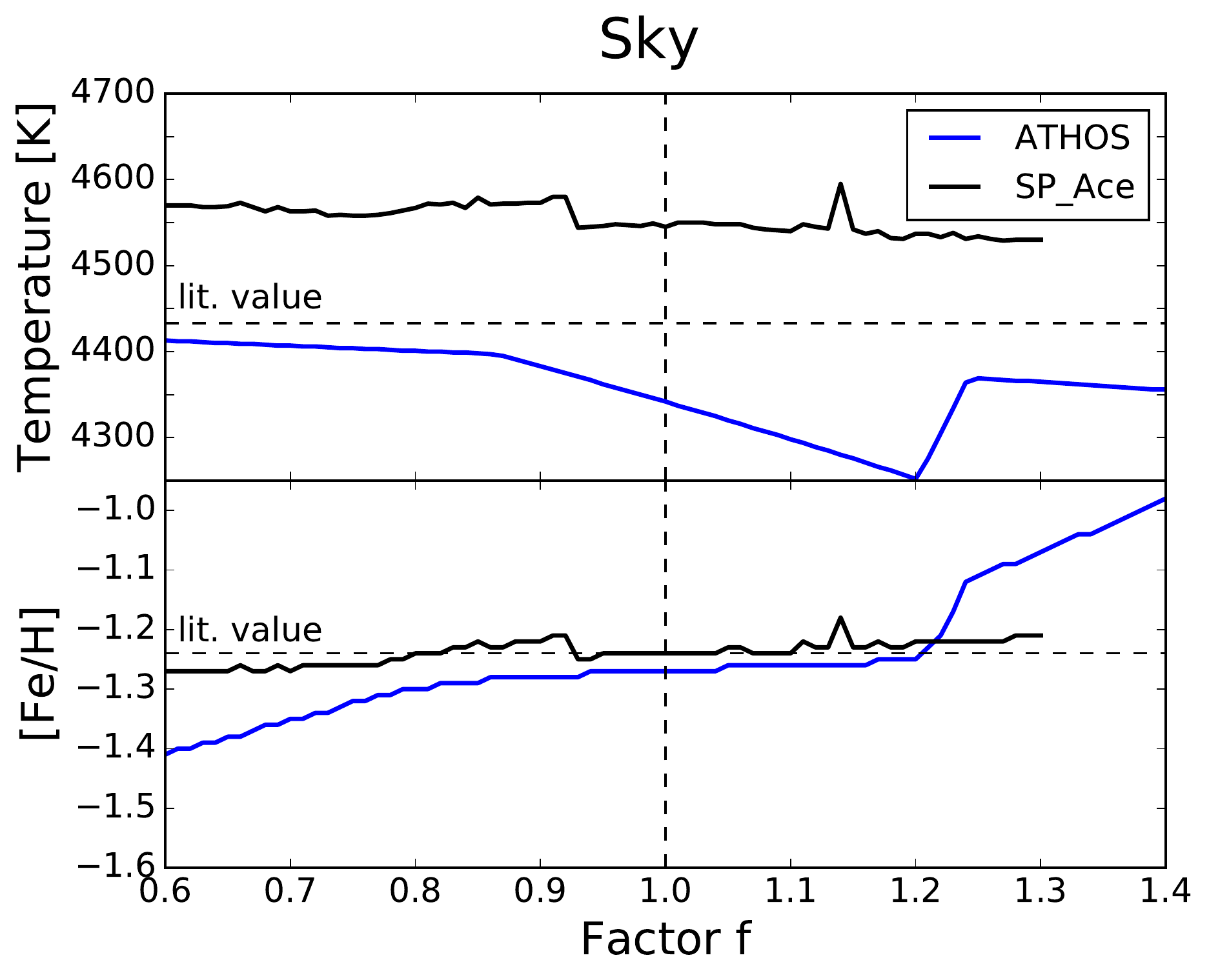}%
   \includegraphics[width=\hsize/3]{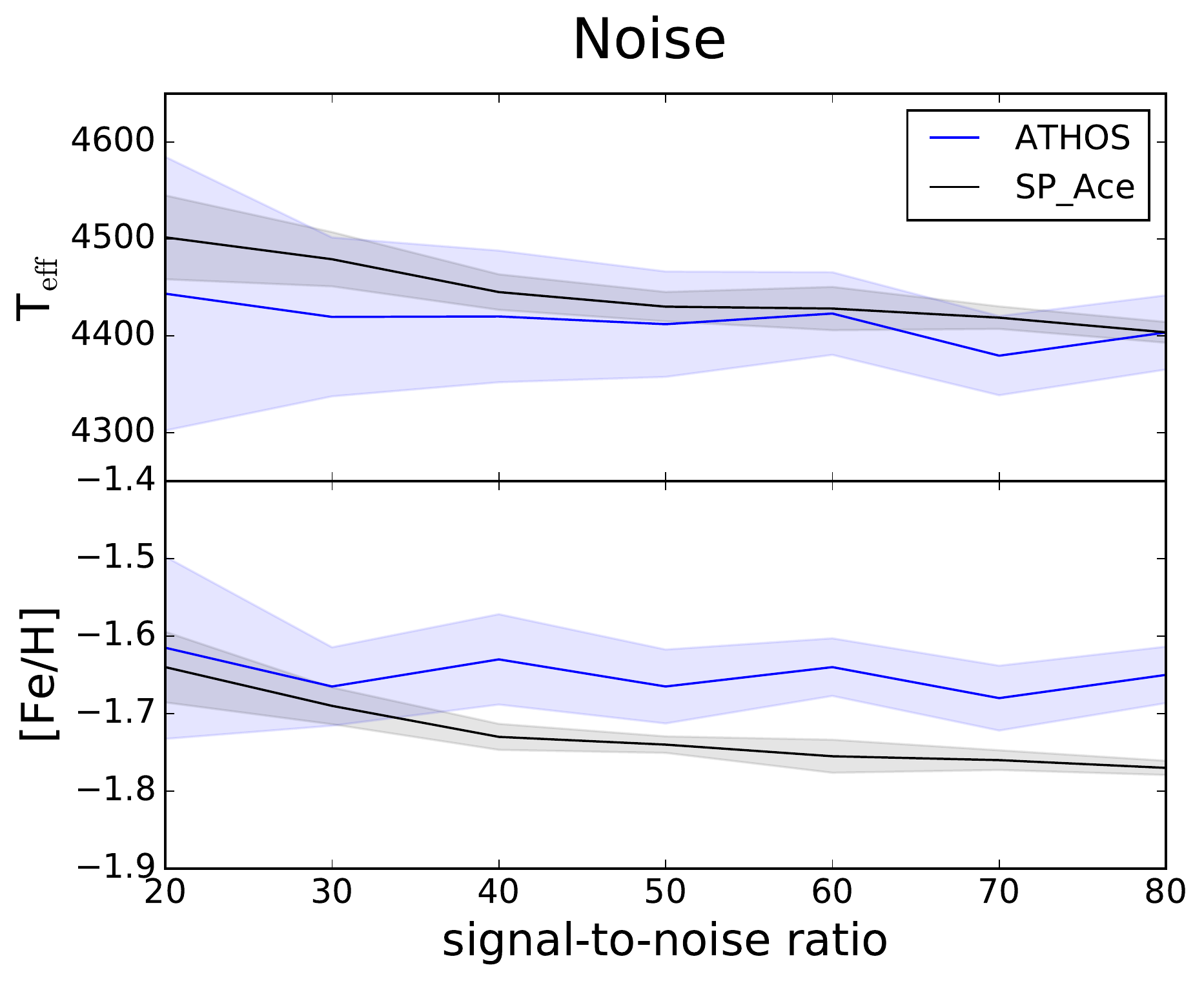}%
   \includegraphics[width=\hsize/3]{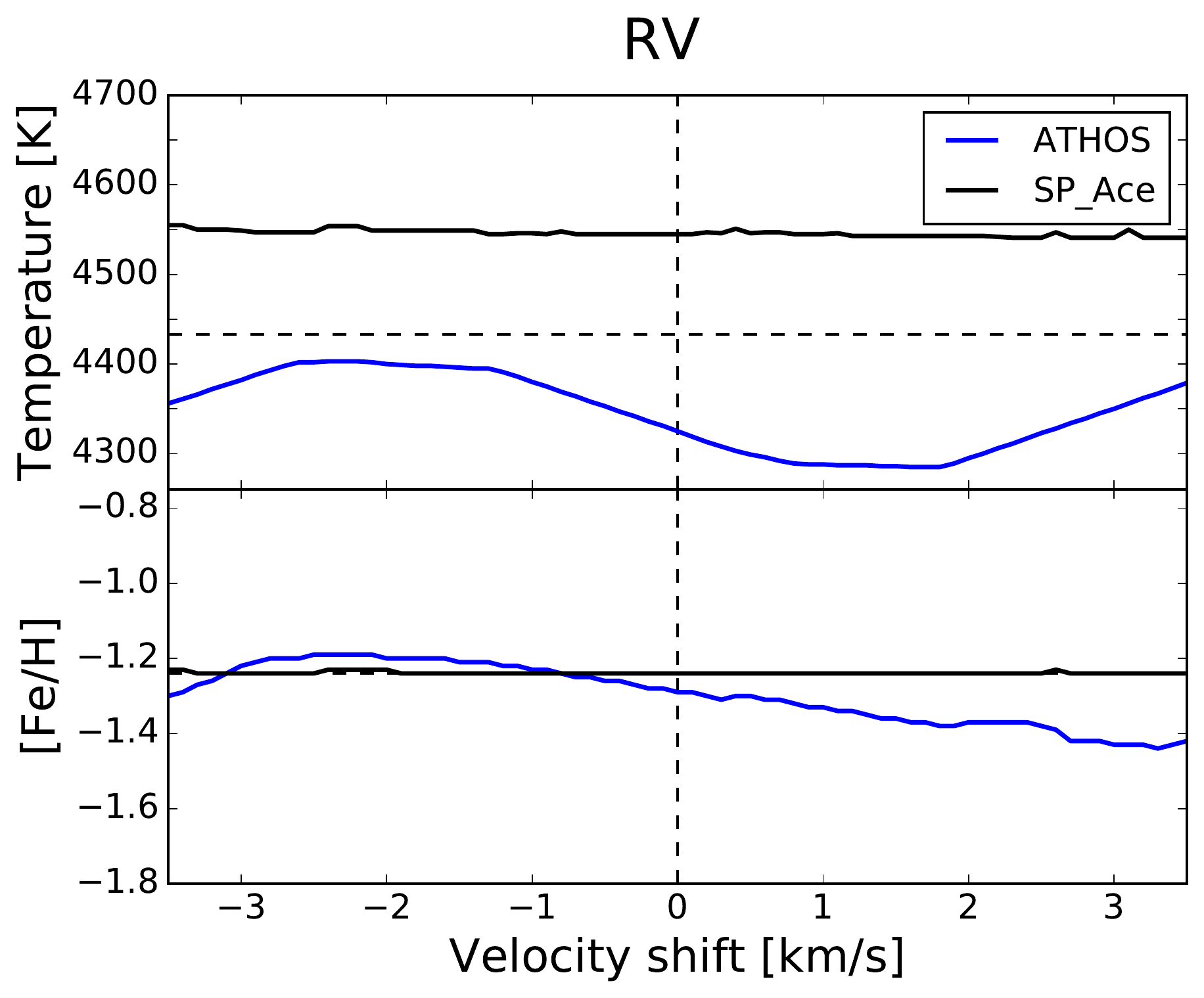}%
      \caption{Left panel: Impact of sky subtraction for \object{2MASS J00595966-3339319} on temperature and metallicity. The dashed line indicates the literature value of \citet{Skuladottir2015}. Middle panel: Impact of noise on the stellar parameters for the benchmark star HD26297. The shaded bands indicate a one sigma spread after 20 repetitions of analysis. The signal-to-noise ratio (S/N) is given per resolution element. Right panel: Uncertainty due to different radial velocity shifts for star \object{2MASS J00595966-3339319}.}
        \label{fig:tests}
   \end{figure*}

\subsection{Velocity shift}
\label{sct:vel_shift}
The radial velocity of a star makes it necessary to correct the spectrum for the Doppler effect. Figure~\ref{fig:tests} (right panel) shows the impact of different velocity shifts on temperature and metallicity. 
For a precise determination of the radial velocity, we perform a cross-correlation on various synthetic templates. 
These templates include different wavelength ranges and stellar parameters. The choice of the template for each star relies on literature values that we obtained from SIMBAD. For our sample, a typical error of the radial velocity is $0.8 \, \rm km\, s^{-1}$.

\begin{table}
\caption{
Number of stars per galaxy in our sample with S/N at $5528\, \rm \AA$ per resolution element. Total stellar masses of the target dwarf galaxies are from \citet{McConnachie2012}.}             
\label{tab:sample}      
\centering                          
\begin{tabular}{llrrr}        
\hline\hline                 
Galaxy & Abbreviation  & $\mathrm{M}_*$ [$10^6 M_\odot$]& S/N &\# Stars  \\
\hline                        
Sagittarius&Sgr &21&141& 32 \\
Fornax & For& 20& 79 &123 \\
Leo I& Leo I& 5.5&52&2  \\
Sculptor&Scl& 2.3 &93& 96 \\
Sextans&Sex& 0.44 &86& 46 \\
Carina&Car&0.38 &47& 47 \\
Ursa Minor& UMi&0.29&88& 13  \\
Draco & Dra &0.29&92& 7 \\
Bootes I& Boo I&0.029&38 & 3 \\
Ursa Major II& UMa II&0.0041&98 & 3 \\
Segue I& Seg I & 0.00034&300 &1 \\
Reticulum II& Ret II&- &47& 6 \\
Triangulum II& Tri II &-&173& 1 \\
Total & & & & 380\\
\hline                                   
\end{tabular}
\end{table}

\subsection{Solar abundances}
The choice of solar abundances is not uniform across the literature, and, depending on the element, this causes offsets on the order of $\Delta\mathrm{[X/Fe]}$\footnote{$\mathrm{[X/Fe]}=\log{\left(N_X/N_\mathrm{Fe}\right)}-\log{\left( N_X/N_\mathrm{Fe}\right)}_\odot$}$=0.2\, \rm dex$ as shown in Fig.~\ref{fig:solarabu}. We use solar abundances from \citet{Asplund2009} throughout this work and whenever comparing to the literature, we correct the abundances to the same scale. If the adopted solar abundance scale is not known, we do not attempt to correct for it.

\section{Stellar parameters}
\label{sct:stellar_pars}
The stellar parameters, that is to say, effective temperature ($T_\mathrm{eff}$), surface gravity ($\log g$), and metallicity ($\mathrm{[Fe/H]}$), 
determine the structure of a star. They are therefore key input when deriving chemical abundances.
To determine the stellar parameters, we use the two automated codes SP\_Ace \citep{Boeche2016} and ATHOS \citep{Hanke2018}.
We adopt surface gravities derived from the distance to each galaxy and microturbulences as derived through Eq.~\ref{eq:micro}. As outlined above, the temperature and metallicities tend to rely on measuring equivalent widths, which must be automatized in samples as big as our. However, most automated codes are somewhat restricted in the parameter space they can be applied to. Hence, effective temperatures and metallicities are determined with SP\_Ace for stars with $\mathrm{[Fe/H]}>-2.3$ and ATHOS for our most metal-poor stars ($-4.18 \le \mathrm{[Fe/H]}\le -1.91$). There is a slight overlap in metallicity of $10$ stars, where SP\_Ace returned a lower metallicity and was therefore out of its range of validity. 
We use the Kurucz atmosphere models ODFNEW for $\mathrm{[Fe/H]} \ge -1.5$ and the atmosphere model with enhanced alpha optical potentials AODFNEW for $\mathrm{[Fe/H]}<-1.5$ \citep{Castelli2003}. All stellar parameters are given online in electronic form in Table~o1.
The details of this choice are discussed in the following sections.

 Small changes in the stellar parameters can have a strong impact on the derived abundances. We thus derive these parameters in several different ways to test this and to explore offsets with respect to the literature. Table~\ref{tab:sensitivity_abundances} exemplarily shows the influence on the abundance of a star when varying stellar parameters. 
A homogeneous method to derive the stellar parameters is therefore crucial to reveal internally unbiased information for the investigated galaxies. In addition, the large sample size creates a need to automatize the extraction of stellar parameters. Most studies with a sample size $\gtrsim 30$ use the automated code DAOSPEC \citep{Stetson2008} to measure equivalent widths. These equivalent widths are then used to determine the stellar parameters (mostly only [Fe/H]) and abundances. However, DAOSPEC requires at least $\mathrm{S/N} > 30$, which is not available for the majority of the stars in our sample. Therefore, we use ATHOS (Sect. 4.1) and SP\_Ace (Sect. 4.2).

\begin{table}
\caption{Abundance sensitivity assuming a change $\Delta$ in the stellar parameters for a representative virtual star with $T_\mathrm{eff}=4400\,\rm K$, $\log g =1$, $\mathrm{[Fe/H]}=-1.5$ and $\xi _t = 1.8\, \rm km\, s^{-1}$. Uncertainties added in quadrature are listed in $\sigma_{\rm tot}$.}             
\label{tab:sensitivity_abundances}      
\centering                          
\begin{tabular}{l c c c c c}        
\hline\hline                 
Element & $\Delta T_\mathrm{eff}$ & $\Delta \log g$ & $\Delta \mathrm{[M/H]}$ & $\Delta \xi_t$&$\sigma_\mathrm{tot}$ \\    
    & $140\, \rm K$ & $0.25\, \rm dex$ & $0.2\, \rm dex$ & $0.15\, \rm km\, s^{-1}$ & dex\\ 
\hline                        
 $\Delta \log \epsilon (\mathrm{Mg})$ & $0.13$ & $0.07$ & $0.06$ & $0.03$& $0.16$\\     
 $\Delta \log \epsilon (\mathrm{Sc})$ & $0.02$ & $0.11$ & $0.07$ & $0.01$& $0.13$\\   
 $\Delta \log \epsilon (\mathrm{Ti})$ & $0.03$ & $0.14$ & $0.05$ & $0.12$& $0.19$\\  
 $\Delta \log \epsilon (\mathrm{Cr})$ & $0.30$ & $0.05$ & $0.05$ & $0.10$& $0.32$\\ 
 $\Delta \log \epsilon (\mathrm{Mn})$ & $0.22$ & $0.03$ & $0.03$ & $0.03$& $0.23$\\ 
 $\Delta \log \epsilon (\mathrm{Ni})$ & $0.19$ & $0.01$ & $0.01$ & $0.06$& $0.20$\\ 
 $\Delta \log \epsilon (\mathrm{Zn})$ & $0.06$ & $0.07$ & $0.03$ & $0.05$& $0.11$\\ 
 $\Delta \log \epsilon (\mathrm{Sr})$ & $0.04$ & $0.04$ & $0.01$ & $0.16$& $0.17$\\ 
 $\Delta \log \epsilon (\mathrm{Y})$  & $0.01$ & $0.09$ & $0.05$ & $0.02$& $0.11$\\ 
 $\Delta \log \epsilon (\mathrm{Ba})$ & $0.03$ & $0.10$ & $0.05$ & $0.07$& $0.14$\\  
 $\Delta \log \epsilon (\mathrm{Eu})$ & $0.08$ & $0.05$ & $0.06$ & $0.07$& $0.13$\\  
\hline                                   
\end{tabular}
\end{table}

\subsection{ATHOS \citep{Hanke2018}}
ATHOS \citep[A Tool for Homogenizing Stellar parameters, ][]{Hanke2018} is an automated code that uses flux ratios to determine stellar parameters\footnote{Code version: https://github.com/mihanke/athos, accessed on 6 December 2018}. This gives the advantage of being independent of normalization. The effective temperature is determined employing flux ratios in the spectral region around the Balmer lines H$\alpha$ \mbox{($\lambda_{\mathrm{H}\alpha}=6562.78 \, \AA$)} and H$\beta$ \mbox{($\lambda_{\mathrm{H}\beta}=4861.32 \, \AA$)}. The metallicity, [Fe/H], is measured by using flux ratios of up to 31 Fe lines. 
Finally, log $g$ is measured, using 11 gravity sensitive regions. 
\begin{table}[htbp]
\caption{Number of covered flux ratios of ATHOS for different setups and number of stars measured with the setups. HR denotes the high-resolution mode of FLAMES/GIRAFFE.}
\centering
\begin{tabular}{lrrrrr}
\hline  
Setting & $\lambda _\mathrm{min}\, \mathrm{[\AA]}$ & $\lambda _\mathrm{max}\, \mathrm{[\AA]}$ & $T_\mathrm{eff}$ & Fe & $\log g$\\
\hline
HR7A & $4700$ & $4974$ & 5 & 2  & 1 \\
HR10 & $5340$ & $5620$ & 0 & 11 & 1 \\
HR13 & $6120$ & $6400$ & 0 & 0  & 0 \\
HR14A& $6390$ & $6620$ & 4 & 2  & 0 \\
HR15 & $6610$ & $6960$ & 0 & 0  & 0 \\
UVES & $4800$ & $6800$ & 9 & 31 & 11\\
\hline
\end{tabular}
\label{tab:setups_coverage}
\end{table}
\begin{table*}
\caption{Range of validity for our different automated methods used to derive stellar parameters.}        
\label{tab:setups_validity_stell_pars}      
\centering                          
\begin{tabular}{ccccccccc}        
\hline\hline                 
Method & $T_\mathrm{eff, min}$ [K]& $T_\mathrm{eff, max}$ [K]&$\log g_\mathrm{min}$&$\log g_\mathrm{max}$& $[\mathrm{Fe}/\mathrm{H}]_\mathrm{min}$& $[\mathrm{Fe}/\mathrm{H}]_\mathrm{max}$\\
\hline  
ATHOS & $4000$ & $6500$ & $1.0$ & $5.0$ & $-4.5$ & $0.3$ \\
SP\_Ace & $3600$ & $7400$ & $0.2$ & $5.4$ & $-2.4$ & $0.4$ \\
\hline   
\end{tabular}
\end{table*}
Owing to our limited spectral coverage, the number of flux ratios is reduced (see Table~\ref{tab:setups_coverage}).
ATHOS is trained on a sample of high-resolution and high S/N spectra of stars that span a range of  stellar parameters given in Table~\ref{tab:setups_validity_stell_pars}. Telluric absorption lines are automatically masked for each star by providing the radial velocity.
Furthermore, the spectra have to be exactly at rest frame (velocity corrected) and the continuum sky emission must have been properly subtracted from the spectrum (for consequences, see Fig.~\ref{fig:tests}). 
For our sample, ATHOS operates at or even out of the bounds of which it was trained (see Table~\ref{tab:setups_validity_stell_pars} and Fig.~\ref{fig:logg_comparison}). In total, 22 out of 380 stars lay outside the calibrated regions of ATHOS. Some galaxies suffer more than others from the limitation in the trained stellar parameter ranges. In each dSph galaxy this accounts for 1 out of 32 stars in Sagittarius, 4 out of 123 stars in Fornax, 5 out of 96 stars in Sculptor, 1 out of 46 in Sextans, 2 out of 47 stars in Carina, 5 out of 13 stars in Ursa Minor, and 4 out of 7 stars in Draco. For the most metal-poor stars in our sample ($\mathrm{[Fe/H]<-3}$) ATHOS is used outside the trained parameter range in 8 out of 21 metal-poor stars. However, these metal-poor stars do not show an obvious offset of the stellar parameters in comparison with literature values. Based on this we deem the possible parameter offset due to limited training ranges minor.

\subsection{SP\_Ace \citep{Boeche2016}}
SP\_Ace \citep[Stellar Parameters And Chemical abundances Estimator, ][]{Boeche2016}\footnote{Code version 1.3: http://dc.zah.uni-heidelberg.de/sp\_ace/q/dist/static/, accessed on 9 March 2019} is an automated code that computes stellar parameters as well as abundances. It needs normalized and radial velocity corrected spectra (shifted to an accuracy of the full width at half maximum, FWHM, of the absorption lines). Overall, for stars that we processed with SP\_Ace, the determined radial velocity of SP\_Ace scatters with a standard deviation of $\sigma = 0.8\, \rm km\, s^{-1}$ compared to the templates that we applied (Sect.~\ref{sct:vel_shift}). SP\_Ace starts by computing equivalent widths, which it converts via a curve of growth into a first abundance estimate. In a precomputed library of synthetic spectra, it finds a best fit, which it matches to the observed spectra. This is an iterative process that is repeated until a satisfactory match has been found. In this way, abundances of up to 30 elements can be estimated. 

The code is limited to dealing with $32000$ pixels owing to the usage of $16-$bit integers. Spectra obtained with UVES or HIRES exceed this limit and we therefore re-bin all spectra to this maximum number of pixels. To test a possible resolving-power dependence, we down-sampled the resolution of the star HD26297 to a typical resolution of UVES and FLAMES/GIRAFFE HR10, HR13, HR14A and HR15. For these different resolutions, the stellar parameters determined by SP\_Ace agreed within $\sim 30\, \mathrm{K}$ and $0.04\, \rm dex$ for the effective temperature and metallicity, respectively.

\subsection{Effective temperature}
Because red giant stars in nearby dSph galaxies are relatively faint ({\it V} $\gtrsim 15\, \rm mag$), most of the spectra have relatively low S/N (as shown in Table~\ref{tab:sample}). Therefore, the majority of literature studies relies on photometrically determined temperatures, while high-resolution or high S/N studies tend to rely on spectra, where the effective temperatures are determined from excitation equilibrium. 

Depending on metallicity, we use ATHOS and SP\_Ace, which derive effective temperatures differently. ATHOS relies on flux ratios, whereas SP\_Ace fits an entire spectral region. The derived temperatures from each code are compared to the literature values in Table~\ref{tab:temp_lit_comp}. Furthermore, we determined photometric temperatures based on the calibration of \citet{Alonso1999,Alonso2001}. We transformed the colors according to \citet{Bessell1979} as well as \citet{Alonso1998} and applied a dereddening according to \citet{Schlafly2011} using values from the IRSA Dust Database\footnote{https://irsa.ipac.caltech.edu/applications/DUST/, accessed on 13 March 2019}. In contrast to most of the literature studies, we use an individual extinction for every star rather than relying on one value per galaxy. However, most galaxies are located far from the Galactic plane (Fig.~\ref{fig:loc_dsph}) and the reddening correction is therefore small. This results in a negligible effect on the temperature (for $T_\mathrm{eff}(B-V)$ in, e.g., Sculptor the difference is $\sim 5 \, \mathrm{K}$) caused by the variation in the local vs. global dereddening. The photometry was taken from various literature studies based on the SIMBAD Database \citep{Simbad}. The spread caused by using different methods is $\sim 150 \, \rm K$, which is the average error we expect on the temperature. Photometric temperatures are on average lower than derived spectroscopic temperatures (see Table~\ref{tab:temp_lit_comp}). 

ATHOS has a higher spread compared to SP\_Ace due to a higher sensitivity to velocity shifts, sky subtraction, and S/N, which may arise owing to the reduced number of covered flux ratios. However, the stars may also be out of the trained parameter range \citep[cf. Fig.~2 of][ and Table~\ref{tab:setups_validity_stell_pars}]{Hanke2018}. Table~\ref{tab:temp_lit_comp} gives the average residual temperature $\langle \Delta T_\mathrm{eff} \rangle$ of SP\_Ace and ATHOS compared to the literature values. In addition, we list the standard deviation of this residual, $\sigma (\Delta T_\mathrm{eff}) $. Some stars show a large discrepancy between our methods caused by noisy spectra around the Balmer lines or in a few cases strong stellar winds as in the Sculptor star \object{2MASS J00592830-3342073}.
\begin{figure*}[h!]
   \centering
   \includegraphics[width=\hsize]{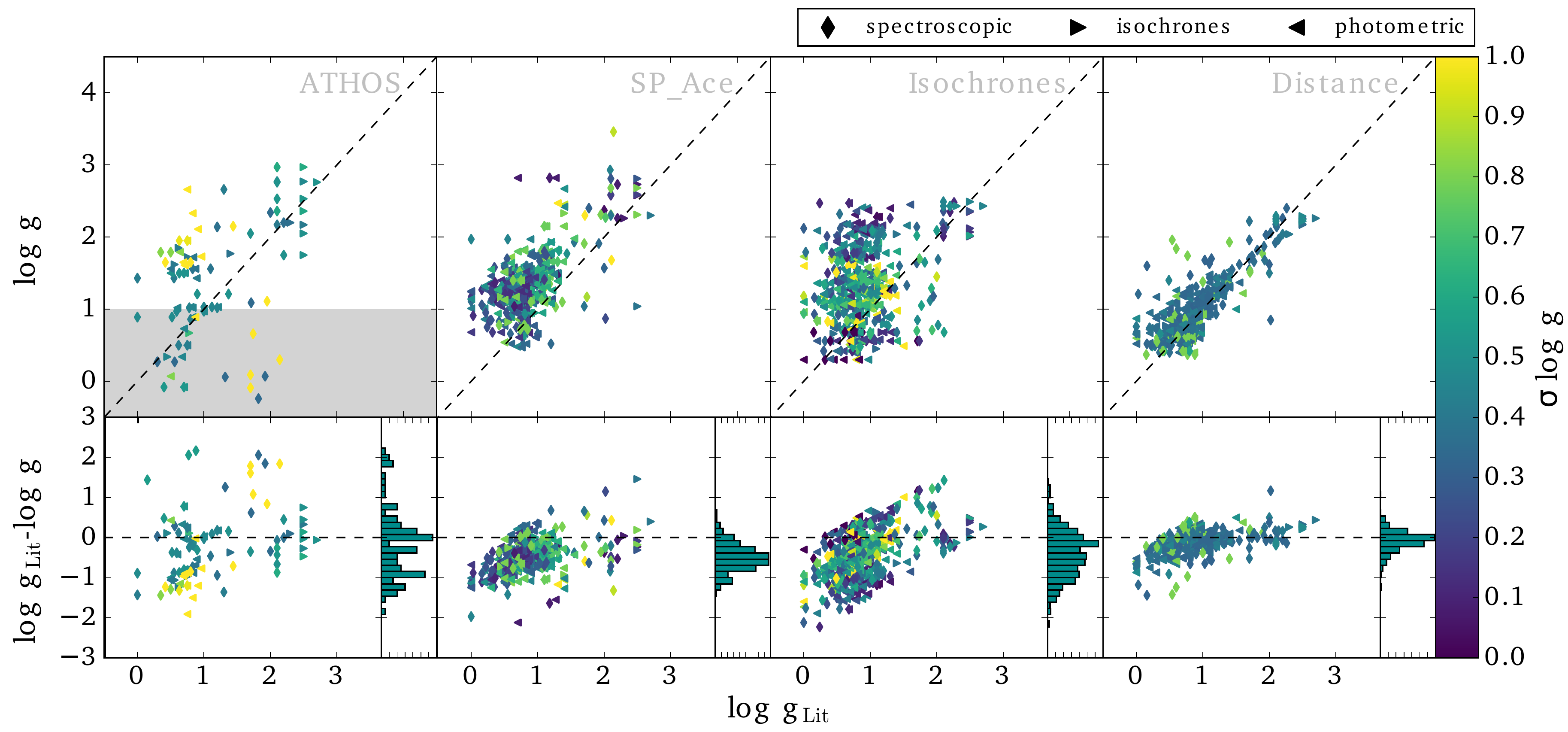}
      \caption{Comparison between literature surface gravities and our calculated values. The upper panels show the surface gravity calculated using different methods versus the literature surface gravity. Symbols indicate the method that was used to derive the literature value and colors indicate the error of the here calculated value. Lower panels indicate the difference between the here derived values and literature values together with a histogram. Shaded areas indicate regions where ATHOS is out of the trained range. We adopt surface gravities shown in the upper right panel based on distances to the respective galaxy.}
         \label{fig:logg_comparison}
   \end{figure*}

Owing to the higher sensitivity to velocity shifts, reduced number of covered flux ratios, low S/N, and sky subtraction of ATHOS, we decided to adopt stellar parameters including effective temperatures determined by SP\_Ace for all stars within its valid metallicity range respecting the boundary ($\mathrm{[Fe/H]} >-2.3$). The temperatures of the more metal-poor stars were determined with ATHOS. The choice of using spectroscopically determined effective temperatures was made as the different photometric temperatures typically have an offset of $\sim 100\, \mathrm{K}$ (tested on $43$ stars in Sculptor, see also Fig.~\ref{fig:temperature_comp}). A similar discrepancy between photometric temperatures of different colors was also noted by \citet{Letarte2010}, \citet{Hansen2012} and \citet{Hill2019}. In the study of \citet{Letarte2010}, constant offsets of $\sim 100\, \mathrm{K}$ are applied to their adopted $T_\mathrm{eff}(V-\{J,H,K_s\})$ to account for systematics between the colors. In \citet{Hill2019}, this discrepancy is on average $\sim 200\, \rm K$ for $T_\mathrm{eff}(V-I)$ compared to the photometric temperatures of other colors. In addition, our analysis revealed differences in temperature, depending on the photometric sources. \citet{Hill2019} rely on the photometry of \cite{Battaglia2008a} for $V$ and $I$ and that of \citet{Babusiaux2005} for $J$ and $K_s$, whereas \citet{Kirby2010} rely on the photometry of \citet{Westfall2006} for Sculptor. These various photometric calibrations easily lead to differences on the order of $\sim 200\, \rm K$ for $T_\mathrm{eff}(V-I)$. In total, we determined $T_\mathrm{eff}(B-V)$, $T_\mathrm{eff}(V-K_s)$, and $T_\mathrm{eff}(V-I)$ for $145$, $207$, and $233$ dwarf galaxy stars, respectively. Not all colors are available for all stars, and we therefore decided to use spectroscopic temperatures to avoid these systematic offsets between colors and photometric sources. This increases the homogeneity of our sample, with the disadvantage of slightly larger uncertainties.
\begin{table*}[htbp]
\caption{Average offset and spread of adopted effective temperatures compared to the literature values, where $\Delta T_\mathrm{eff} = T_\mathrm{eff,lit}-T_\mathrm{eff,SP\_Ace/ATHOS}$. ``Method $T_\mathrm{lit}$'' indicates the method used in the literature to determine the effective temperature: photometrically (Phot), spectoscopically (Spec), or with fits over spectral regions (Fit).}\label{tab:temp_lit_comp}
\centering
\resizebox{2\columnwidth}{!}{
\begin{tabular}{lclrrrrrr} 
\hline\hline             
Study & Galaxy & Method $T_\mathrm{lit}$& $\langle \Delta T_\mathrm{eff} \rangle _\mathrm{ATHOS}$&$\sigma( \Delta T_\mathrm{eff} )_\mathrm{ATHOS}$&  \# $\mathrm{ATHOS}$ &  $\langle \Delta T_\mathrm{eff} \rangle _\mathrm{SP\_Ace}$&$\sigma( \Delta T_\mathrm{eff} )_\mathrm{SP\_Ace}$ & \# $\mathrm{SP\_Ace}$\\
\hline
\citet{Letarte2010}      & For & Phot      & $261$     & -      & $1$    & $-99$  & $51$  & $72$ \\
\citet{Kirby2010}        & For & Fit/Spec  & -         & -      & -      & $-87$  & $65$  & $16$ \\
\citet{Kirby2013}        & For & Fit/Spec  & -         & -      & -      & $-217$ & $73$  & $20$\\
\citet{Lemasle2014}      & For & Phot      & $68$      & $56$   & $2$    & $-235$ & $66$  & $36$\\
\citet{Skuladottir2015}  & Scl & Phot      & $-57$     & $163$  & $7$    & $-157$ & $82$  & $72$ \\
\citet{Hill2019}         & Scl & Phot      & $-57$     & $163$  & $7$    & $-158$ & $79$  & $71$ \\
\citet{Geisler2005}      & Scl & Phot      & -         & -      & -      & $-191$ & $74$  & $2$ \\
\citet{Simon2015}        & Scl & Spec      & $3$       & -      & $1$    & -      & -     & - \\
\citet{Kirby2010}        & Scl & Fit/Spec  & $-45$     & -      & $1$    & $-100$ & $87$  & $6$\\
\citet{Kirby2013}        & Scl & Fit/Spec  & $-41$     & $80$   & $4$    & $-100$ & $226$ & $44$ \\
\citet{Kirby2010}        & Sex & Fit/Spec  & -         & -      & -      & $-172$ & -     & $1$ \\
\citet{Kirby2013}        & Sex & Fit/Spec  & $205$     & $249$  & $7$    & $58$   & $251$ & $18$\\
\citet{Shetrone2001}     & Sex & Spec      & -         & -      & -      & $-172$ & -     & $1$ \\
\citet{Sbordone2007}     & Sgr & Phot      & -         & -      & -      & $-96$  & $114$ & $12$ \\
\citet{Hansen2018}       & Sgr & Spec      & $-89$     & $367$  & $2$    & $-130$ & $81$ & $11$\\
\citet{Bonifacio2004}    & Sgr & Phot      & -         & -      & -      & $26$   & $66$ & $9$\\
\citet{Koch2008}         & Car & Phot      & -         & -      & -      & $-167$ & -     & $1$\\
\citet{Norris2017}       & Car & Phot      & $-50$     & -      & $1$    & $-61$  & $115$ & $9$\\
\citet{Fabrizio2012}     & Car & Phot      & $-110$    & -      & $1$    & $-78$  & $119$ & $17$\\
\citet{Lemasle2012}      & Car & Phot      & $174$     & $341$  & $4$    & $-112$ & $150$ & $29$\\
\citet{Ural2015}         & UMi & Spec      & $129$     & -      & $1$    & $-23$  & $21$  & $2$  \\
All studies              & -   & -         & $132$     & $260$  & $62$   & $-119$ & $137$ & $462$  \\
\hline
\end{tabular}
}
\end{table*}

\subsection{Surface gravity}  

Due to a large variety of methods present in the literature when determining the surface gravity, we compare the following different approaches:\\
\\
\textbf{ATHOS}: This code calculates surface gravities based on up to $11$ flux ratios involving \ion{Fe}{II} lines. We compared the ATHOS results to  the literature in Fig.~\ref{fig:logg_comparison} for all stars in our sample, which contain more than eight flux ratios. The error on the surface gravity is split into a systematic error that is set to a constant value of $\Delta \log g = 0.36$ and a statistical error based on the scatter of the individual flux ratios. The derived surface gravity is partly based on a training sample where ionization equilibrium was enforced. 
Applying ionization balance in LTE can lead to a lower $\log g$ (for our sample on average $\sim 0.3 \rm \, dex$) owing to the lower \ion{Fe}{I} abundance derived under this assumption \citep{Lind2012}. 

The majority of our sample consists of spectra from the FLAMES/GIRAFFE HR10 setting, which only covers one of these flux ratios (see Table~\ref{tab:setups_coverage} and Fig.~\ref{fig:wavcover}). In addition, most of the spectra have low S/N (Fig.~\ref{tab:sample}) leading to an uncertain surface gravity and some stars are out of the valid stellar parameter range (cf. Table~\ref{tab:setups_validity_stell_pars}). 
Hence, we did not adopt this method to determine surface gravities in our study due to the large uncertainty. 
As indicated in Fig.~\ref{fig:logg_comparison}, most literature studies are based on photometric surface gravities. Therefore, the residual depicted in the lower panel of Fig.~\ref{fig:logg_comparison} has a slight offset. 

\textbf{SP\_Ace}: Similar to \citet{Kirby2010,Kirby2013}, SP\_Ace creates a synthetic spectrum and compares this to a large wavelength range of the spectrum ($5212-6860 \, \AA$). The derived surface gravity is therefore based on ionization balance. In comparison to ATHOS, SP\_Ace relies on more lines. The surface gravity is systematically offset by $\sim 0.3 \, \rm dex$ compared to the literature values, but has in contrast to ATHOS a lower scatter of $\sigma \left( \Delta \log g \right) \sim 0.3\, \rm dex$.\\
\newline
\textbf{Isochrones}: Surface gravities can be obtained by the position in the Hertzsprung-Russel-diagram (HR-diagram) and knowing the age of the star. We determined surface gravities using Yonsei-Yale (YY) isochrones \citep{Demarque2004} employing a nearest neighbor interpolation in a grid of previously calculated isochrones. We applied a constant age of $10\, \rm Gyr$, an enhanced $\alpha$-abundance, $[\alpha/\mathrm{Fe}] = 0.3$, and assumed that our sample consists of giants only. The error was determined by applying an age error of $\Delta t_\mathrm{Age} = 1\, \rm Gyr$ including the errors from the derived temperatures and metallicities. Compared to the literature values, the surface gravity spreads around $\sim 1\, \rm dex$, with the majority of the stars showing higher surface gravities. \\
\newline
\textbf{Distance}: Due to the low S/N in the spectra of these faint, distant stars, most studies use surface gravities determined by the distance to the galaxy (see Fig.~\ref{fig:logg_comparison}).
The distance can be used to determine the surface gravity by the relation:
\begin{equation}
\log g_* = \log g_{\odot} +\log \frac{M_*}{M_\odot} + 4\, \log \frac{T_{\text{eff},*}}{T_{\text{eff},\odot}}+0.4\left( M_{\text{bol},*} - M_{\text{bol},\odot} \right),
\end{equation}
with $\log g_{\odot}=4.44$, $T_{\text{eff},\odot}=5780\, \rm K$, $M_*=0.8\pm 0.2 M_\odot$ and $M_{\text{bol},\odot}=4.72\, \rm mag$. We apply the distance to the dwarf galaxy (Table~\ref{tab:distances_dwarf}) to derive bolometric magnitudes according to the calibration of \citet{Alonso1999}. Most of the distances given in Table~\ref{tab:distances_dwarf} were derived by using RR Lyr-type variable stars as standard candles. For stars where the $V$ magnitude is not known, we use {\it Gaia}'s broadband $G$ magnitude \citep{Gaia2016} and calibrations of \citet{Andrae2018}. 

We compared the surface gravity derived with the {\it G} and {\it V} magnitude for stars where both magnitudes are available. These surface gravities were in excellent agreement within $0.05\,\rm dex$. We note, however, that we did not apply any dereddening for the $G$ magnitudes, whereas we used values from IRSA Dust for correcting $V$ magnitudes. Compared to the literature values, assuming the distance to each galaxy to derive the surface gravities has the lowest deviation since most of these studies rely on the same method to derive surface gravities. For all stars that deviate by more than $\sim 0.5 \, \rm dex$ from our value, the surface gravity was derived by assuming ionization balance in the original publication (see Fig.~\ref{fig:logg_comparison}). In general, there is no offset compared to the literature values and the spread of values lies within $\sim 0.3 \, \rm dex$ caused by assuming different temperatures. The distance to our target galaxies is well known and the diameter of a galaxy is negligible compared to their heliocentric distances. As a consequence, the error on the surface gravity is relatively small. 
Therefore, we derive surface gravities from the distances to the respective galaxy throughout this work.
\begin{table}[htbp]
\caption{Distances used to derive absolute bolometric magnitudes.}
\centering
\begin{tabular}{lrr}
\hline  
Galaxy & Distance [kpc] & Reference  \\
\hline
Seg I & $23 \pm 2$ & \citet{Belokurov2007} \\
Sgr & $27 \pm 1$ & \citet{Hamanowicz2016}\\
Tri II&$30 \pm 2$ & \citet{Laevens2015}\\
Ret II & $32 \pm 2$ & \citet{Mutlu2018} \\
UMa II  & $35\pm 2$ & \citet{Dall2012}\\
Boo I & $66 \pm 2$ & \citet{Dall2006} \\
UMi   & $76 \pm 4$ & \citet{Bellazzini2002} \\
Dra  & $79\pm 6$ & \citet{Kinemuchi2008} \\
Scl    & $86 \pm 3$ & \citet{Pietrzynski2008}\\
Sex    & $86 \pm 6$ & \citet{Pietrzynski2008} \\
Car    & $106 \pm 6$ & \citet{Karczmarek2015} \\
For    & $146 \pm 8$ & \citet{Karczmarek2017} \\
Leo I & $269 \pm 12$ & \citet{Stetson2014} \\
\hline
\end{tabular}
\label{tab:distances_dwarf}
\end{table}

\subsection{Metallicity}
As for the temperature, we determined the metallicity via SP\_Ace and ATHOS. On average, ATHOS and Sp\_Ace provide values that are approximately $0.1 \, \rm dex$ more metal-rich than the literature values, with a standard deviation of $0.38 \, \rm dex$ and $0.28\, \rm dex$ for ATHOS and Sp\_Ace, respectively. This systematic difference can be explained by higher temperatures compared to the literature values. We analyzed the offset between ATHOS and Sp\_Ace for stars in Sculptor, where both codes are operating within their calibrated regions of stellar parameters (see Table~\ref{tab:setups_validity_stell_pars}). For these $38$ stars, both codes agree within a standard deviation of $\sigma=0.15$ and the average offset is $\mathrm{[Fe/H]_{ATHOS}}-\mathrm{[Fe/H]_{Sp\_Ace}}=0.09$. We want to stress that the fairly large scatter is mainly driven by large errors in metallicity obtained with ATHOS ($\sim 0.3 \, \rm dex$), which uses only a fraction of its flux ratios. This is caused by the narrow wavelength coverage of FLAMES/GIRAFFE spectra ($11$ out of $31$ for most stars, see Table~\ref{tab:setups_coverage}). At lower metallicities, most stars are observed with UVES or HIRES, which cover more flux ratios implemented in ATHOS. An average offset of $0.09\, \rm dex$ is the trade-off in homogeneity that we have to consider when using these two different methods. This value, however, is much lower than our error on the metallicity. We additionally note that ATHOS was trained on metallicities determined with MOOG and the equivalent width database of SP\_Ace relies on MOOG as well thus increasing the homogeneity slightly between the codes. We therefore do not attempt to reanalyze iron abundances but adopt the value determined by the corresponding code.  
We note that our metallicities are biased by not taking NLTE effects into account. Like most other studies we rely on \ion{Fe}{i} absorption lines as most strong \ion{Fe}{ii} lines are unfortunately outside the covered wavelength range.
According to \citet{Lind2012}, a typical star in our sample ($T=4500\, \rm K$, $\log g=1$, $\mathrm{Fe/H}=-1.5$) would have a higher abundance/metallicity in NLTE by $\sim 0.1\, \rm dex$. This correction may differ with varying stellar parameters and is in general stronger at lower metallicities. It can reach a maximum value of $0.4-0.5\, \mathrm{dex}$ adopting low surface gravities and metallicities  \citep[][]{Lind2012,Amarsi2016}.

\subsection{Microturbulence}
We use a formula to calculate the microturbulence from \citet{Kirby2009}:
\begin{equation}
\label{eq:micro}
\xi _\mathrm{t} = ((2.13\pm 0.05) - (0.23\pm 0.03)\cdot \log g) \, \rm km \, s^{-1}.
\end{equation}
\citet{Mashonkina2017} present another empirical formula that involves temperature as well as metallicity:
\begin{equation}
\label{eq:micro_mashonkina}
\xi _\mathrm{t} = (0.14 -0.08 \cdot \mathrm{[Fe/H]}+4.90\cdot (T_\mathrm{eff}/10^4) -0.47 \cdot \log g) \, \rm km \, s^{-1}.
\end{equation}
These calibrations yield results that differ by $\sim 0.3 \, \rm km\, s^{-1}$ for our sample stars. 
We use Eq.~\ref{eq:micro} for determining the microturbulence, however, their internal error seems unrealistically low in our sample. Hence, we calculate the error assuming Gaussian error propagation of Eq.~\ref{eq:micro_mashonkina} throughout this work. The adopted equation from \citet{Kirby2009} was derived based on stars with $4000 \, \rm K \lesssim T_\mathrm{eff} \lesssim 5500 \, \rm K$, $0.4 \lesssim \log g \lesssim 3.5$, and metallicities $-2.4 \lesssim \mathrm{[Fe/H]} \lesssim -1.4$. A similar formula was derived in \citet{Marino2008}, where the metallicity reaches $\sim -0.9 \, \rm dex$. Equation~\ref{eq:micro} has been used in \citet{Kirby2009,Kirby2010}, and \citet{Duggan2018} for dSph member stars in the same stellar parameter range. 
   \begin{figure*}[!ht]
   \centering
   \includegraphics[width=\hsize]{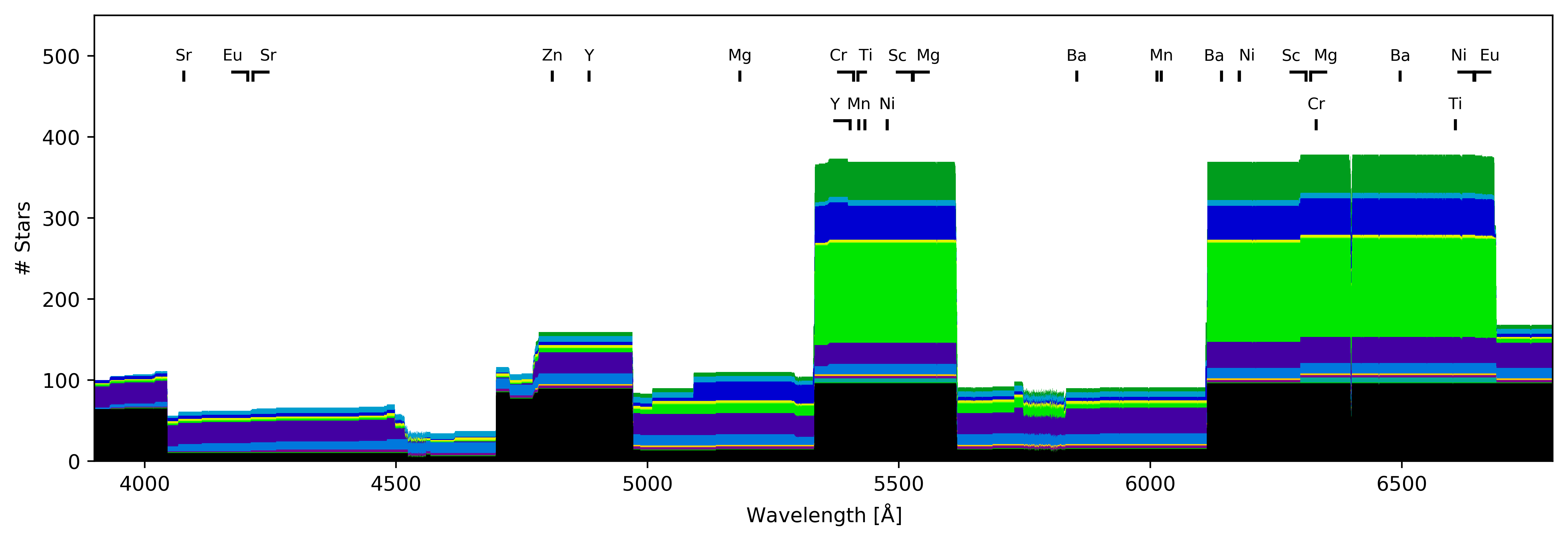}
      \caption{Wavelength coverage and position of the individual spectral lines versus the number of stellar spectra per galaxy. Colors indicate different galaxies as in Fig.~\ref{fig:loc_dsph}. The ionization state of each individual line is given in Table~\ref{tab:linelist} and the wavelength coverage of different FLAMES/GIRAFFE settings is given in Table~\ref{tab:setups_coverage}.}
         \label{fig:wavcover}
   \end{figure*}
To enable a more direct comparison, we adopt the same relation. We stress that the microturbulence becomes relevant for strong lines. We varied the microturbulence for three stars with metallicities of $-0.5$, $-1.5$, and $-3.14$ dex by $0.3\, \rm km \, s^{-1}$. For the most metal-rich star, the abundance of titanium, strontium, and barium is most affected resulting in a difference of $\sim 0.2\, \rm dex$. The difference for all other elements is less than $0.1\, \rm dex$. At lower metallicities the impact of the microturbulence ($\pm 0.3\, \rm km \, s^{-1}$) is less striking yielding a maximum abundance difference of $\sim 0.1 \, \rm dex$ \citep[see also][]{McWilliam1995b, Lai2008, Hansen2016}.

\section{Abundance analysis}
\label{sct:abundance_analysis}
To derive stellar abundances we use the LTE spectral synthesis code MOOG \citep[][version 2014]{Moog} together with the aforementioned 1D Kurucz atmosphere models.

Using different spectral lines for the analysis can cause systematic uncertainties (e.g., due to uncertainties in atomic data), which makes it difficult to exclude biases. We try to minimize this systematic uncertainty by attempting to measure the same lines for all stars. This is not always possible, because the lines wavelength may not be covered by the instrument (Fig.~\ref{fig:wavcover}). We therefore try to use lines that are covered by most of our sample and avoid lines that are only covered in a minority. In addition, different absorption lines may be either too weak or too strong, depending on the metallicity of the star. However, stars with similar stellar parameters should be similarly affected by systematics. 

Another source of uncertainty is the applied method to determine abundances -- that is, either by measuring equivalent widths or by running a spectral synthesis. The first method requires a technique to deblend the absorption lines, which can be uncertain and most studies avoid these blended lines. We therefore perform spectrum synthesis using MOOG and manually inspect all lines that may be blended. We claim a successful detection if the absorption line is found to be at the $2\sigma$ significance level, and the line furthermore has be covered by at least three pixels. We did not attempt to automatize this step. In fact, we measured equivalent widths for chromium and titanium as they had clean lines; all other elemental abundances were synthesized. All abundances are available online in electronic form in Table~o2 and o3.
  
  \begin{figure}[!h]
   \centering
   \includegraphics[width=\hsize]{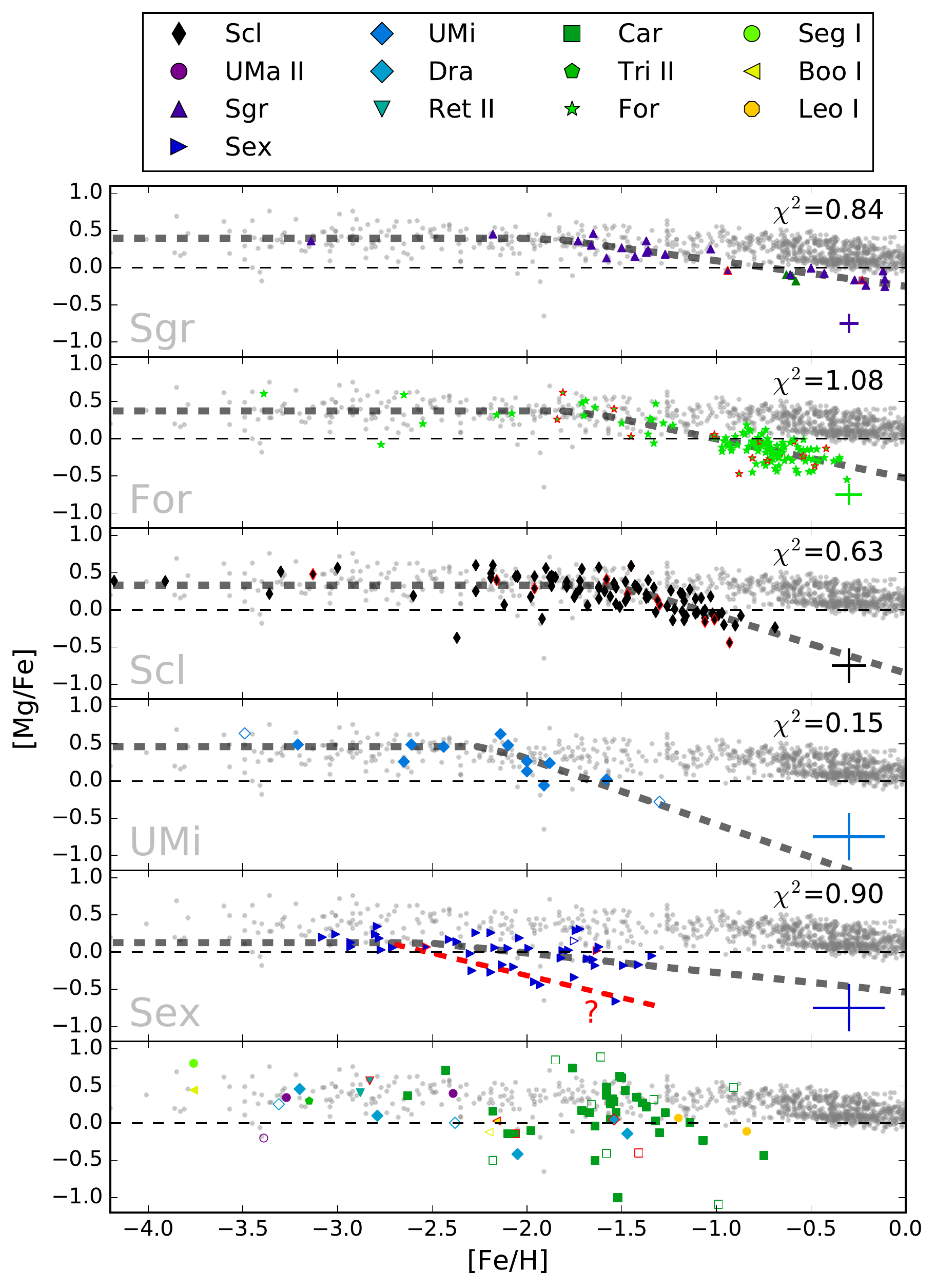}
      \caption{Derived LTE abundances of [Mg/Fe] over metallicities for individual dSph galaxies ordered by decreasing magnitude. The bottom panel shows the remaining dSph and UFD galaxies. Green triangles indicate stars of the globular cluster Terzan 7. Gray dots show MW stars of \citet{Gratton1988}, \citet{Edvardsson1993}, \citet{McWilliam1995}, \citet{Ryan1996}, \citet{Nissen1997}, \citet{Hanson1998}, \citet{Prochaska2000}, \citet{Fulbright2000}, \citet{Stephens2002}, \citet{Ivans2003}, \citet{Bensby2003}, and \citet{Reddy2003}. All stars where the parallax differs significantly from the distance to the host galaxy are marked with red symbols. Stars with an error of $\mathrm{[Mg/H]}>0.4$ are indicated with open symbols. The reduced chi-square of the least-square fit (Sect.~\ref{sct:alphaknee}) is given in the upper right corner of each panel. The median error for each galaxy is denoted in the lower right corner of each panel.}
         \label{fig:mg_evolution}
   \end{figure} 
The following sections present the choice of adopted spectral lines as well as a brief discussion on NLTE corrections. 
  
\subsection{Magnesium}
\begin{figure}[!h]
   \centering
   \includegraphics[width=\hsize]{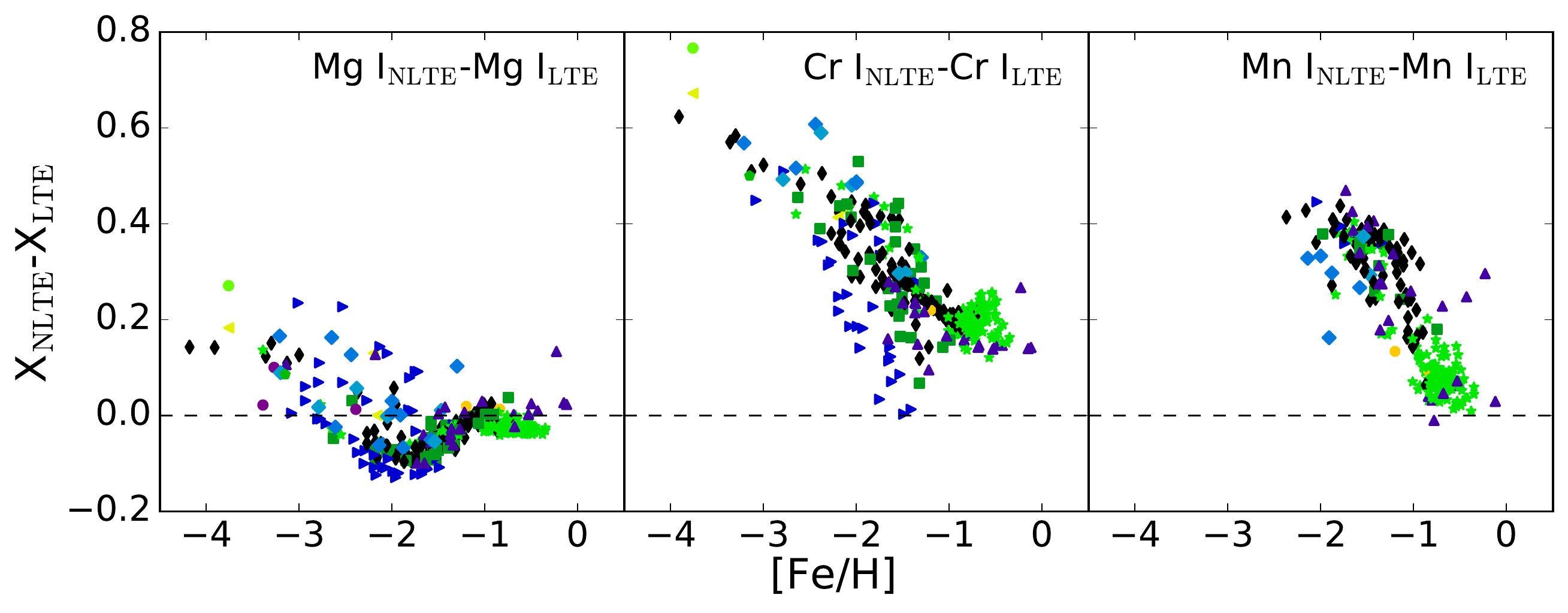}
      \caption{NLTE corrections for \ion{Mg}{I}, \ion{Cr}{I}, and \ion{Mn}{I} taken from \citet{Bergemann2008,Bergemann2010}, and \citet{Bergemann2015} \citep[accessed via the interface of][]{NLTE_MPIA}. The symbols were chosen as in Fig.~\ref{fig:mg_evolution}. }
         \label{fig:nlte_corr_cr}
   \end{figure}

\begin{figure*}
   \centering
   \includegraphics[width=\hsize]{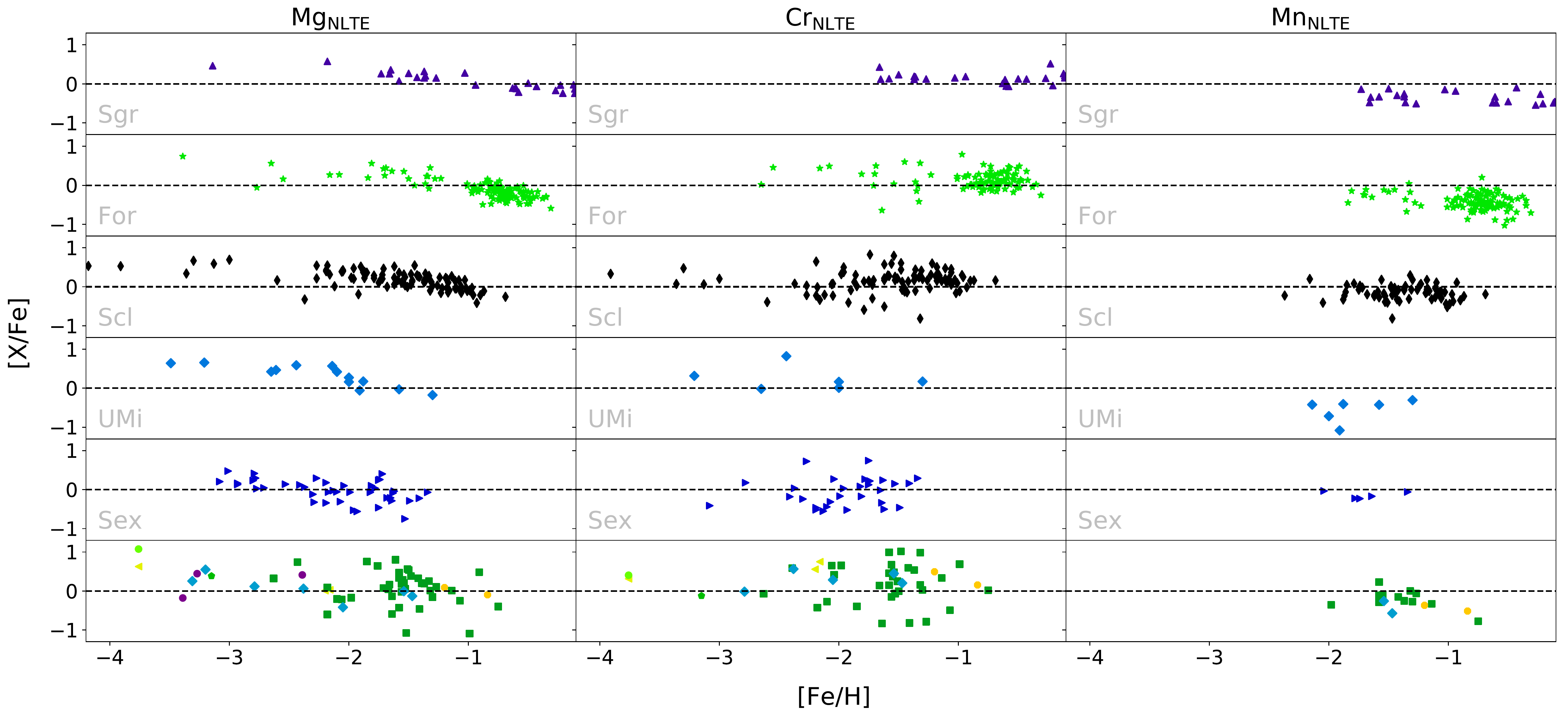}
      \caption{NLTE corrected abundances of Mg, Cr, and Mn. Corrections are taken from \citet{Bergemann2015,Bergemann2010,Bergemann2008} accessed via \citet{NLTE_MPIA}. The symbols were chosen as in Fig.~\ref{fig:mg_evolution}.}
         \label{fig:nlte_mg_cr_mn}
   \end{figure*}

For magnesium, we determined the abundance from the \ion{Mg}{I} feature at $\lambda = 5528.48 \, \AA$ (see Fig.~\ref{fig:mg_evolution}). This line does not saturate at the highest metallicities and is detectable also in extremely metal-poor stars. Furthermore, we analyze the weakest line of the magnesium triplet at $\lambda = 5183.27 \, \AA$ and the region of the relatively weak Mg lines at $\lambda = 6318.71$, $6319.24$, and $6319.50\, \AA$ (see Table~\ref{tab:linelist}). We synthesized the convolved features centered around 6319.24\, \AA, and provide only one best-fitting Mg abundance for this blend.
The atomic data of all \ion{Mg}{I} lines are the recommended (theoretical) values from \citet{Pehlivan2017}. Neutral magnesium abundances are predominantly affected by temperature and their uncertainties are shown in Table~\ref{tab:sensitivity_abundances}. 
We also investigated the impact of this assumption by applying NLTE corrections from \citet{Bergemann2015} \citep[accessed via the interface of][see Fig.~\ref{fig:nlte_corr_cr} and ~\ref{fig:nlte_mg_cr_mn}]{NLTE_MPIA}. In all galaxies, this leads to higher Mg abundances at low metallicities with corrections of $\sim 0.1 \, \rm dex$ (see Fig.~\ref{fig:nlte_corr_cr}). We note that due to the limited coverage of the NLTE correction grid, all stars with $\mathrm{[Fe/H]}<-2$ and simultaneously $\log g<0.5$ are extrapolated from the grid\footnote{the grid coverage can be accessed at http://nlte.mpia.de/grids.png}. 

   \begin{figure*}
   \centering
   \includegraphics[width=\hsize]{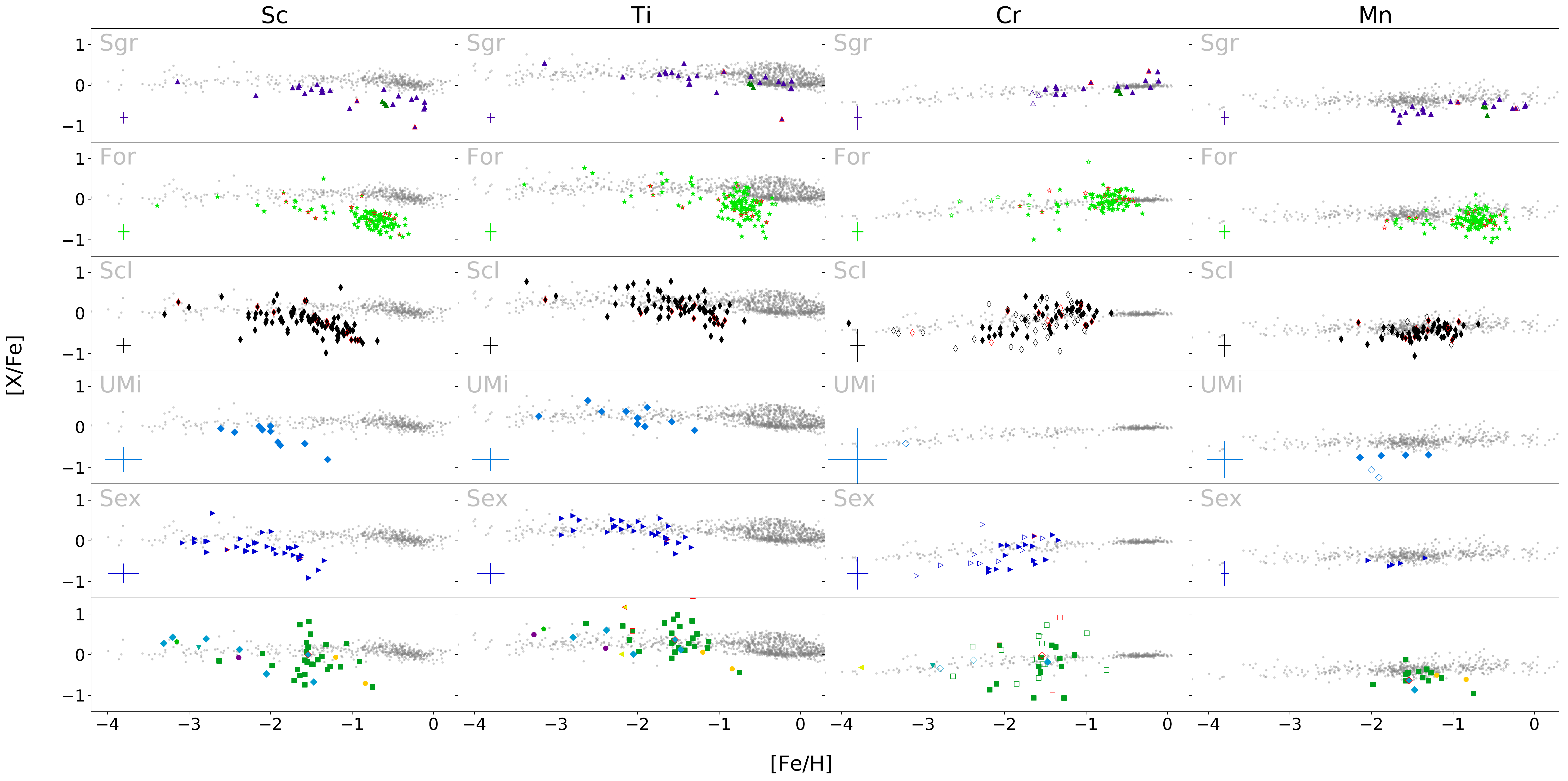}
      \caption{Derived abundances of scandium, titanium, chromium, and manganese versus $\mathrm{[Fe/H]}$ for individual dSph galaxies. The bottom panels show multiple galaxies with low amount of stars. Green triangles indicate stars of the globular cluster Terzan 7. Gray dots show MW stars of \citet{Reddy2003}, \citet{Cayrel2004}, \citet{Reddy2006}, \citet{Ishigaki2013}, \citet{Fulbright2000}, \citet{Nissen1997}, \citet{Prochaska2000}, \citet{Stephens2002}, \citet{Ivans2003}, \citet{McWilliam1995}, \citet{Ryan1996}, \citet{Gratton1988}, \citet{Edvardsson1993}. The symbols were chosen as in Fig.~\ref{fig:mg_evolution}, stars marked in red indicate stars with close distances according to the parallax (Appendix~\ref{sct:appendix_surf_gravs}). }
         \label{fig:sc_ti_cr_mn}
   \end{figure*}
   
   \begin{figure*}
   \centering
   \includegraphics[width=\hsize]{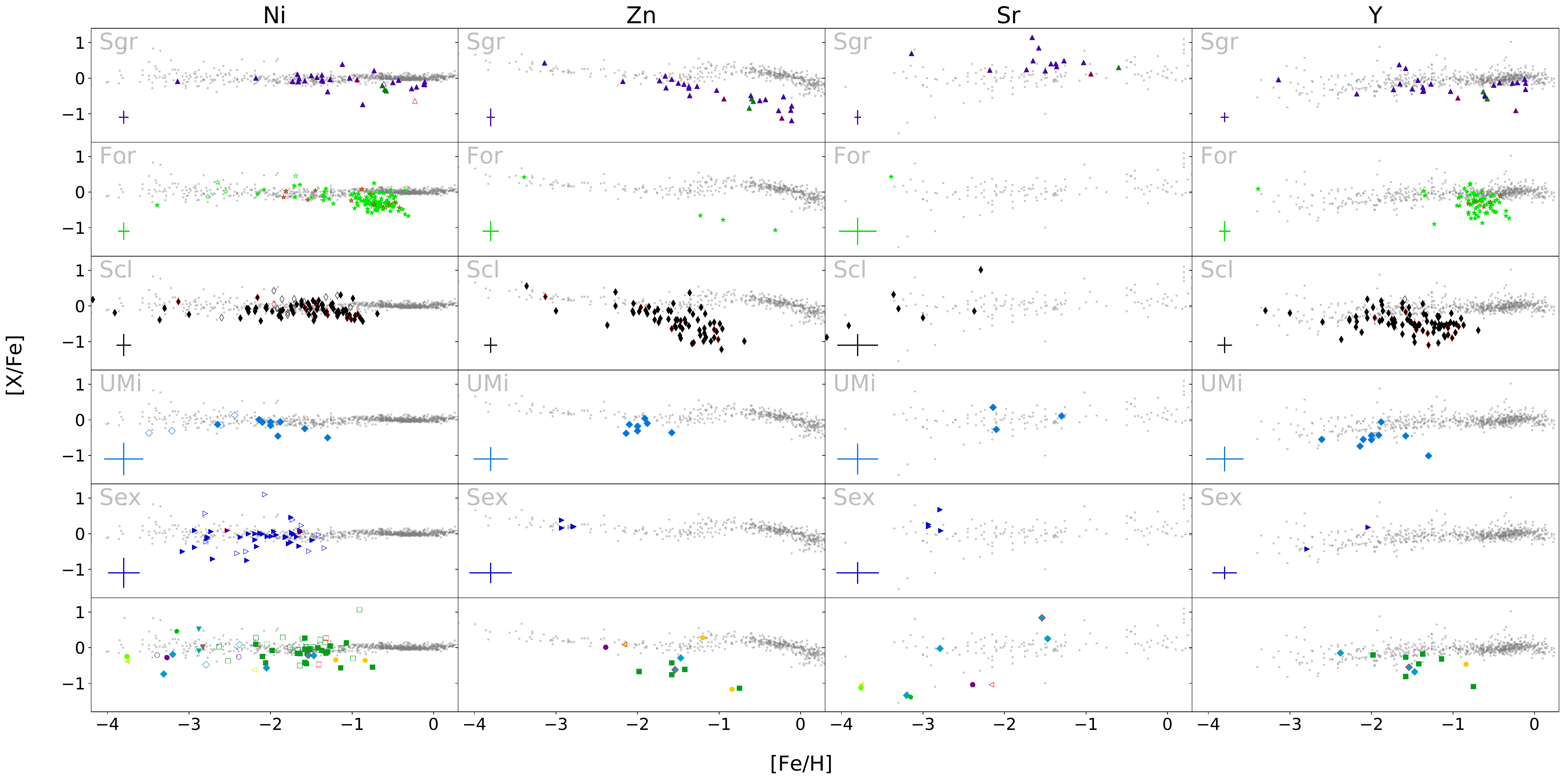}
      \caption{Same as Fig.~\ref{fig:sc_ti_cr_mn}, but for nickel, zinc, strontium, and yttrium versus $\mathrm{[Fe/H]}$ for individual dSph galaxies. Gray dots show MW stars of \citet{Reddy2003}, \citet{Cayrel2004}, \citet{Reddy2006}, \citet{Ishigaki2013}, \citet{Fulbright2000}, \citet{Nissen1997}, \citet{Prochaska2000}, \citet{Stephens2002}, \citet{Ivans2003}, \citet{McWilliam1995}, \citet{Ryan1996}, \citet{Gratton1988}, \citet{Edvardsson1993}. }   \label{fig:ni_zn_sr_y}
   \end{figure*}

      \begin{figure*}
   \centering
   \includegraphics[width=\hsize]{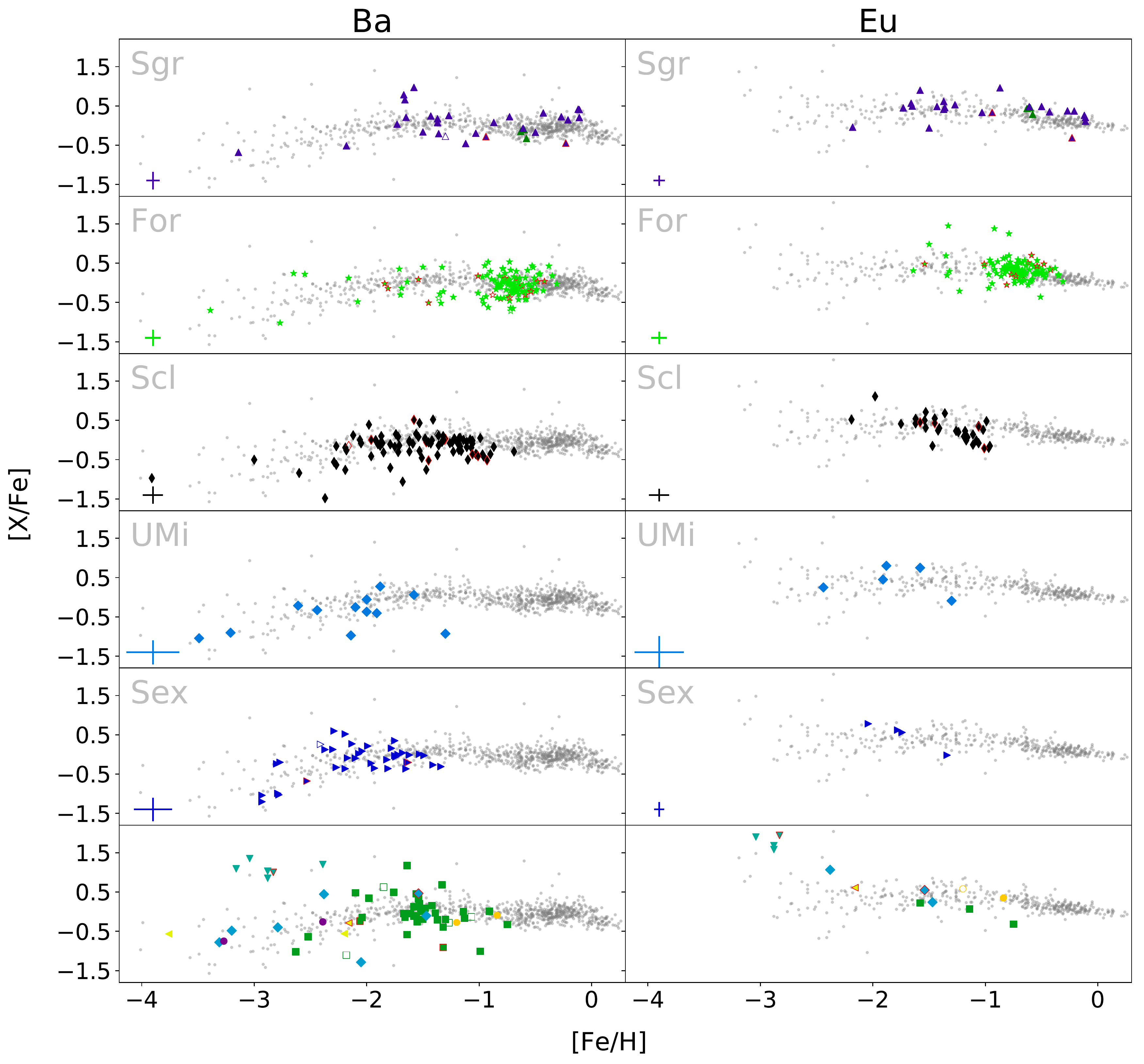}
      \caption{Same as Fig.~\ref{fig:sc_ti_cr_mn}, but for [Ba/Fe] and [Eu/Fe] versus $\mathrm{[Fe/H]}$.}
         \label{fig:ba_eu}
   \end{figure*}

\subsection{Scandium}
 \ion{Sc}{II} abundances are derived from the spectral line at $\lambda = 5526.80 \, \AA$. We also consider a weak line at $\lambda = 6309.90\, \AA$. For both lines, we adopted atomic data and hyperfine splitting (HFS) from \citet[][see Table~\ref{tab:linelist}]{Lawler1989}. The latter is a relatively weak line and only detectable in stars with $\mathrm{[Fe/H]} \gtrsim -2$. It tends to yield lower ($\sim0.2$\,dex) abundances compared to the bluer line. The standard deviation of both lines throughout our sample is $0.37\, \rm dex$.
Since both Sc lines are singly ionized, they are less affected by our LTE assumption, as confirmed by 
\citet{Zhang2008,Zhang2014}, who calculate almost negligible NLTE corrections for \ion{Sc}{II}. The Sc LTE abundances can be found in Fig.~\ref{fig:sc_ti_cr_mn}.

\subsection{Titanium}
We measured the equivalent widths of \ion{Ti}{II} at $\lambda = 5418.77\, \AA$ and $6606.95\, \AA$ (see Table~\ref{tab:linelist}). The Ti LTE abundances can be found in Fig.~\ref{fig:sc_ti_cr_mn}. We obtained similar results as \citet{Skuladottir2017} and \citet{Hill2019}, but lower values than \citealt{Letarte2018} (see Fig.~\ref{fig:sc_ti_cr_mn}).

\subsection{Chromium}
We measured the equivalent widths of two \ion{Cr}{I} lines depicted in Fig.~\ref{fig:wavcover}. Most of the stellar spectra only include one line at $\lambda = 5409.77\, \AA$, since the line at $\lambda = 6330.09\, \AA$ is often too weak.  The scatter in $\mathrm{[Cr/Fe]}$ is significant and in addition driven by the temperature uncertainty (Fig.~\ref{fig:sc_ti_cr_mn}). Chromium can be highly affected by NLTE effects, especially when covering a large metallicity range \citep[see, e.g.,][]{Mashonkina2017}. We apply NLTE corrections from \citet{Bergemann2008} to the Cr I lines \citep[accessed via the interface of ][]{NLTE_MPIA}. The correction, $\Delta$NLTE, reaches extreme values between $0.77 \le \Delta\rm{NLTE} \le 0.00$. Applying the corrections as illustrated in Fig.~\ref{fig:nlte_corr_cr}, we computed abundances that are slightly supersolar with average values of $\langle\mathrm{[Cr/Fe]}_\mathrm{NLTE}\rangle= 0.15 $ in Fornax, Sagittarius and Sculptor, and $\langle\mathrm{[Cr/Fe]}_\mathrm{NLTE}\rangle= 0.0 $ in Sextans (Fig.~\ref{fig:nlte_mg_cr_mn}).

\subsection{Manganese}
We synthesized four manganese absorption lines (Table~\ref{tab:linelist}). For these we adopt the HFS from \citet{Kurucz2011}. \cite{North2012} reveal a large discrepancy between the lines at $\lambda = 5420.36\, \AA$ and $\lambda = 5432.55\, \AA$ for stars in Fornax that may be caused by LTE effects. The Mn LTE abundances can be found in Fig.~\ref{fig:sc_ti_cr_mn}. Manganese is highly affected by LTE assumptions. We therefore apply NLTE corrections from \citet{Bergemann2008} as shown in Fig.~\ref{fig:nlte_corr_cr}.

\subsection{Nickel}
We synthesized two weak Ni lines at $\lambda = 6176.82, 6177.25 \, \AA$ and two stronger lines at $\lambda = 5476.92, 6643.56 \, \AA$. We adopt excitation potentials and oscillator strengths from \citet{Wood2013}, \citet{Martin1988b}, \citet{Kostyk1982} and \citet{Lennard1975}. We assign different weights for the individual lines as indicated in Table~\ref{tab:linelist}. The Ni LTE abundances can be found in Fig.~\ref{fig:ni_zn_sr_y}.

\subsection{Zinc}
 We detected the zinc line at $\lambda = 4810.54 \, \AA$ in a subsample of our stars. This line is only covered by X-shooter, FLAMES/GIRAFFE HR7A, UVES, and the HIRES setup (see Fig.~\ref{fig:wavcover}). We therefore measured zinc mainly for Sculptor and Sagittarius.  The Zn LTE abundances are depicted in Fig.~\ref{fig:ni_zn_sr_y}. We get the same trends as already presented in \citet{Skuladottir2017} and \cite{Sbordone2007}. For Sagittarius, we extend the current knowledge of zinc by also adding abundances from stars presented in \cite{Bonifacio2004,Monaco2005}, and \citet{Hansen2018}.

\subsection{Strontium}
We used the \ion{Sr}{II} spectral line at $\lambda = 4077.71 \, \AA$, which can be blended with \ion{La}{II} and \ion{Dy}{II}. For high metallicities, this line is saturated, resulting in a lower limit of the abundance. In addition, we included the slightly weaker line at $\lambda = 4215.52 \, \AA$. For both lines we assume HFS and isotopic shifts according to \citet{Bergemann2012}. Unfortunately, this range is not covered in the FLAMES/GIRAFFE setups and we can therefore only extract \ion{Sr}{II} in 38 stars. The assumption of LTE can affect the abundance of Sr, especially at $\mathrm{[Fe/H]}<-3$ \citep{Hansen2013,Mashonkina2017}. The LTE Sr abundances are illustrated in Fig.~\ref{fig:ni_zn_sr_y}. In addition, our MOOG version does not include the effect of Rayleigh scattering \citep[see, e.g.,][]{Sobeck2011}. This effect is increasingly important for blue absorption lines in metal-poor stars and would lead to slightly higher Sr abundances if properly treated. 
We tested this for the bluest Sr II line and found that scattering on average increases the abundance by 0.03\,dex for a typical metal-poor, cool giant ($T_\mathrm{eff}=4875\, \mathrm{K}$, $\log g=1.68$, $\mathrm{[Fe/H] = -1.73}$). In addition, we tested the impact of scattering for a more metal-poor star ($T_\mathrm{eff}=4600\, \mathrm{K}$, $\log g=1.5$, $\mathrm{[Fe/H] = -2.5}$) and found a difference in the strontium abundances of $0.12 \, \mathrm{dex}$. This effect can therefore introduce a weak trend in strontium. Strontium is the lightest n-capture element that we analyzed. 

\subsection{Yttrium}
We analyzed two \ion{Y}{II} lines, one strong line in the blue part of the spectra $\lambda =4883.68\, \AA$ and a weak redder line at $\lambda =5402.78\, \AA$. Similar to zinc, these lines are included in a minority of our sample owing to limited wavelength coverage. The $\log gf$ values for these lines are uncertain \citep[see e.g.,][]{Ruchti2016}. For the most metal-rich stars in Sculptor, we were able to measure both \ion{Y}{II} lines. The comparison between the lines yields a constant offset of $\sim 0.2\, \rm dex$, where the redder line results in higher abundances. 
The LTE yttrium abundances are presented in Fig.~\ref{fig:ni_zn_sr_y}.
So far there is no NLTE grid available, but we note that we analyze a spectral line of singly ionized Y that is formed in the lower photosphere where collisions are likely dominating. 

\subsection{Barium}
\label{ssct:barium}
We synthesized the three \ion{Ba}{II} lines at $\lambda=5853.69$, $6141.73$, $6496.90\,\AA$ given in Table~\ref{tab:linelist} and the resulting LTE abundances are shown in Fig.~\ref{fig:ba_eu}. We assign a lower weight to the line at $\lambda = 6141.73\, \AA$, because it has been demonstrated to be heavily affected by the assumption of LTE \citep{Korotin2015}. We use HFS for all lines from \citet{Gallagher2012}. According to \citet{Mashonkina2017}, we expect NLTE effects on the order of $0.3\, \rm dex$ over the entire metallicity range \citep[see also][]{Mashonkina2019}. We derived much lower values for Ba in Fornax than previous studies. This difference is discussed in Appendix~\ref{sct:appendix_ba_fnx}. 
  
\subsection{Europium} 
We derived abundances from one blue \ion{Eu}{II} line at $\lambda = 4205.05\, \AA$, which is present in high-resolution spectra only (UVES and HIRES) and a red \ion{Eu}{II} line at $\lambda = 6645.21\, \AA$. We were able to detect \ion{Eu}{II} down to metallicities of $\mathrm{[Fe/H]} \sim -3$ and the abundances can be found in Fig.~\ref{fig:ba_eu}. This was possible due to a large enhancement of europium in the Reticulum II stars. Both investigated europium lines do not yield a significantly different abundance (with a standard deviation of $\sigma = 0.12 \, \rm dex$ and an average offset of $0.01\, \rm dex$). Throughout all galaxies, [Eu/Fe] is slightly enhanced compared to the sun. NLTE corrections could increase the abundances by $\sim 0.15 \, \rm dex$ \citep{Mashonkina2017}. Similar to Sr, we tested the impact of scattering on the blue Eu line, and found 0.03\,dex (for a star with $T_\mathrm{eff}=4875\, \mathrm{K}$, $\log g=1.68$, $\mathrm{[Fe/H] = -1.73}$) and 0.07\,dex ($T_\mathrm{eff}=4600\, \mathrm{K}$, $\log g=1.5$, $\mathrm{[Fe/H] = -2.5}$) higher abundances when scattering was included.

\subsection{Error determination}
\label{ssct:error_det}
\begin{figure*}
   \centering
   \includegraphics[width=\hsize]{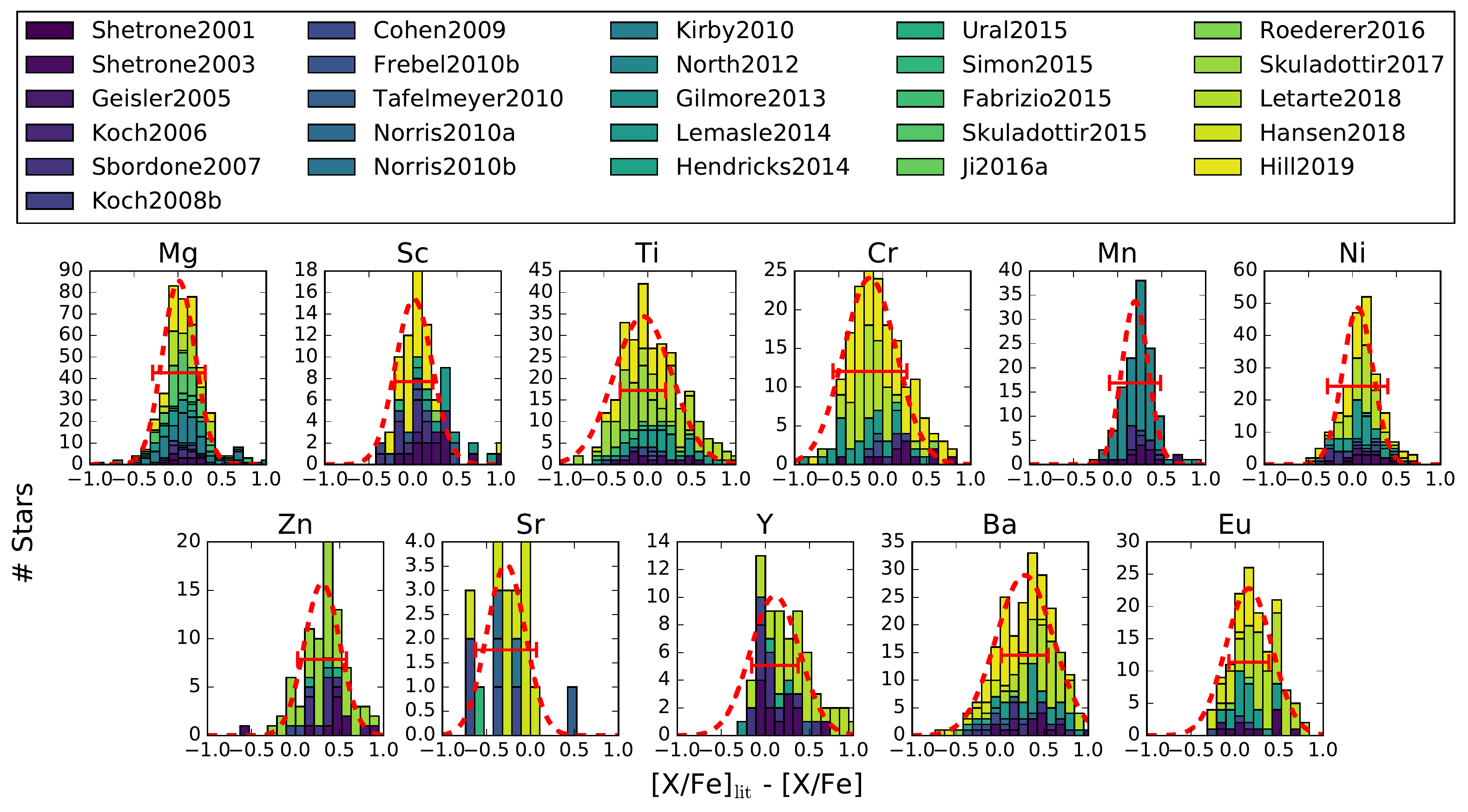}
      \caption{Comparison of derived [X/Fe] abundances with various literature values. The dashed line depicts a Gaussian fit and the horizontal line indicates the FWHM one would expect due to the average error of the abundances. Literature values are taken from: \citet{Skuladottir2015,Skuladottir2017}, \citet{Hendricks2014}, \citet{Cohen2009}, \citet{Hill2019}, \citet{Letarte2018}, \citet{Shetrone2003}, \citet{Gilmore2013}, \citet{Ural2015}, \citet{Geisler2005}, \citet{Koch2006,Koch2008}, \citet{Roederer2016}, \citet{Norris2010a,Norris2010b}, \citet{North2012}, \citet{Sbordone2007}, \citet{Simon2015}, \citet{Ji2016a}, \citet{Tafelmeyer2010}, \citet{Kirby2010}, \citet{Fabrizio2015}, \citet{Shetrone2001}, \citet{Lemasle2014}, \citet{Frebel2010b}, and \citet{Hansen2018}.}
         \label{fig:comp_lit}
   \end{figure*}

We determined errors on the abundances by first converting our abundances to equivalent widths and we investigated how uncertainties in individual stellar parameters propagate into the abundances. We assumed the errors to be uncorrelated and computed a new model atmosphere with a new temperature that was offset by its adopted uncertainty. We then computed new abundances, which were compared to the adopted abundances. This resulted in $\sigma_{T_\mathrm{eff}}$. This was repeated for each of the stellar parameters and the differences were then added in quadrature resulting in a total uncertainty:
\mbox{$\sigma _{\mathrm{tot}} = \sqrt{\sigma_{T_\mathrm{eff}}^2+\sigma_{\mathrm{[Fe/H]}}^2+\sigma_{\log g}^2+\sigma_{\xi _t}^2+\sigma_{\mathrm{stat}}^2+\sigma_\mathrm{noise}^2}$}.\\ Different elements have different sensitivity due to the stellar parameters (see Table~\ref{tab:sensitivity_abundances}). We note that estimating the error this way does not take into account the effects of possible blends. For elements where we measured only one absorption line, we weigh the error by the S/N per pixel at this line by assuming an additional uncertainty of 
\begin{equation}
\sigma_\mathrm{noise} =
    \begin{cases}  -0.01\cdot \mathrm{S/N} +0.4 & , -0.01\cdot \mathrm{S/N} +0.4>0\\ 0 &\text{, else} \end{cases}
\end{equation}
The total calculated error can be directly tested by comparing the derived abundances with literature values. The residual deviation of the abundances is illustrated in Fig.~\ref{fig:comp_lit}. The histogram reveals an approximate Gaussian shape that we fit (dashed line). The FWHM of this Gaussian can also be predicted by taking the estimated errors of the abundances into account. We calculate the FWHM that we would expect from the errors as $\mathrm{FWHM} = 2.354\cdot \sqrt{\sigma_{\mathrm{lit}}^2+\sigma_\mathrm{tot}^2}$, where the value is illustrated as a horizontal line. Here $\sigma_{\mathrm{lit}}^2$ is the average variance that the literature provides and $\sigma_\mathrm{tot}^2$ is the average error on the abundance derived within this study. A smaller FWHM indicates an underestimation of the error whereas a larger FWHM indicates an overestimation or large systematic uncertainties between the studies. Nearly all elements show a perfect agreement of the theoretical and determined value of the FWHM. We therefore conclude that our derived errors are in a reasonable range. 

\section{A recap on nuclear formation sites and abundance trends}
\label{sct:abundance_trends}
Our homogeneous set of stellar abundances across different galaxies includes stars covering a broad range of metallicities, $-4.18\le \mathrm{[Fe/H]} \le -0.12$. This gives us the unique possibility to study the characteristics of the individual systems. Each of the derived elements can trace different processes and production sites. In the following, we refer to two different types of core-collapse supernovae (CC-SNe), namely the ``standard'' CC-SNe (CC-SNe) and a rare type of CC-SNe (rare CC-SNe). 

The $\alpha$-elements are particularly interesting when studying the star formation history (SFH) and the IMF \citep[e.g.,][]{Tinsley1980,McWilliam1997}. We therefore investigate the evolution of $\alpha$-elements in Sect.~\ref{sct:alphaknee} to learn about the chemical history of these galaxies.
Other elements like n-capture elements give insight to nuclear processes, such as the s-process hosted by AGB stars, and the r-process (e.g., in rare CC-SNe or NSMs).

Figure~\ref{fig:mg_evolution} presents magnesium, which is created during hydrostatic burning \citep[e.g., ][]{Truran2003,Cayrel2004,Kobayashi2006,Nomoto2013}. This element is predominantly ejected by CC-SNe and provides insight into the IMF and star formation history (SFH). Moreover, one has to consider the iron synthesized in type Ia SNe.
A low [Mg/Fe] at low metallicities that is derived for some dSph galaxies (lower than the one in the MW) indicates that more low-mass stars may have contributed to the chemical enrichment, whereas higher values can be caused by the chemical contribution from more massive stars. Sextans seems to have a lower plateau value of [Mg/Fe] than other dwarf galaxies \citep[cf., e.g., Fig.~2 of][]{Aoki2009,Shetrone2001,Theler2015phd,Theler2019}. This deficiency of Mg at low metallicities is less striking when including NLTE corrections (Fig.~\ref{fig:nlte_mg_cr_mn}) and may therefore be a consequence of LTE assumptions. Another possible explanation might be a poor stochastic sampling of the IMF for small systems \citep{francois2016,Applebaum2018}, resulting in lower $\mathrm{[}\alpha\mathrm{/Fe]}$ \citep{Tolstoy2003,Carigi2008}. Alternatively, less efficient chemical enrichment could cause that type Ia SNe may have already contributed to even the lowest metallicity stars in this galaxy.

Titanium on the other hand traces explosive burning and forms predominantly in CC-SNe \citep[e.g.,][]{Cayrel2004,Nomoto2013}. We observe a similar evolution for Ti as for the MW with the exception of the position of the $\alpha$-knee, which is discussed in Sect.~\ref{sct:alphaknee}. In Fig.~\ref{fig:sc_ti_cr_mn} we see a clear $\alpha$-knee in Ti. Despite being produced in both CC-SNe and type Ia SNe, the two sites form different fractions of Ti with respect to Fe which ensures that the knees remain detectable. 

The iron-peak elements, chromium, manganese, and nickel are produced by a mixture of type Ia SNe as well as CC-SNe in the sun \citep[e.g.,][]{Bergemann2010,North2012,Nomoto2013,Kirby2019}.
The heavier element zinc shows a decrease in Sagittarius and Sculptor at approximately the same position as magnesium. In addition, it has a plateau at low metallicities. This, in combination with a low abundance scatter, may point toward CC-SNe as main production channel (Fig.~\ref{fig:ni_zn_sr_y}, see also \citealt{Skuladottir2017}). Both chromium and manganese show decreasing trends with decreasing metallicities in LTE (Fig.~\ref{fig:sc_ti_cr_mn}). However, when correcting for LTE effects, both elements follow flat trends (Fig.~\ref{fig:nlte_mg_cr_mn}). Similarly, nickel shows a flat trend with [Fe/H] (Fig.~\ref{fig:ni_zn_sr_y}) and only Fornax and possibly Sculptor possess a large amount of clumping at higher [Fe/H], which would indicate a deviation from this flat trend. We note that compared to the literature, this conclusion is different. In \citet{Hill2019} and \citet{Kirby2019}, a clear decreasing trend in nickel is visible, which is probably caused by the choice of different Ni absorption lines.

The abundances of the lighter n-capture elements, strontium and yttrium, are presented in Fig.~\ref{fig:ni_zn_sr_y}. According to \citet{Sneden2003} and \citet{Bisterzo2014}, strontium is mainly produced by the s-process in the sun. However at low metallicities, it may also have contributions from other processes such as the light element primary process (LEPP; \citealt{Travaglio2004,Montes2007} see also \citealt{Hansen2013} or \citealt{Cristallo2015,Prantzos2018}, for a discussion on the need of the LEPP.).
 
In our Solar System, the heavy n-capture element barium was dominantly formed by the s-process \citep[85\% according to][]{Bisterzo2014}. This makes it an excellent tracer of the s-process. On the other hand, in the solar system europium was to $94 \, \%$ produced by the r-process \citep{Bisterzo2014}, making it our best r-process tracer. In Fig.~\ref{fig:ba_eu} we see the chemical evolution of these two heavy elements. Most dSphs show a large star-to-star scatter and a tendency of decreasing heavy element to iron ratios below [Fe/H]$=-2$. This is further discussed in Sect.~\ref{ssct:ncapture}.

\subsection{The IMF and the alpha knee}
\label{sct:alphaknee}
To study the alpha knee we focus on two main production sites, namely CC-SNe forming the $\alpha$-elements (i.e., Mg, Si, Ca, Ti) and type Ia SNe creating in large amounts the iron-peak elements (i.e., Fe, Ni, Zn).
The former results from massive stars and can occur early; the latter includes white dwarfs in a binary system and appears later in the galactic history. 
Early on, the chemical enrichment is dominated by CC-SNe whose yields have high ratios of magnesium versus iron ($\mathrm{[\alpha/Fe]}_\mathrm{CC-SNe}$), while type Ia SNe yield lower ratios of magnesium versus iron later on (see Fig.~\ref{fig:mg_evolution}). Observationally, the onset of type Ia SNe is seen as a decrease in $\alpha$ to iron, and it is known as the ``$\alpha$-knee'' ($\mathrm{[Fe/H]}_\mathrm{\alpha,knee}$).
The location of this knee is the most striking difference between the investigated dwarf galaxies. To quantitatively describe the position, we assume a toy model with a plateau for low metallicities and a linear decrease after $\mathrm{[Fe/H]}_\mathrm{\alpha,knee}$ \citep[for a similar approach see, e.g., ][]{Cohen2009,Vargas2013,deBoer2014,Hendricks2014b,Kirby2019}.

\begin{figure*}
 \centering
 \includegraphics[width=\hsize]{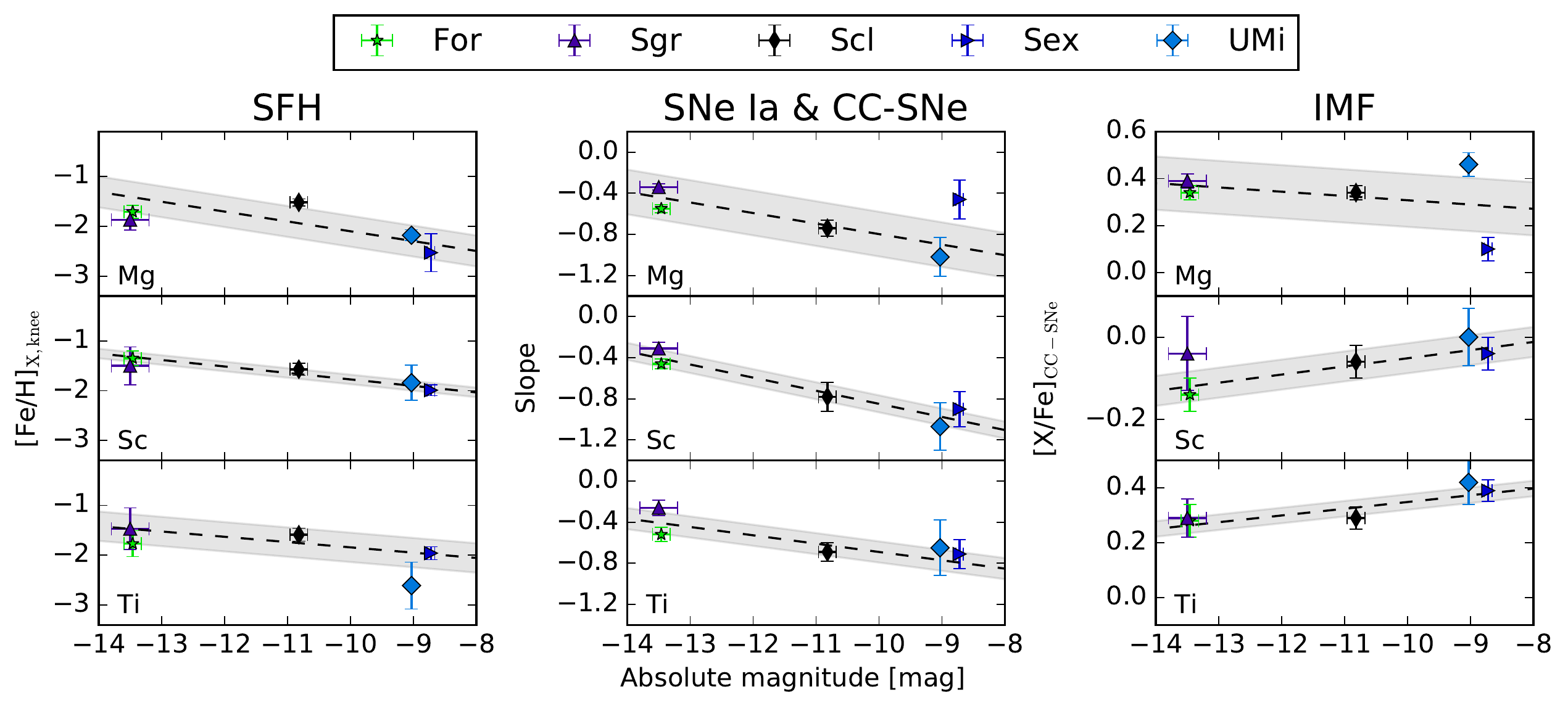}
 \caption{Left panel: Measured position of the $\alpha$-knee in five classical dSphs as a function of metallicity and absolute $V$ magnitude. Middle panel: Slope of the decrease of $\mathrm{[X/Fe]}$ over $\mathrm{[Fe/H]}$ at metallicities higher than the locus of the $\alpha$-knee. Right panel: Constant value of the plateau in $[\alpha/\mathrm{Fe}]$ at low [Fe/H], before the onset of type Ia SNe for different $\alpha$-elements in the individual systems. Symbols are the same as in Fig.~\ref{fig:mg_evolution}.}
 \label{fig:alpha_knee_fit}
\end{figure*}

First, we determined the position of the $\alpha$-knee for magnesium, scandium, and titanium individually (left panel of Fig.~\ref{fig:alpha_knee_fit}). We use Mg, Sc, and Ti, because they show a decrease at the same metallicity, even if they are not all typical $\alpha$-elements. The $\alpha$-knees were fit by an ordinary least squares fit\footnote{via the Python package scipy.optimize.curve\_fit, \citealt{scipy}}.

In a next step, we calculated the error weighted mean and standard deviation of the knee position shown in Table~\ref{tab:alphaknee}. \citet{Hendricks2014b} found a much lower value of the $\alpha$-knee in Fornax ($\mathrm{[Fe/H]}_\mathrm{Mg,knee} = -1.88$ and $\mathrm{[Fe/H]}_\mathrm{\alpha,knee} = -2.08$), based on different elements (Mg, Si, and Ti). The higher values of our work are related to high scandium values, whereas low values in \citet{Hendricks2014b} are mainly driven by silicon ($\mathrm{[Fe/H]}_\mathrm{Si,knee} = -2.49$). Chemical evolution models that successfully reproduce the metallicity distribution function in Fornax predict a knee at $\mathrm{[Fe/H]}_\mathrm{\alpha,knee} \approx -1.4$ \citep{Kirby2011,Hendricks2014b}. This is close to our values, however, \citet{Kirby2011} note that the applied chemical evolution model does not reflect the complete complex behavior of Fornax and the agreement between the model and our value may be a coincidence (see, e.g., \citealt{Lanfranchi2003,Lanfranchi2004,Lanfranchi2010} for other chemical evolution models of dSph galaxies or \citet{Cohen2009,Cohen2010,deBoer2012,Starkenburg2013,Kirby2019} for other observational derivations of the location of the knee). 

\begin{figure}
 \centering
 \includegraphics[width=\hsize]{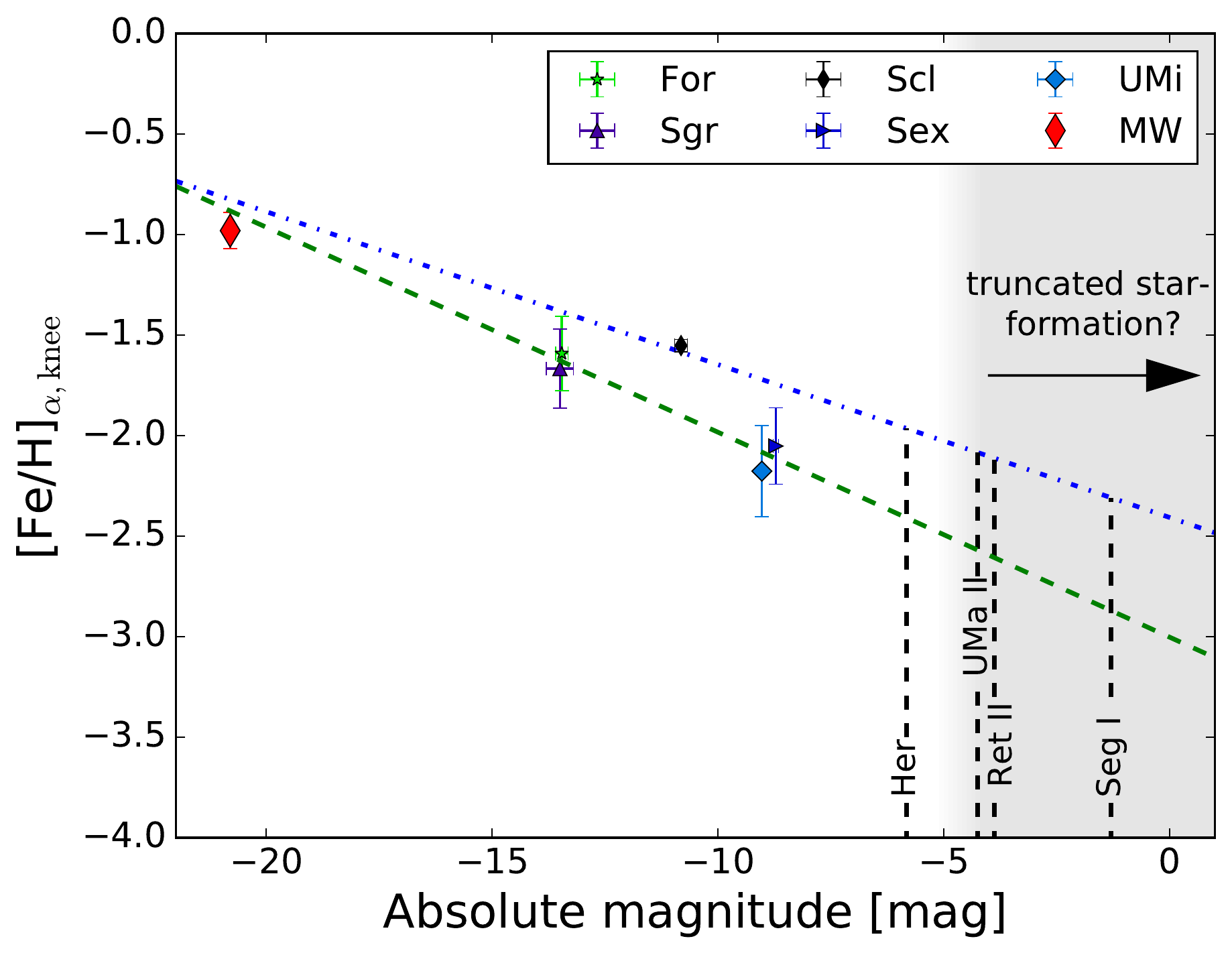}
 \caption{Metallicity of the $\alpha$-knee versus the absolute magnitudes of the parent dSph galaxy and the MW. The lines indicate an orthogonal-distance regression (ODR) including (blue dotted-dashed line) and excluding (green dashed line) Sculptor.}
 \label{fig:alpha_knee_fit_mw_rel}
\end{figure}

Previous studies \citep[e.g., ][]{Tinsley1980,Matteucci1990,Gilmore1991,Venn2004,Tolstoy2009,Hendricks2014b} proposed a relation between the mass (and therefore the luminosity) of a galaxy and the position of the $\alpha$-knee. In the following, we use the luminosity of a galaxy as a proxy for its mass. We note, however, that for tidally disrupted galaxies, this may not be a good substitute \citep[e.g.,][]{Tolstoy2009,McConnachie2012}. Nevertheless, a large deviation from the mass and luminosity relation should only apply for a strong tidal disruption, because the galaxies dark matter is thought to get disrupted before the stars and the impact on the luminosity may therefore be minor\citep[see e.g.,][]{Smith2016}. More massive systems are expected to keep or synthesize their metals more efficiently, leading in most cases to the appearance of the $\alpha$-knee at higher metallicities due to an extended SFH. By assuming a simple linear dependency of these quantities, we get a rough estimate of the knee (in Fornax, Sagittarius, Sextans, and Ursa Minor, see Table~\ref{tab:alphaknee}):
\begin{equation}
\label{eq:knee_formula}
\mathrm{[Fe/H]}_\mathrm{\alpha,knee,estimate} \approx (-0.10\pm 0.03)\cdot M_V +(-3.00 \pm 0.25).
\end{equation}
Here we included x- as well as y- errors in the fit adopting an orthogonal distance regression (ODR)\footnote{via the python package scipy.odr}. We note that the fit contains large uncertainties due to the limited amount of data points. Nevertheless, this relation also fits reasonably well for the MW at $M_\nu = -20.8 \, \rm mag$ \citep[][see Fig.~\ref{fig:alpha_knee_fit_mw_rel}]{Karachentsev2004}. In addition to the knee position, the slope of each [$\alpha$/Fe] also depends on the galaxy mass (middle panel of Fig.~\ref{fig:alpha_knee_fit}).
\begin{table*}[!h]
\caption{Approximate position of the $\alpha$-knee and absolute magnitudes taken from \citet{McConnachie2012} for Sagittarius and from \citet{Munoz2018} for the other four dSphs.}             
\label{tab:alphaknee}      
\centering                          
\begin{tabular}{lcccrr}
\hline\hline                 
Galaxy & $\mathrm{[Fe/H]}_\mathrm{Mg, knee}$ & $\mathrm{[Fe/H]}_\mathrm{Sc, knee}$ & $\mathrm{[Fe/H]}_\mathrm{Ti, knee}$ & $\mathrm{[Fe/H]}_\mathrm{\alpha, knee}$ & Absolute magnitude [mag]\\
\hline                        
Sgr & $-1.87 \pm 0.21$ & $-1.48 \pm 0.38$ & $-1.47 \pm 0.42$ &  $-1.67 \pm 0.19$ &$-13.50 \pm 0.30 $ \\
For      & $-1.71 \pm 0.13$ & $-1.35 \pm 0.15$ & $-1.77 \pm 0.26$ & $-1.59 \pm 0.18$ &$-13.46 \pm 0.14$ \\
Scl    & $-1.52 \pm 0.07$ & $-1.57 \pm 0.12$ & $-1.59 \pm 0.13$ & $-1.55 \pm 0.03$ &$-10.82 \pm 0.14$ \\
UMi         & $-2.18 \pm 0.11$ & $-1.84 \pm 0.35$ & $-2.61 \pm 0.47$ & $-2.18 \pm 0.23$ &$-9.03  \pm 0.05 $ \\
Sex    & $-2.53 \pm 0.39$ & $-1.99 \pm 0.11$ & $-1.96 \pm 0.13$ & $-2.05 \pm 0.19$  &$-8.72  \pm 0.06$ \\
\hline                                   
\end{tabular}
\end{table*}
Simplified, this linear trend can be explained with a combination of the SFH and the relative amount of CC-SNe and type Ia SNe yields retained within each system at any given time. Imagine a smaller (absolute) amount of Fe in less massive systems due to an earlier halt of star formation, a lower gravitational potential, and the implied lower escape velocities. A single type Ia SN event, therefore, lowers the $\alpha$/Fe-ratio by a larger factor. The slopes of the decreasing trends of the elements depend on the rate of CC-SNe, the rate of type Ia SNe, the yields, and the SFHs. We note that this is a simplified view on the evolution of elements, as there are outliers from the here presented mass relation such as Carina, which may have undergone a merger event. 

The Sextans dSph galaxy seems to show an indication of having two knees when looking at the magnesium abundances (dashed line in Fig.~\ref{fig:mg_evolution}, cf. Fig.~3.6 of \citealt{Theler2015phd} and \citealt{Theler2019}). For more information of the two of our most metal-poor stars in Sextans, we refer the reader to \citealt{Lucchesi2020}). 
In order to test\footnote{Using a Gaussian mixture clustering algorithm via the python package scikit-learn, \citealt{scikit-learn}} whether a model with two knees should be favored over a model with one, we apply the Akaike information criterion \citep[AIC, ][]{Akaike1974} and the Bayesian information criterion \citep[BIC,][]{Schwarz1978}. These criteria take the goodness of the fit as well as the amount of parameters into account. 
For Sextans, the BIC favors a model with one cluster only, whereas the AIC favors a model with two clusters. However, the difference of the two models is negligible. 
From the current data it is not possible to get a clear answer or evidence, but future observations may provide better statistics to improve this. In addition, we are aware that titanium does not show a signature of a second knee. Nevertheless, we briefly discuss the implication of this finding.

One knee lies around $\mathrm{[Fe/H]}_\mathrm{Mg,knee}=-2.5$ and one around $\mathrm{[Fe/H]}_\mathrm{Mg,knee}=-2.0$. \citet{Cicuendez2018} found that Sextans may have undergone a galaxy accretion or merger event. They note that simulations \citep[e.g., ][]{Benitez2016} indicate that older stars (early population) may assemble in mergers, whereas the younger ones are formed in-situ. Following this idea, the knee at $\mathrm{[Fe/H]}_\mathrm{Mg,knee}=-2.0$ may therefore be consistent with the similar luminous Ursa Minor that has not undergone a merger event (Fig.~\ref{fig:alpha_knee_fit_mw_rel}), whereas the more metal-poor knee may be formed before the accretion. According to Eq.~\ref{eq:knee_formula}, a knee at $\mathrm{[Fe/H]}_\mathrm{Mg,knee} \sim -2.5$ would translate to $M_V \sim -5.2 \, \rm mag$ and therefore $M_* \sim 10^4 \mathrm{M}_\odot$. This agrees with masses of globular clusters or UFD galaxies comparable to Ursa Major I, which was previously pointed out to show remarkable similarities to Sextans \citep{Willman2005}. The lack of metals before the onset of type Ia SNe in Sextans could be also explained by a short main episode of star formation. \citet{Bettinelli2018} found that the main episode of star formation lasted only $\sim 0.6 \, \rm Gyr$ and ended $\sim 12.9\,\rm Gyr$ ago, which supports the hypothesis that Sextans quickly drained most of its gas reservoir.

The IMF can be investigated by the value of the plateau of $\mathrm{[\alpha/Fe]}$ at low metallicities \citep[e.g.,][]{Tinsley1980,McWilliam1997}. 
The yields of more massive stars are predicted to have higher $\mathrm{[\alpha/Fe]}$ \citep[e.g.,][]{Tsujimoto1995,Woosley1995,Kobayashi2006}. The value of this plateau is shown in the right panel of Fig.~\ref{fig:alpha_knee_fit}. A linear fit of the plateau value results in slopes of $-0.02\pm 0.03$, $0.02\pm 0.01$, and $0.02\pm 0.01$ for Mg, Sc, and Ti, respectively. These almost negligible slopes indicate that there is no dependence of the IMF on the absolute magnitude of the system. Finally, we extrapolate the knee position (see Table~\ref{tab:theo_alphaknee}) for small systems to test if the relation breaks down. 
We expect this relation to break down at the lowest galaxy masses, if they stopped forming stars before the onset of type Ia SNe \citep[see e.g.,][]{Weisz2019}, shown by a gray band in Fig.~\ref{fig:alpha_knee_fit_mw_rel}. \citet{Vargas2013} demonstrated with medium resolution DEIMOS spectra that this is the case for Ursa Major II and Segue I, but most UFDs are able to form stars long enough to maintain the signatures of type Ia SNe. 

\begin{figure}
 \centering
 \includegraphics[width=\hsize]{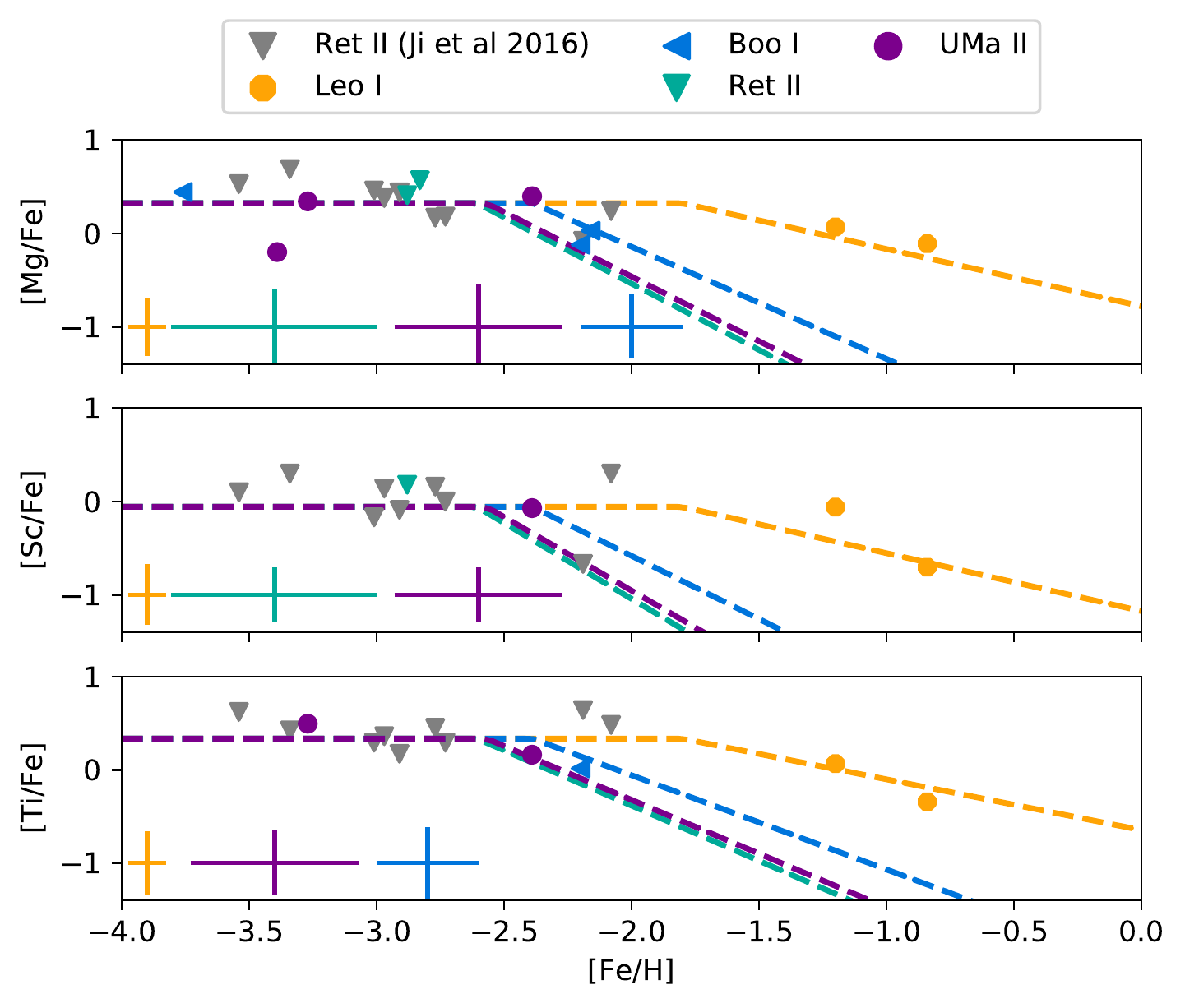}
 \caption{Estimate of [Mg/Fe], [Sc/Fe], and [Ti/Fe] for Reticulum II, Leo I, Bootes I, and Ursa Major II. The median error of each galaxy is shown in the lower left of every panel.}
 \label{fig:constr_gal}
\end{figure}

For small galaxies (low stellar masses), with insufficient data to directly fit a knee position, we can extrapolate the relation given in Eq.~\ref{eq:knee_formula} and Appendix~\ref{sct:construction_galaxies}. Figure~\ref{fig:constr_gal} indicates the knee position estimates for Leo I, Bootes I, Reticulum II, and Ursa Major II. In addition, we added literature values for the UFD galaxy Reticulum II \citep{Ji2016b}. The star \object{[KCB2015] Reti 15} has an unusual low ratio of $\mathrm{[Sc/Fe]}$. Only a few stars at low metallicities show such a deficiency in Sc \citep{Casey2015}. 
Our relation strengthens the hypothesis that this star (although low in metallicity) was formed in a pocket of the ISM where type Ia SNe already contributed \citep[see also, e.g.,][]{Ji2016b}. This is supported by the low ratio of $\mathrm{[Mg/Fe]}$, but stands in contradiction to the high ratio of $\mathrm{[Ti/Fe]}$. 

The overall good agreement between all galaxies suggests that SFH, and the fraction of type Ia-, and CC-SNe yields are strongly dependent on the absolute magnitude (and therefore the stellar mass) of the galaxy. Having different SFHs and varying fractions of dark matter \citep[for Sculptor, see, e.g.,][]{Massari2018} in these systems such a strong relation (Eq.~\ref{eq:knee_formula} and depicted in Fig.~\ref{fig:alpha_knee_fit}) is not obvious.  

\begin{table}
\caption{
Estimates of the $\alpha-$knee, using Eq.~\ref{eq:knee_formula} for different galaxies, with absolute magnitudes taken from \citet{Munoz2018}.}       
\label{tab:theo_alphaknee}     
\centering                         
\begin{tabular}{lrc}
\hline\hline                
Galaxy & $M_V\, \rm [mag]$ & $\mathrm{[Fe/H]}_\mathrm{\alpha,knee,estimated}$\\
\hline                       
Leo I & $-11.78 \pm 0.28$ & $-1.8 \pm 0.1 $ \\
Leo II & $-9.74 \pm 0.04$ & $-2.0 \pm 0.1 $ \\
Car & $-9.43 \pm 0.05$ & $-2.0 \pm 0.1 $\\
Dra & $-8.71 \pm 0.05$ & $-2.1 \pm 0.1 $\\
Boo I & $-6.02 \pm 0.25$ & $-2.4 \pm 0.1 $\\
Her & $-5.83 \pm 0.17$ & $-2.4 \pm 0.1 $\\
Com & $-4.38 \pm 0.25$ & $-2.6 \pm 0.2 $\\
UMa II & $-4.25 \pm 0.26$ & $-2.6 \pm 0.2 $ \\
Ret II & $-3.88 \pm 0.38$ & $-2.6 \pm 0.2 $\\
Seg I & $-1.3 \pm 0.73$ & $-2.9 \pm 0.2 $\\
\hline                                   
\end{tabular}
\end{table}

\subsection{Neutron-capture elements in dwarf galaxies}
\label{ssct:ncapture}
 Here we start by determining when the s-process first contributes to the chemical enrichment and later study possible hosts of the r-process in dSph systems. Furthermore, we discuss whether the stellar mass of the galaxies has an impact on the enrichment in heavy elements.
\subsubsection{The s-process} 
The s-process is mainly hosted in AGB stars, but for the r-process there are multiple possible pathways. Since the gravitational wave detection of a binary neutron star merger GW170817 \citep{Abbott2017} and the optical counterpart AT 2017gfo, these events have shown to be able to produce heavy elements, indicated by the transient \citep[see e.g.,][]{Cowperthwaite2017,Kasliwal2017,Chornock2017,Drout2017,Shappee2017,Pian2017,Waxman2018,Arcavi2018,Watson2019}. In the following, we want to recap arguments against and for NSMs being the dominant production site for the r-process elements in various dSph galaxies or whether an additional source is necessary (e.g., rare types of CC-SNe) to explain the abundance trends that were derived within this study. 

In order to trace contributions from various nuclear formation processes we search for abundance correlations. An indication of two elements being formed by the same process can be obtained if the element abundance ratio [A/B] versus [B/H] shows a flat trend, indicative of the two elements, A and B, growing at the same rate as a function of the element B \citep{Hansen2012}. Alternatively, a very clean trace can be obtained by simply plotting the absolute abundances of A versus B where a 1:1 ratio (with low scatter) shows a correlation and process co-production \citep{Hansen2014b}. We stress that these two elements do not necessarily have to form in exactly the same event, but with the same time delay. A delayed production of A results in a more positive slope of the absolute abundances of A versus B \citep{Duggan2018,skuladottir2019}. 

\begin{figure}
 \centering
 \includegraphics[width=\hsize]{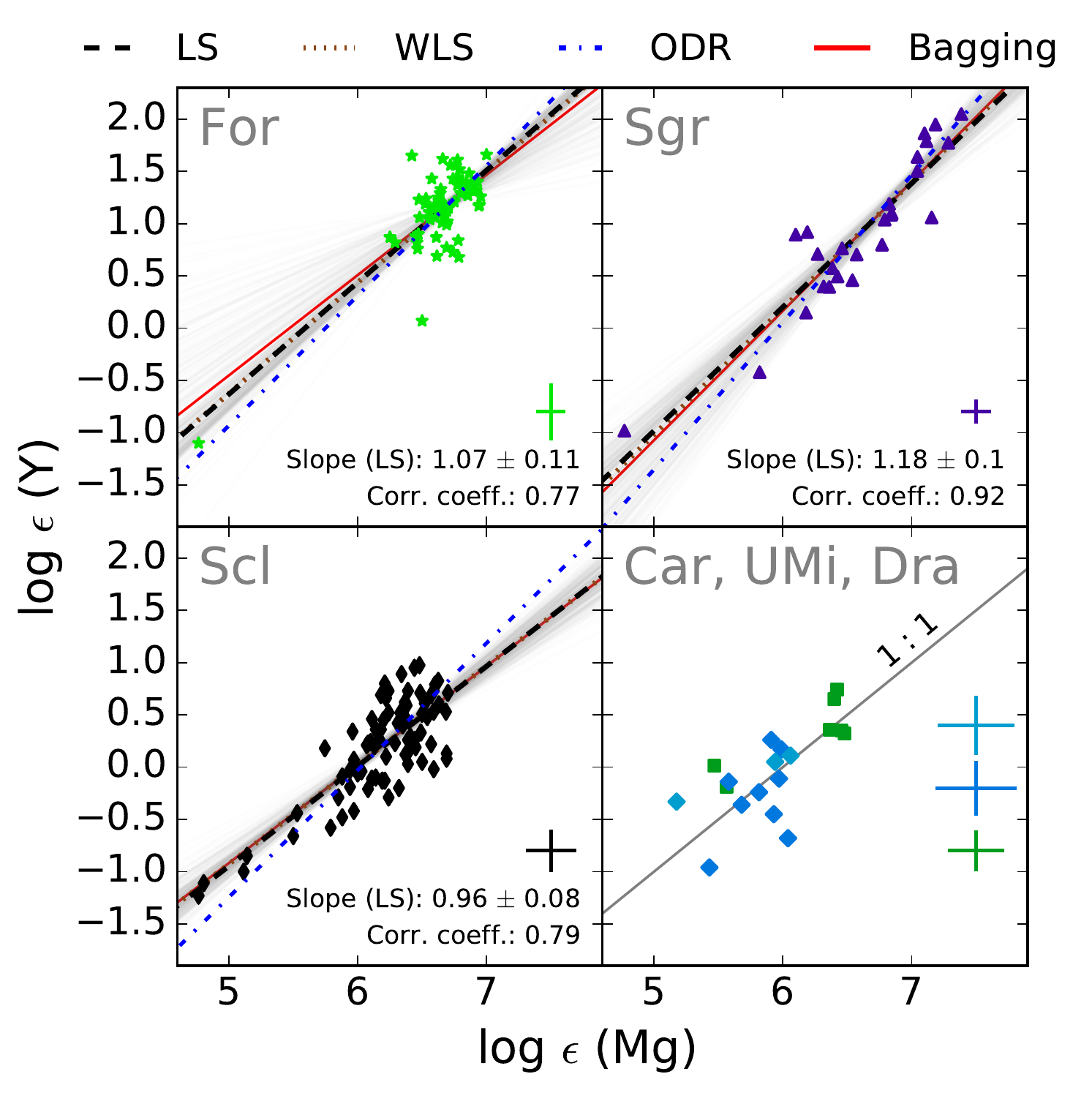}
 \caption{Absolute yttrium abundance versus absolute magnesium abundance together with a linear fit. Symbols are the same as in Fig.~\ref{fig:mg_evolution}. Different line styles and colors indicate different fitting techniques. LS denotes a least squares fitting without any weights. The slope given in each panel is calculated with this fitting technique. WLS indicates a weighted least squares fit, ODR stands for an orthogonal distance regression, and Bagging for a bootstrapping aggregation. Light gray lines indicate different slopes that are involved in the bootstrapping algorithm. The line in the lower right panel indicates a 1:1 correlation without performing a fit.}
 \label{fig:alphas_Y}
\end{figure}

A linear least squares fit of the lighter heavy element yttrium ($\mathrm{Z}=39$) over magnesium reveals a slope close to 1, indicating a similar production site of these elements (Fig.~\ref{fig:alphas_Y}) in all cases with sufficient amount of data (i.e., for Fornax, Sagittarius, and Sculptor). Plotting together Carina, Ursa Minor, and Draco shows a good agreement with a 1:1 correlation (lower right panel of Fig.~\ref{fig:alphas_Y}). This is a strong indication that yttrium may be produced by fast rotating massive stars \citep[e.g.,][]{Pignatari2008,Frischknecht2012} or CC-SNe (e.g., \citealt{Arcones2011,Hansen2014,Arcones2014}). At high metallicities, the s-process also contributes to the production of yttrium, best visible in the case of Fornax and Sculptor (and only weak indications in Sagittarius, see Fig.~\ref{fig:alphas_Y}), where the increased scatter at $\log \epsilon (\mathrm{Mg})>6$ indicates a contribution from a different process.

The details of the fitted slope depend to some extent also on the fitting procedure. We have applied four different fitting algorithms, one ordinary least squares fit, a weighted least squares fit with each point weighted by \mbox{$w=1/\sqrt{x_\mathrm{err}^2 +y_\mathrm{err}^2}$} (using the python package statsmodel.api.WLS), an orthogonal distance regression including errors in x- and y-direction (using the python package scipy.odr), and a bootstrapping aggregation in combination with a weighted least squares fit. For this, we used the same weights $w$, taking $80\%$ of all stars per estimator into account (using the python packages sklearn.ensemble.BaggingRegressor and sklearn.linear\_model.LinearRegression). The obtained slopes vary only slightly for Fornax, Sagittarius, and Sculptor. Therefore, we only give the value of the slope of a simple unweighted least squares fit (Fig.~\ref{fig:alphas_Y}). We note, however, that an orthogonal distance regression lead to a constant offset of the slope of $\sim 0.2$ toward higher values. 

\begin{figure}
   \centering
           \includegraphics[width=\hsize]{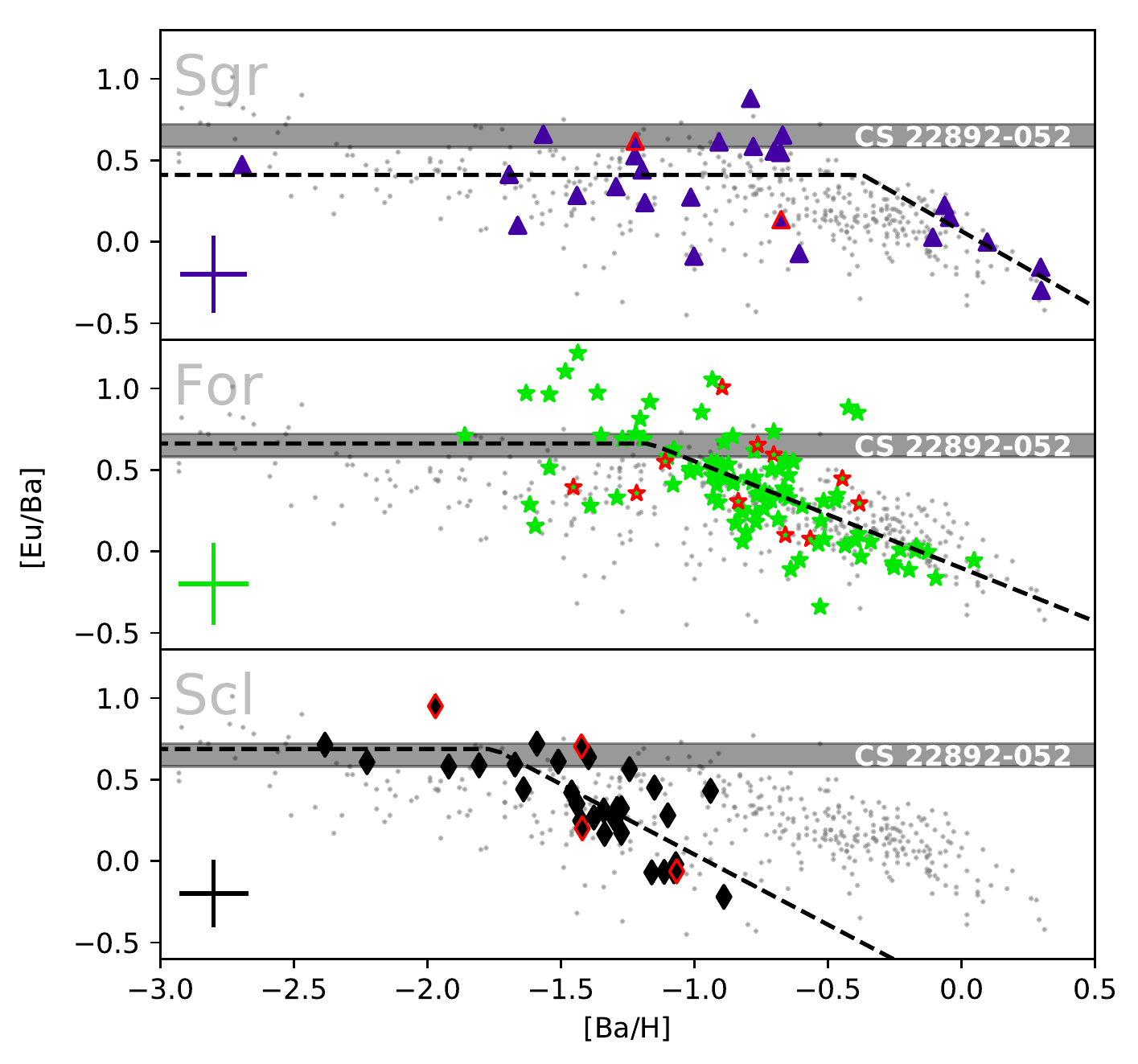}
      \caption{[Eu/Ba] as an indicator of the s-process contribution to Ba. Red symbols indicate stars with unreasonable parallax. The gray band shows the highly r-process enriched star CS 22892-052 \citep{Sneden2003}. Gray dots indicate stars of the MW \citep{Reddy2003,Cayrel2004,Reddy2006}. The median error is indicated in the lower left of each panel.}
         \label{fig:agb_start}
   \end{figure}

In Fig.~\ref{fig:agb_start}, we identify the point where barium is no longer predominantly produced by the r-process but rather by the s-process by using the trend of [Eu/Ba] versus [Ba/H] (cf., e.g., Fig.~20 in \citealt{Tolstoy2003} or Fig.~14 in \citealt{Tolstoy2009}). This is similar to, but less clearly seen in [Eu/Y] versus [Y/H] \citep[see, e.g., ][for a chemical evolution model of this trend]{Lanfranchi2008}. Barium is created in large amounts in the s-process. 
Hence, we expect the onset of the s-process to be at  metallicities that are similar to those of the $\alpha$-knee because the s-process is hosted by AGB stars, which are low-mass, slowly-evolving stars in a late phase of their evolution and thus may experience a time delay similar to type Ia SNe (as seen in Sculptor, \citealt{Hill2019}). The location on the [Ba/H]-axis, where [Eu/Ba] decreases translates to a metallicity (by fitting a linear function with an ordinary least squares fit to [Ba/H] over [Fe/H] of $-0.57\, \rm dex$, $-1.04\, \rm dex$, and $-1.57\, \rm dex$ for Sagittarius \citep[cf. Fig.~10 of][]{Hansen2018}, Fornax (cf. Fig.~20 of \citealt{Letarte2018} and Fig.~19 of \citealt{Lemasle2014}), and Sculptor \citep[cf. Fig.~12 of][]{Hill2019}, respectively. Thus, the onset of the s-process seems to occur at metallicities increasing with the stellar mass of the galaxy. For Sculptor it coincides with the position of the $\alpha$-knee, while for Sagittarius and Fornax this happens at a higher metallicity. This may be affected by uncertainties in determining the locus of the knee. A larger sample size would constrain the onset with more precision, especially in the case of Sagittarius. For low values of [Ba/H], the [Eu/Ba] ratio comes close to a constant value of the r-II\footnote{using the definition of \citet{Beers2005}} star CS 22892-052 \citep{Sneden2003}. In most dSphs, this level is reached at low metallicity. We therefore assume that barium is a clean trace of the r-process for $[\mathrm{Fe/H}]<-2$ and thus $\mathrm{Ba}_r=\mathrm{Ba}$. However, there may be individual stars at low metallicities that still have an s-process contribution from other sites such as rapidly rotating spin stars  \citep[e.g., ][]{Pignatari2008,Frischknecht2012,Chiappini2013}. 

\subsubsection{The r-process}
Recent Galactic chemical evolution (GCE) calculations indicate that the trend of r-process elements with respect to $\alpha$-elements or iron is thought to be flatter for production sites that have a similar delay time as CC-SNe (for GCE models of the MW see, e.g., \citealt{Matteucci2014,Cescutti2006,vandevoort2020}). GCE models that implement NSMs as the only source of the r-process produce positive slopes [r/$\alpha$]\footnote{with ``r'' being a short writing for an element that traces the r-process.} versus [$\alpha$/H] as the production of r-process elements sets in later than the production of $\alpha$-elements. This can be explained by the larger delay time of NSMs in comparison with CC-SNe. The expected slope is model-dependent and can be influenced by many parameters such as, for example, the delay time distribution and the rate of the event \citep[cf., Fig.~7 of][]{vandevoort2020}. In addition, the r-process may be hosted by several production sites, which would influence the slope as well. A final answer to distinguish different astrophysical production sites is therefore difficult to obtain and beyond the scope of this work. We can, however, estimate the range of slopes from different galaxies. For this, we look again for correlations of $\alpha$-elements (represented by magnesium) with $\log \epsilon (\mathrm{Eu})$. 

Due to the vanishing europium line at $\lambda = 6645.21$\,\AA\ at lower metallicities, the Pearson correlation coefficient \citep{Pearson1895} is low for all galaxies with more than five stars ($r_\mathrm{Pearson}<0.37$), except for Sagittarius (Fig.~\ref{fig:Eu_alpha_corr}). Supported by the large amount of high-resolution UVES spectra in Sagittarius, we were able to measure europium at lower metallicities. The high correlation coefficient in Sagittarius together with the good metallicity coverage of our sample this results in a highly correlated value of $r_\mathrm{Pearson}=0.92$. Sagittarius is therefore the only galaxy where a linear fit provides meaningful results. A linear least squares fit of $\log \epsilon (\mathrm{Eu})$ over $\log \epsilon (\mathrm{Mg})$ has a slope of $1.28 \pm 0.13$, close to a one-to-one correlation, indicating a main production of europium and $\alpha$-elements on similar timescales at least for Sagittarius (see Fig.~\ref{fig:Eu_alpha_corr}). Rare CC-SNe such as MR-SNe \citep{Winteler2012,Nishimura2015,Nishimura2017,Moesta2018} or collapsars \citep{Siegel2019} may be possible candidates (for a discussion see also \citealt{skuladottir2019}, cf. Fig.~11 of \citealt{vandevoort2020}). However, as the slope is not 1 within the uncertainty, but more positive, we cannot exclude a delayed contribution to Eu from NSMs.

\begin{figure}
   \centering
           \includegraphics[width=\hsize]{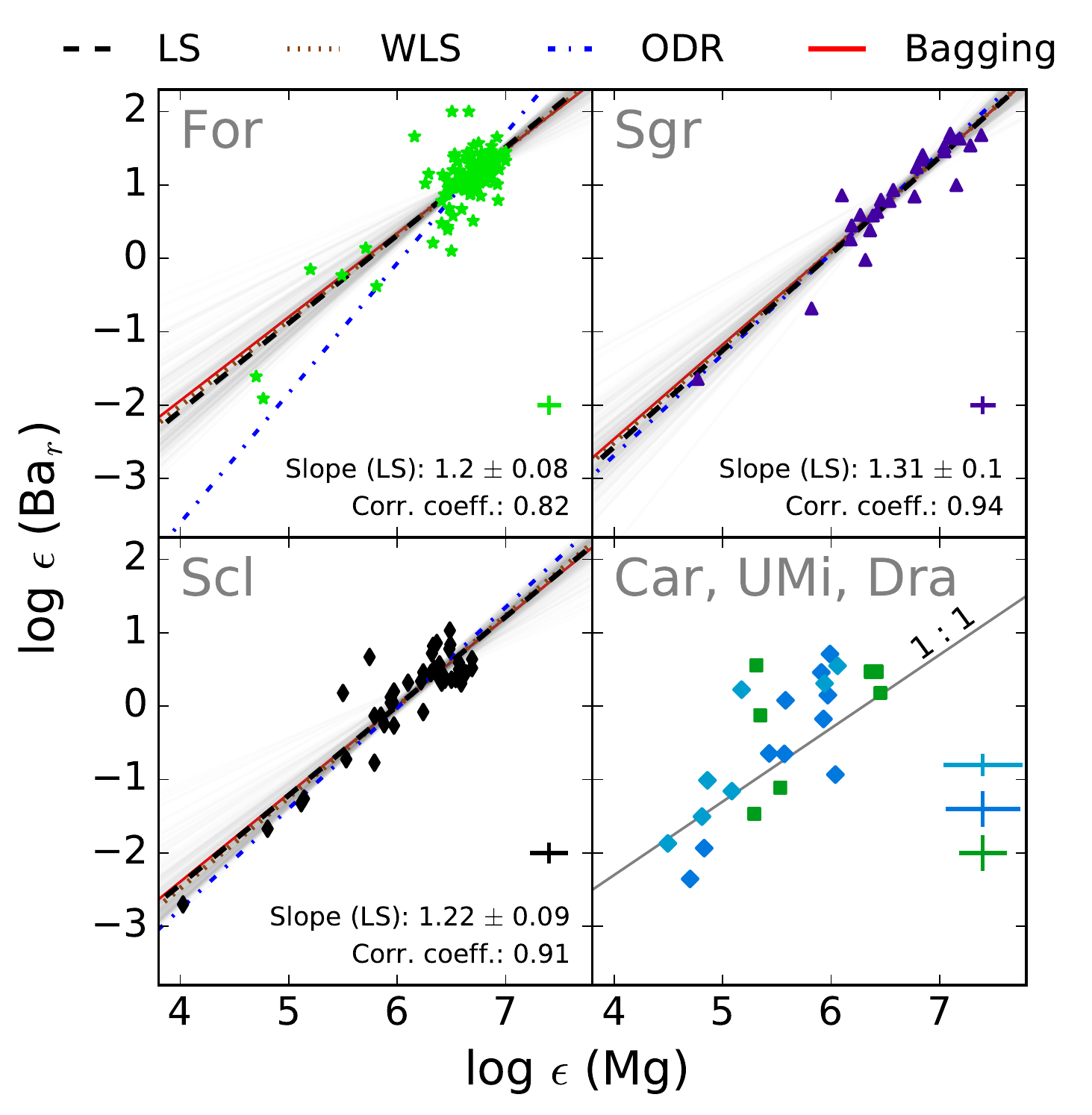}
      \caption{Same as Fig.~\ref{fig:alphas_Y}, but for $\log \epsilon (\mathrm{Ba_r})$ and $\log \epsilon (\mathrm{Mg})$ in six dSphs. We note that the subscript "r" denotes that Eu was used for high metallicites and Ba for low metallicities trying to filter out the s-process contribution as described in the text.}
         \label{fig:bar_alpha_corr}
   \end{figure}

For barium, we have a larger sample size due to the stronger absorption lines. As previously discussed, we can use barium as a tracer for the r-process below $\mathrm{[Fe/H] < -2}$, while at higher metallicities we use europium abundances. To bring both elements to the same scale, we assume that the r-process is robust and europium is only produced by the r-process. We can then convert our europium measurements at $[\mathrm{Fe/H}]>-2$ to the r-process fraction of barium abundances by assuming solar r-process residual ratios between these elements \citep{Sneden2008}. We calculate Ba$_r$ as \citep[similar to, e.g., ][]{Hansen2014}:
\begin{equation}
    \log \epsilon (\mathrm{Ba_r}) = 1.02 + \log \epsilon (\mathrm{Eu}).
\end{equation}
We define 
\begin{equation}
    \mathrm{[Ba_r/Fe]}=\left(\log \epsilon (\mathrm{Ba_r})-\log \epsilon (\mathrm{Fe})\right) - \left(\log \epsilon (\mathrm{Ba})_\odot-\log \epsilon (\mathrm{Fe})_\odot\right).
\end{equation}
A linear least squares fit of $\log \epsilon(\mathrm{Ba_r})$ and $\log \epsilon (\mathrm{Mg})$ in Fig.~\ref{fig:bar_alpha_corr} reveals a slope close to $1$, but with a much better correlation coefficient than in the case of $\log \epsilon (\mathrm{Eu})$. This holds for all dSph galaxies where we had sufficient data and therefore in a stellar mass range of $2.1\cdot 10^7 \, \mathrm{M}_\odot\,  -\, 2.9 \cdot 10^5\, \mathrm{M}_\odot $ for Sagittarius and Ursa Minor, respectively. We note that Fig.~\ref{fig:bar_alpha_corr} does not show Sextans. However, this galaxy is with a slope of $1.39 \pm 0.31$ also in agreement with the other galaxies. Similar results have been obtained using $\log \epsilon (\mathrm{Mg_{NLTE}})$ instead of $\log \epsilon (\mathrm{Mg})$, which demonstrates that the employed NLTE correction does not have a major impact on the derived slopes. Literature values of Reticulum II from \citet{Ji2016b} lead to a slope of $1.11\pm 0.37$ indicating a production of $\alpha$-elements and Ba$_r$ on comparable delay times also in UFD. 

The occurrence of n-capture elements at very low metallicities causes a problem when arguing for NSMs as the only source of r-process elements. Previous studies \citep[e.g., ][]{Argast2004,Matteucci2014,Cescutti2015,Wehmeyer2015,Haynes2019,Wehmeyer2019} highlighted the problem of producing an early enrichment of n-capture elements at low metallicities $\mathrm{[Fe/H]}<-3$ due to the delay of NSM events as well as the inferred iron abundance that will be mixed in by the previous CC-SNe (since neutron stars are the final product of CC-SNe). Possible solutions to this problem have been suggested such as the pollution of the interstellar medium (ISM) with material from proto-galaxies \citep[e.g., ][]{Komiya2016}. Furthermore, neutron star kicks may move the NSM event far away from the birthplace of the neutron stars and therefore escape the previously synthesized iron \citep{Wehmeyer2019}. \citet{skuladottir2019} argued that this would, however, indicate that the MW is as efficient as lower mass dwarfs to maintain the neutron star in the system, which may be unlikely. Another solution could be the occurrence of a black hole-neutron star merger \citep{Wehmeyer2019}.

We confirm the enhancement of [Eu/Mg] for Fornax and Sagittarius at high metallicities as discussed in \citet{skuladottir2019} and \citet{Skuladottir2020}. These supersolar values are still an outstanding problem. We note, however, that the metal-rich stars in Sagittarius and Fornax are extremely similar in their stellar parameters. Therefore this issue may be related to a systematic uncertainty when analyzing these stars (e.g., LTE or 3D effects when modeling the structure of the photosphere).

The MW shows a decreasing trend of [Eu/Fe] for high metallicities. \citet{Cote2019} and \citet{Simonetti2019} argued that this decrease is directly related to a different delay time distribution of NSMs or that there must be an additional source of r-process elements that dominantly contributed in the early universe causing the plateau followed by the declining trend \citep[cf. Fig.~2 of][]{Cote2019}. GCE models that assume a similar delay time distribution (e.g., $t^{-1}$) for both events (i.e., NSMs and type Ia SNe) are unable to reproduce this decreasing trend. 
All dSph galaxies with sufficient data, namely Sagittarius, Fornax, Sculptor, and Sextans (Fig.~\ref{fig:ba_eu}) show this decreasing trend as well. This decrease occurs at similar metallicities as the $\alpha$-knee (see Table~\ref{tab:alphaknee}) and is therefore connected to the additional iron contribution of type Ia SNe. The presence of this decrease in the dSph galaxies favors a similar production site of n-capture elements as in the MW.
  \begin{figure}
   \centering
           \includegraphics[width=\hsize]{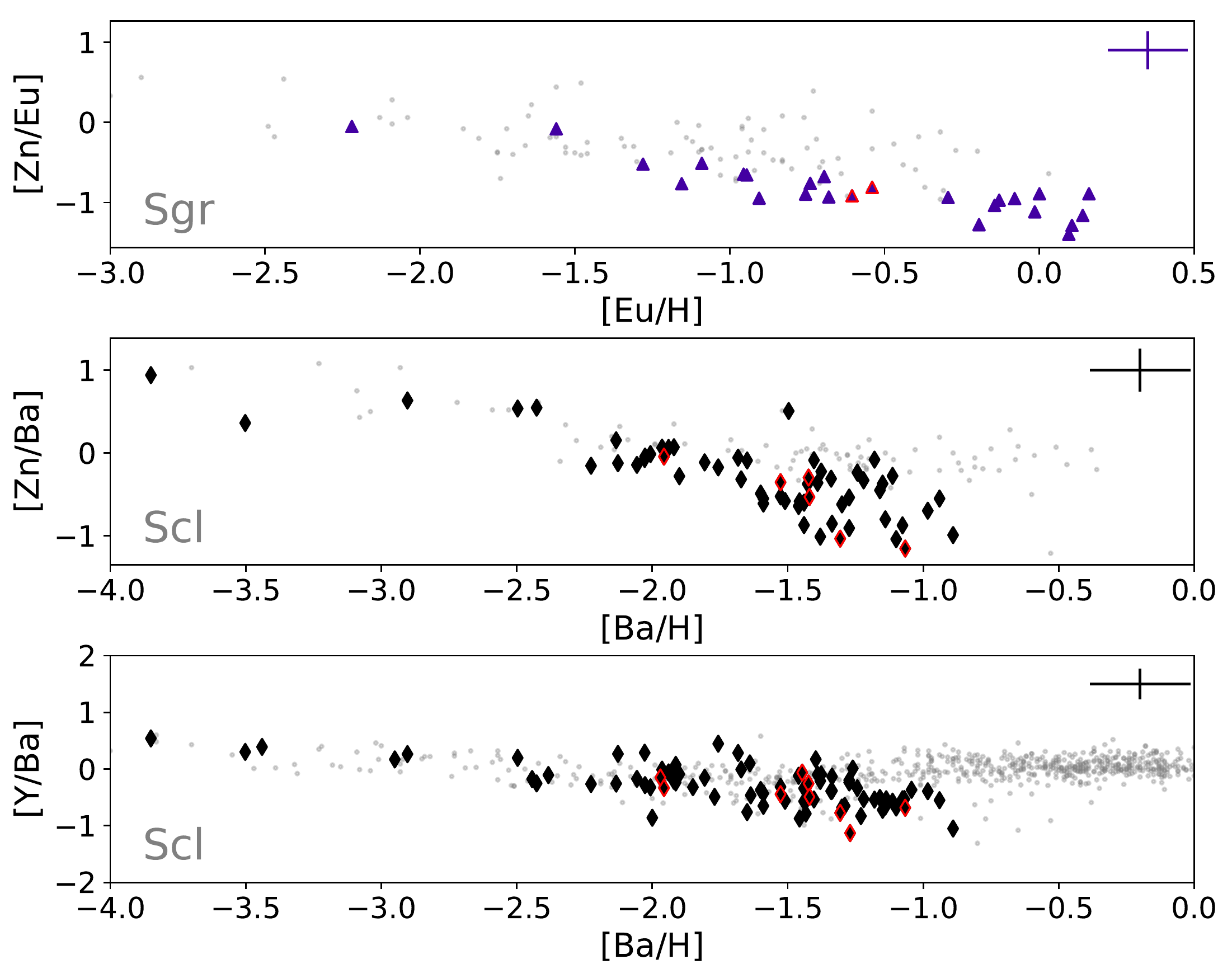}
      \caption{$\mathrm{[Zn/Eu]}$ ratio over $\mathrm{[Eu/H]}$ for Sagittarius (upper panel), $\mathrm{[Zn/Ba]}$ over $\mathrm{[Ba/H]}$ for Sculptor (middle panel), and $\mathrm{[Y/Ba]}$ over $\mathrm{[Ba/H]}$ for Sculptor (lower panel). Gray dots indicate stars of the MW \citet{Reddy2003}, \citet{Cayrel2004}, \citet{Reddy2003}, \citet{Ishigaki2013},  \citet{Fulbright2000}, \citet{Nissen1997}, \citet{Prochaska2000}, \citet{Stephens2002}, \citet{Ivans2003}, \citet{McWilliam1995}, \citet{Ryan1996}, \citet{Gratton1988}, \citet{Edvardsson1993}, \citet{Johnson2002}, and \citet{Burris2000}. The symbols were chosen as in Fig.~\ref{fig:mg_evolution}. Stars marked in red indicate stars with close distances according to the parallax (Appendix~\ref{sct:appendix_surf_gravs}). Median errors are indicated at the upper right corner of each panel.}
         \label{fig:zn_ba_eu}
   \end{figure}
   
In our study, we determined the abundance of the heavy iron-peak element zinc. The ratios [Zn/Ba] and [Zn/Eu] indicate a plateau at low metallicities, followed by a decreasing trend (Fig.~\ref{fig:zn_ba_eu}). As for the $\alpha$-elements, the ratio is indistinguishable from the MW for low metallicities and drops to subsolar values for higher metallicities. Though being limited to only two stars in Sagittarius and four stars in Sculptor, the existence of this plateau may suggest a production of zinc, europium, and barium on the same timescales (for values below [Ba/H]$\sim -2.3$). A decrease is also visible in the ratio [Y/Ba]. The locus of the decrease does not agree with the value where [Eu/Ba] decreases (at $\mathrm{[Ba/H]} \sim -1.6$, see Fig.~\ref{fig:agb_start}), so it is most likely disconnected from an s-process contribution. 
We note, however, that our data may not be sensitive enough to separate between different timescales at this small time interval.

   \begin{figure}
   \centering
          \includegraphics[width=\hsize]{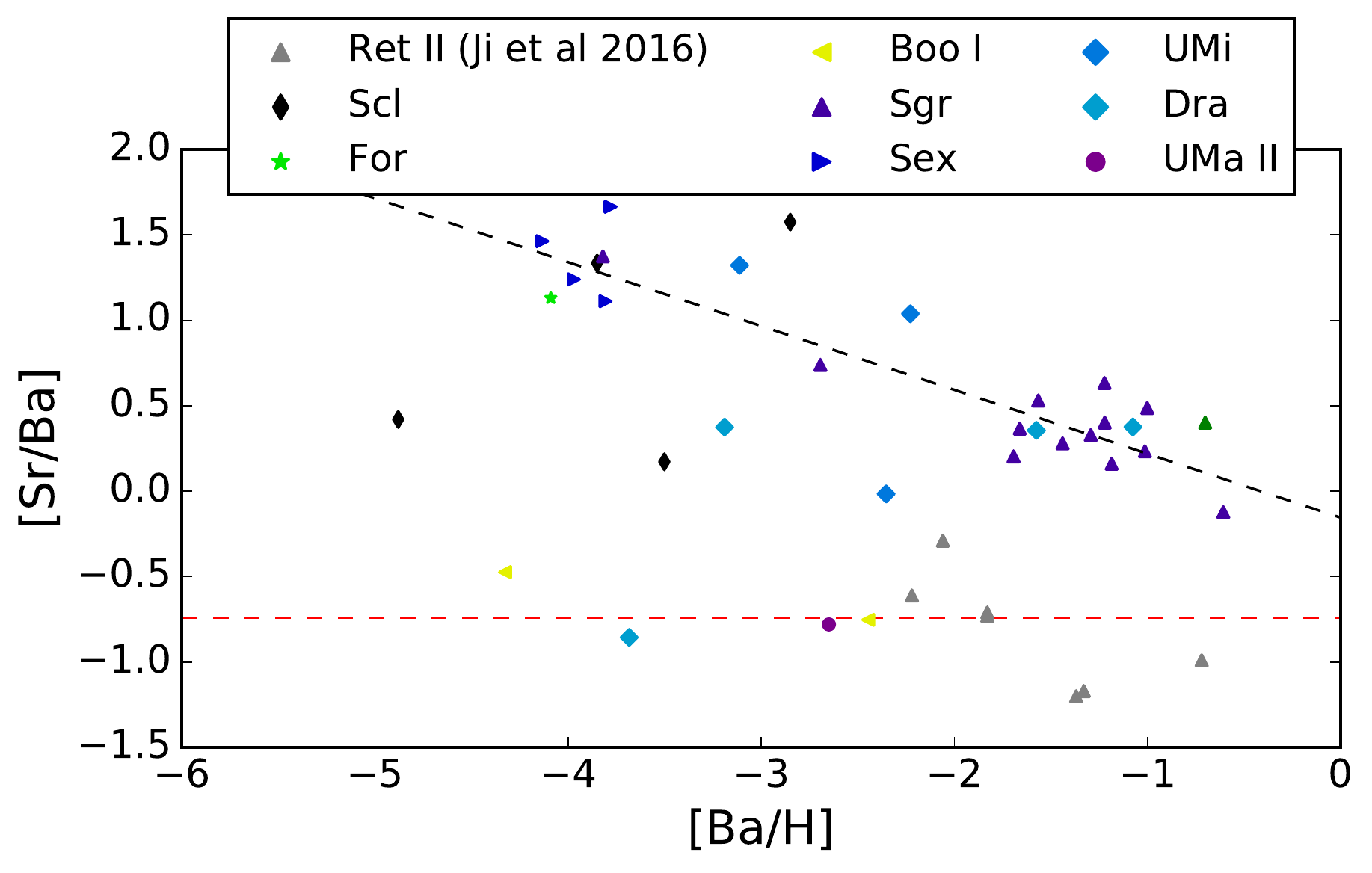}
      \caption{[Sr/Ba] over [Ba/H]. The green triangle indicates a star of the globular cluster Terzan 7. The dashed red line corresponds to a median of all UFD galaxies, the dashed black line is a linear least squares fit to Sagittarius.}
         \label{fig:srbah}
   \end{figure}

We find detectable barium and strontium abundances down to metallicities of $\mathrm{[Fe/H]} \sim -4$ in Sculptor \citep[cf.,][]{Kirby2009, Frebel2010a}, and $\mathrm{[Fe/H]} \lesssim -3$ for nearly all investigated dSph galaxies independent of their mass (Fig.~\ref{fig:ba_eu}).\\ 
The formation of strontium and barium seems to be fundamentally different when comparing UFD galaxies with dSph galaxies or the MW. \citet{Mashonkina2017b} found a deficiency of [Sr/Mg] for UFD galaxies compared to classical dSph galaxies and the MW, whereas [Ba/Mg] follows the same trend in dwarf galaxies of different sizes. They concluded that there are at least two production sites of strontium, one that is coupled to the production of barium and one that is independent from barium. These production sites are present in the MW halo and the dSph galaxies, but the barium-independent production channel of strontium seems to be missing in the UFD galaxies leaving the strontium lower here. Unfortunately, our statistics of strontium in UFD galaxies is limited due to the wavelength coverage of our sample (see Fig.~\ref{fig:wavcover}). 
However, adding the literature values from \citet{Ji2016b} supports this suggestion. 
The UFD galaxies seem to follow a flat trend in $\mathrm{[Sr/Ba]}$ over $\mathrm{[Ba/H]}$ indicated by a red dashed line in Fig.~\ref{fig:srbah} (compare also with Fig.~5 of \citealt{Ji2016b} and Fig.~8 of \citealt{Mashonkina2017b} or for the MW with Fig.~13 of \citealt{Honda2004}). Compared to \citet{Mashonkina2017b}, we derive a slightly lower value of $\mathrm{[Sr/Ba]} = -0.74\, \rm dex$ for pure r-process stars. This may be explained by the LTE assumptions in our analysis. Our data suggest that more massive galaxies follow a decreasing trend of $\mathrm{[Sr/Ba]}$ with increasing [Ba/H], which points to more than one production channel of strontium. Due to the limited amount of data for UFDs, we cannot draw any firm conclusion about the flat trend of $\mathrm{[Sr/Ba]}$ in those galaxies. For this, a larger sample size with strontium measurements is desirable.

\section{Summary and conclusions}
\label{sct:summary_and_conclusion}
We created a large homogeneous spectroscopic abundance catalog out of $380$ stars of the $13$ dSph and UFD galaxies Fornax, Sculptor, Sextans, Carina, Sagittarius, Ursa Minor, Draco, Leo I, Bootes I, Ursa Major II, Segue I, Triangulum II, and Reticulum II. This catalog includes Mg, Sc, Ti, Mn, Cr, Fe, Ni, Zn, Sr, Y, Ba, and Eu abundances, spanning a wide range of metallicities ($-4.18\le \mathrm{[Fe/H]} \le -0.12$). Many abundances of these stars were derived for the first time. 
We investigated common methods and their possible systematic errors when deriving abundances. For a subset of common stars, we compared our derived abundances with previous studies and found an overall agreement, but strongly revised the barium abundances in Fornax. We showed that the previously reported high values in the Fornax dSph galaxy are an artifact of the abundance analysis that had been applied in the past.
Despite applying self-consistent methods within some literature studies, abundance offsets or biases can occur. Hence, when using automated tools, it is important to apply a careful treatment of blends in the crowded spectra of cool giants. 

We identified a relation of $\alpha$-elements and the absolute magnitude (and therefore the mass) of the galaxy. With this relation we are able to predict the evolution of $\alpha$-elements by knowing only the absolute magnitude of the system. This demonstrates that the SFH dominantly depends on the galaxy mass for Fornax, Sagittarius, Sculptor, Sextans, and Ursa Minor, whereas the IMF stays fairly similar. 

Following this idea, the scatter in Mg in the Sextans dSph may be a possible fingerprint of a galaxy merger or accretion event. We speculate that this smaller galaxy should have had a stellar mass of $\sim 10^4 \, \mathrm{M}_\odot$. 

We investigated the onset of the s-process for Sagittarius ([Fe/H]$=-0.57$), Fornax ([Fe/H]=$-1.04$) and Sculptor ([Fe/H]$=-1.57$). [Eu/Ba] versus [Ba/H] shows a decreasing trend for high Ba contents similar to the MW. The onset of this decrease marks the point where the s-process contribution of Ba starts to dominate over the r-process contribution.  Somewhat intuitively but excitingly, we detect an onset of the s-process which seems to occur at metallicities increasing with the galaxy's stellar mass. We note a late onset in Sagittarius and Fornax, relative to that of type Ia SNe. 

Finally, we discussed the implication of the measured Eu and Ba abundances on the production site of the r-process. We investigated trends of $\mathrm{\log \epsilon (Eu)}$ and $\mathrm{\log \epsilon (Ba_r)}$ versus $\mathrm{\log \epsilon (\mathrm{Mg})}$, which favors rare CC-SNe (or other events with a small time delay compared to standard CC-SNe) as production site for heavy elements in these systems. The rare CC-SNe might have had strong jets, which could explain the correlation of scandium with the heavy elements in the more massive galaxies but probably not in Sextans and Ursa Minor. We analyzed the trend in all investigated galaxies separately to reveal a possible mass dependence of the production of r-process elements. The trends did not show any mass dependence within the statistics of our study in a stellar mass range of $2.1\cdot 10^7 \, \mathrm{M}_\odot\,  -\, 2.9 \cdot 10^5\, \mathrm{M}_\odot $ spanned by Sagittarius and Ursa Minor, respectively. 

All investigated dSph galaxies are remarkably similar to the MW in the following sense. They do not show any indication of missing production sites for heavy n-capture elements (i.e., Ba and Eu). As for the MW, these systems therefore indicate the need for more than one production site of at least the lighter heavy elements. In addition to the known existence of NSMs hosting the r-process, there should be a production site that acts on the timescales of standard CC-SNe in order to explain the flat trend of $\mathrm{[Ba_r/Mg]}$. More observations would further constrain the derived slope and further GCE models may shed light into the expected slopes when assuming NSMs or rare CC-SNe as dominant production sites of heavy elements. 

Future work may extend the present catalog by implementing raw spectra and data of other spectrographs. Furthermore, the here presented spectra still include information of other elements that we have not analyzed so far such as Ca, Si, and Cu. Upcoming, future observations and surveys of the Local Group will provide important additions to this homogeneous catalog and help answer remaining open questions on the chemical enrichment of the 13 studied systems.

\begin{acknowledgements}
The authors thank Maria Bergemann, Marius Eichler, Pascale Jablonka, Chiaki Kobayashi, Andreas Koch, and Mikhail Kovalev for valuable discussions and comments. M.R. and A.A. were supported by the ERC starting grant EUROPIUM (grant No. 677912), Deutsche Forschungsgemeinschaft through SFB 1245. C.J.H. acknowledges support from the Max Planck Society. \'A.S.~acknowledges funds from the Alexander von Humboldt Foundation in the framework of the Sofja Kovalevskaja Award endowed by the Federal Ministry of Education and Research. M.H. and E.K.G. gratefully acknowledge support by Sonderforschungsbereich SFB 881 "The Milky Way System" (Project-ID 138713538, subprojects A03, A08) of the German Research Foundation (DFG).
This work was supported by the “ChETEC” COST Action (CA16117), supported by COST (European Cooperation in Science and Technology).

This research has made use of the Keck Observatory Archive (KOA), which is operated by the W. M. Keck Observatory and the NASA Exoplanet Science Institute (NExScI), under contract with the National Aeronautics and Space Administration. Some of the data presented herein were obtained at the W. M. Keck Observatory, which is operated as a scientific partnership among the California Institute of Technology, the University of California and the National Aeronautics and Space Administration. The Observatory was made possible by the generous financial support of the W. M. Keck Foundation. Furthermore, we made use of the services of the ESO Science Archive Facility under request number mreichert 373937, 376650, 376679, and 377450. This involves data from the program IDs: 074.B-0415, 075.B-0127, 076.B-0146, 079.B-0672, 079.B-0837, 080.B-0784, 081.B-0620, 081.D-0286, 083.B-0254, 083.B-0774, 083.D-0688, 084.D-0522, 085.D-0141, 085.D-0536, 087.D-0928, 091.C-0934, 091.D-0912, 092.B-0194, 092.B-0564, 093.D-0311, 095.D-0184, 095.D-0333, 098.B-0160, 099.D-0321, 171.B-0588, 193.B-0936, 281.B-5022, 291.B-5036, 383.B-0038, 383.B-0093, 385.D-0430, 65.H-0375, 65.N-0378, 66.B-0320, 67.B-0147, 71.B-0146, 71.B-0641. In addition, we made use of the VizieR catalog access tool, CDS, Strasbourg, France , the SIMBAD database, operated at CDS, Strasbourg, France, and of the NASA/IPAC Extragalactic Database (NED) and the Infrared Science Archive (IRSA), which are operated by the Jet Propulsion Laboratory, California Institute of Technology, under contract with the National Aeronautics and Space Administration. Within this work, several python packages, including Astropy, a community-developed core Python package for Astronomy, Pandas, SciPy, NumPy, astroquery, and matplotlib, a Python library for publication quality graphics were used. This research has made use of NASA's Astrophysics Data System. This work has made use of data from the European Space Agency (ESA) mission {\it Gaia} (\url{https://www.cosmos.esa.int/gaia}), processed by the {\it Gaia} Data Processing and Analysis Consortium (DPAC, \url{https://www.cosmos.esa.int/web/gaia/dpac/consortium}). Funding
for the DPAC has been provided by national institutions, in particular
the institutions participating in the {\it Gaia} Multilateral Agreement.  
\end{acknowledgements}

\bibliographystyle{aa}
\bibliography{literature}

\begin{appendix}  

\section{Residual of solar abundances}
\begin{figure}
   \centering
   \includegraphics[width=\hsize]{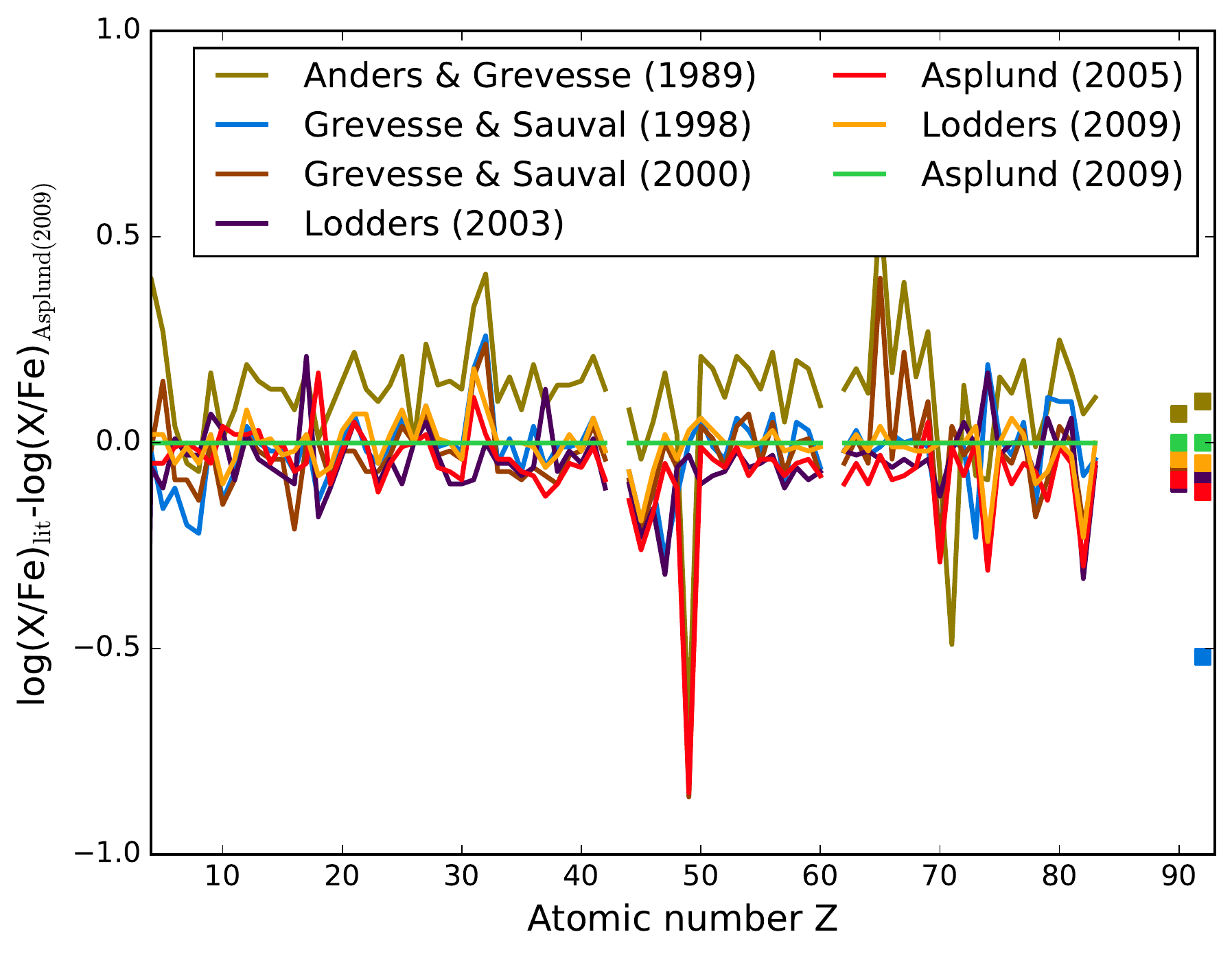}
      \caption{Residuals of solar abundances on the scale of \citet{Asplund2009}. The actinides are shown with symbols. Solar abundances are taken from \cite{Anders1989,Grevesse1998,Grevesse2000,Lodders2003,Lodders2009,Asplund2005,Asplund2009}.}
        \label{fig:solarabu}
   \end{figure}

The choice of solar abundances can introduce additional offsets when not converting them properly. Figure~\ref{fig:solarabu} shows the possible systematic offset when mixing various literature values of solar abundances.  The large difference in, for example, solar indium (with atomic number $Z=49$) abundances is due to the difference in photospheric versus meteoric abundances.
   
\section{Effective temperatures}
\label{appendix:effective_temp}
\begin{figure*}
   \centering
   \includegraphics[width=0.7\hsize]{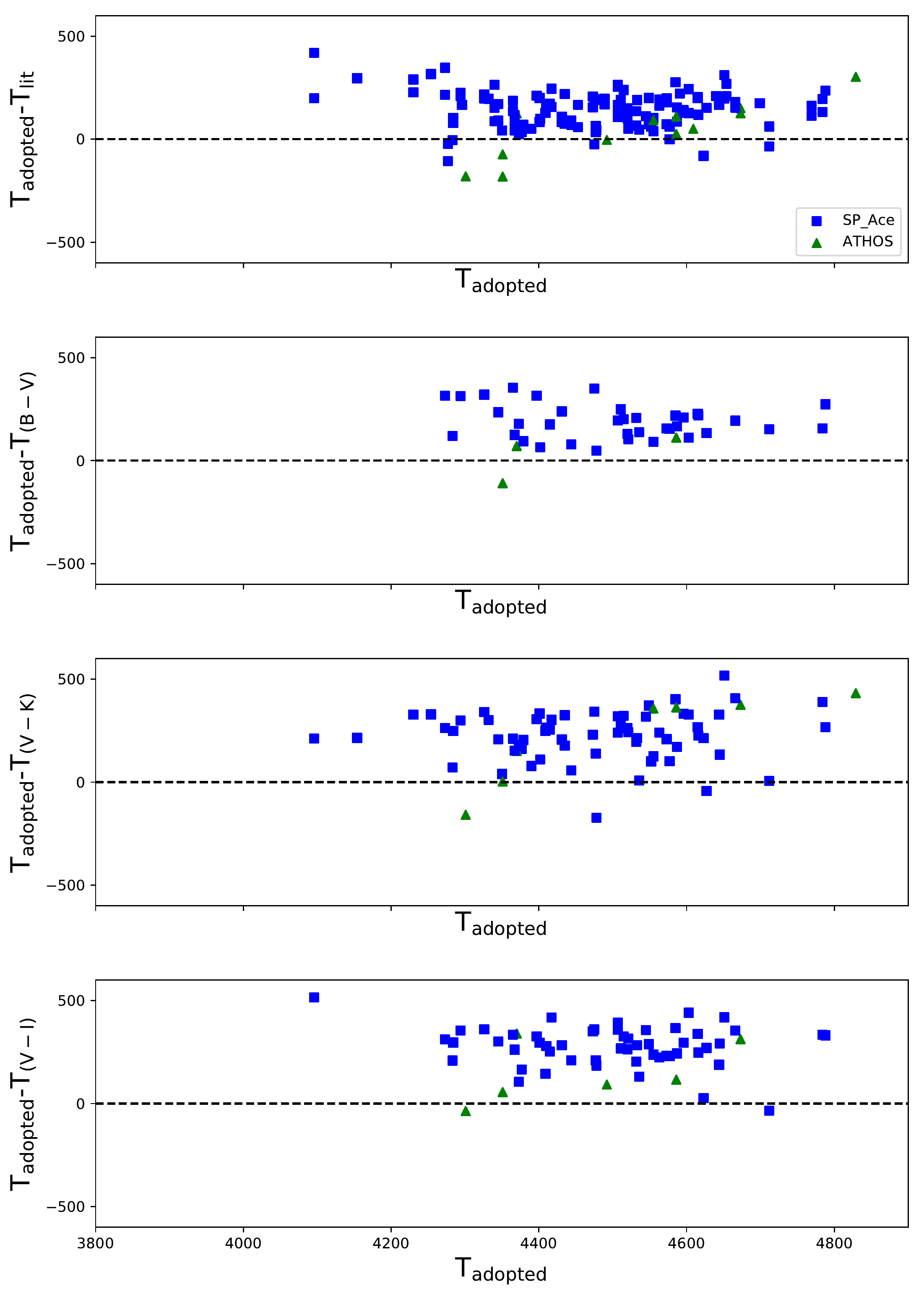}
      \caption{Derived effective temperature in comparison with photometric and literature values for stars in Sculptor. The comparison with literature includes \citet[phot.]{Hill2019}, \citet[phot.]{Kirby2010}, and \citet[spec.]{Simon2015}.}
         \label{fig:temperature_comp}
   \end{figure*}

\begin{figure}[htb]
   \centering
   \includegraphics[width=\hsize]{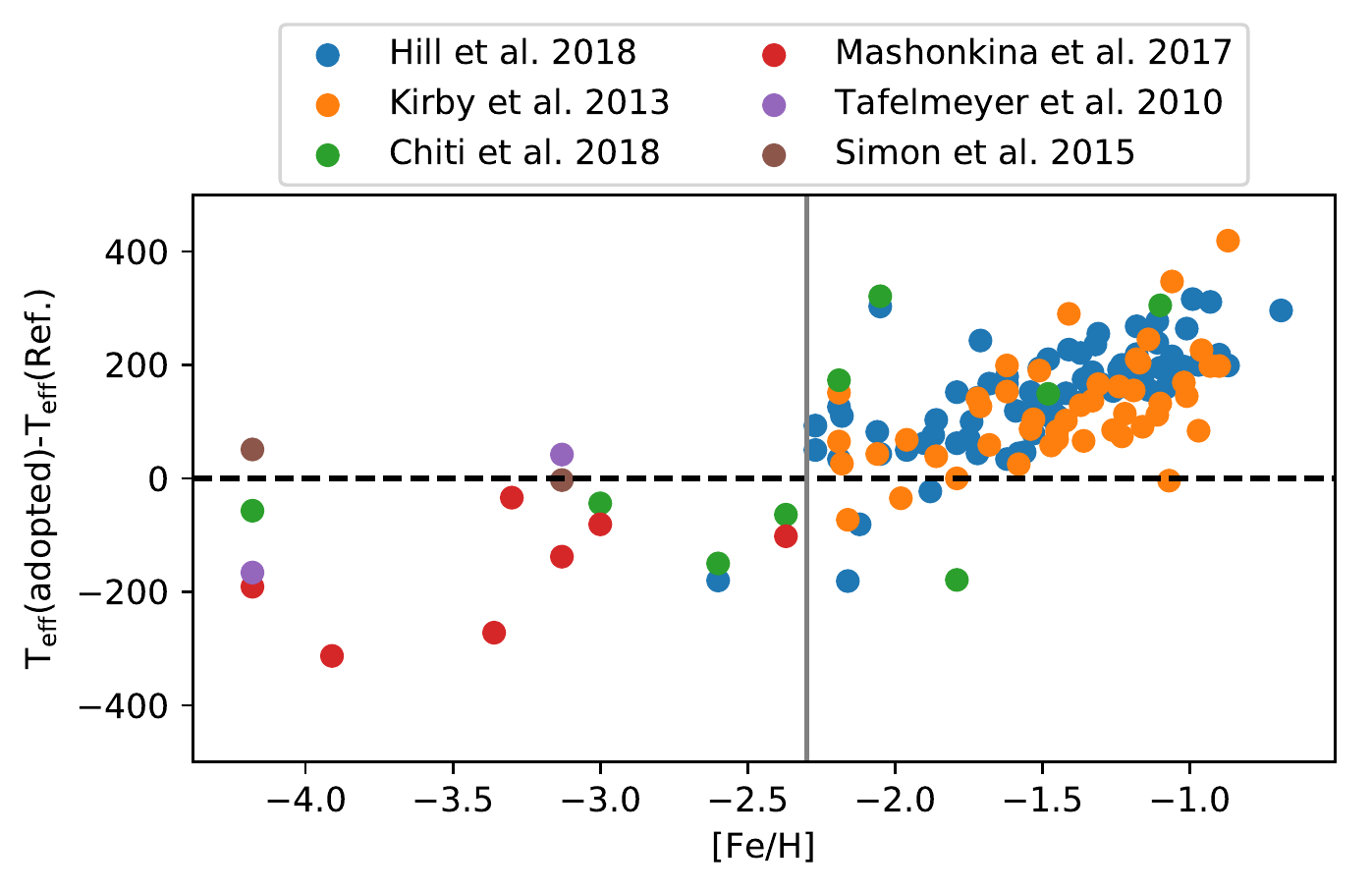}
      \caption{Metallicity versus effective temperature residuals for Sculptor stars. The solid, vertical line indicates the transition in methods from ATHOS (lower [Fe/H]) to SP\_Ace (higher [Fe/H]). The comparison samples ared comprised of \citet{Hill2019}, \citet{Kirby2013}, \citet{Chiti2018}, \citet{Mashonkina2017}, \citet{Tafelmeyer2010}, and \citet{Simon2015} as indicated by the different colors (see legend).}
        \label{fig:sculptor_temp_app}
   \end{figure}

\begin{figure}
   \centering
   \includegraphics[width=\hsize]{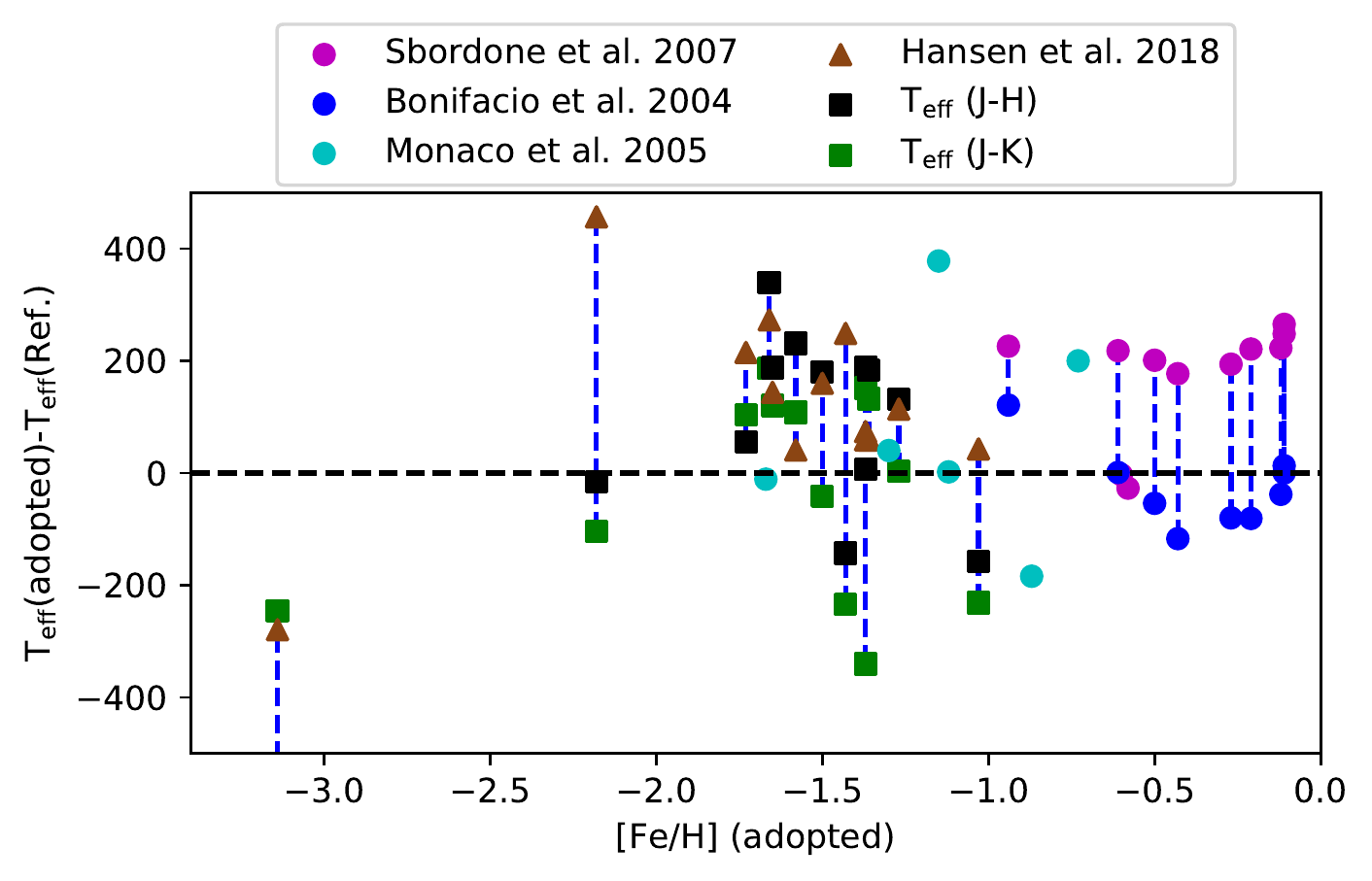}
      \caption{Metallicity versus effective temperature residuals for Sagittarius stars. Literature values of the same star are connected with blue dashed lines. The comparison sample is comprised of \citet{Sbordone2007}, \citet{Bonifacio2004}, \citet{Monaco2005}, and \citet{Hansen2018}. We calculated J-H and J-K effective temperatures for the sample of \citet{Hansen2018} with 2MASS photometric data taken from the catalog of \citet{Zacharias2004} - see legend for details.}
        \label{fig:sag_temp_app}
   \end{figure}

We compared our derived effective temperatures with literature values and photometrically derived temperatures. On average our temperatures are slightly offset to higher values (see Fig.~\ref{fig:temperature_comp}).
We also note a trend in the effective temperatures versus metallicity when comparing the sculptor sample with \citet{Hill2019} and \citet{Kirby2013}. This trend is more pronounced when comparing to \citet{Hill2019} and the difference in effective temperature rises from $\sim 0 \, \mathrm{K}$ to $\sim 250 \, \mathrm{K}$ at $\mathrm{[Fe/H] \approx -2}$ and $-1$, respectively (Fig.~\ref{fig:sculptor_temp_app}). 
Stellar parameters of \citet{Kirby2013} dominantly rely on photometric measurements, but they are changed within a reasonable range ($\sim250$\,K) using the information of the spectra \citep[see Sect. 4.5 of][]{Kirby2010}. This may be the reason that the trend is not as pronounced and the effective temperature reaches on average a smaller deviation of $150\, \mathrm{K}$ at $\mathrm{[Fe/H]\approx -1}$. We note that the large standard deviation of $226 \, \mathrm{K}$ given in Table~\ref{tab:temp_lit_comp} is driven by mainly one star (\object{2MASS J01001759-3346552}) that differs from our temperature by $\sim 1250\, \mathrm{K}$. We do not find any trend with metallicities when comparing Sagittarius stars with stellar parameters from \citet{Bonifacio2004}, \citet{Monaco2005}, \citet{Sbordone2007}, and \citet{Hansen2018} (Fig.~\ref{fig:sag_temp_app}). We note that the difference in effective temperatures between \citet{Bonifacio2004} and \citet{Sbordone2007} is due to a different assumed extinction (see \citealt{Sbordone2007} for a discussion).
In addition, we compared the effective temperatures from SP\_Ace and ATHOS for stars where both codes operate within their calibrated region. For this test, also no trend was visible. \citet{Mucciarelli2020} investigated possible differences between spectroscopic and photometric derived stellar parameters for stars in globular clusters and found deviations for lower metallicity stars. Their spectroscopic parameters were inconsistent with the position of the stars in the color-magnitude diagram likely due to NLTE effects increasing with decreasing metallicity. In contrast, we observe a trend for metal-rich stars, whereas the more metal-poor ones agree. This could point toward blended, strong or saturated lines loosing their temperature sensitivity.

\section{Used linelist}
The used spectral lines together with their atomic input and applied weights are shown in Table~\ref{tab:linelist}. In addition, we list the number of measurements for each absorption line.
 \begin{table*}
\caption{Used linelist. The first column lists the element together with the ionization state, and the second column the wavelength of the absorption line. The third and fourth column show excitation potential and oscillator strength, respectively. The fifth column contains the weight given to the abundance of the absorption feature. The final two columns show the number of stars for that we measured this absorption feature and the source of the atomic data.}             
\label{tab:linelist}      
\centering                          
\begin{tabular}{lrrrccr}
\hline\hline                 
Element & Wavelength [$\AA$] & Excitation potential [eV] & log(gf) & Weights & \# Stars& Lit.\\
\hline                        
Mg I & 5183.60      & 2.700 & -0.17 & 1.0 &49 & 1\\
Mg I & 5528.48      & 4.330 & -0.51 & 1.0 &348& 1\\
Mg I & 6318.72      & 5.104 & -2.02 & 1.0 &138& 1\\
Mg I & 6319.24      & 5.104 & -2.24 & 1.0 &138& 1\\
Mg I & 6319.50      & 5.104 & -2.72 & 1.0 &138& 1\\
Sc II& 5526.80      & 1.767 &  0.13 (HFS) & 1.0 &319& 2\\
Sc II& 6309.90      & 1.496 & -1.52 (HFS) & 1.0 &128& 2\\
Ti II& 5418.77      & 1.582 & -2.13 & 1.0 &293& 3\\
Ti II& 6606.95      & 2.060 & -2.79 & 1.0 &104& 4\\
Cr I & 5409.77      & 1.029 & -0.67 & 1.0 &270& 5\\
Cr I & 6330.09      & 0.941 & -2.92 & 1.0 &147& 5\\
Mn I & 5420.36      & 2.141 &  -1.46 (HFS)  & 1.0 &210& 6\\
Mn I & 5432.55      & 0.000 &  -3.80 (HFS)  & 1.0 &230& 6\\
Mn I & 6013.51      & 3.070 &  -0.35 (HFS)  & 1.0 &36& 7\\
Mn I & 6021.82      & 3.073 &  -0.05 (HFS)  & 1.0 &36& 7\\
Ni I & 5476.92      & 1.825 & -0.89 & 0.3 &351& 8\\
Ni I & 6176.82      & 4.085 & -0.26 & 1.0 &133& 9\\
Ni I & 6177.25      & 1.825 & -3.51 & 1.0 &126& 10\\
Ni I & 6643.56      & 1.676 & -2.30 & 0.5 &335& 11\\
Zn I & 4810.54      & 4.075 & -0.17 & 1.0   &113& 12\\
Sr II &4077.71      & 0.000 & 0.16 (HFS)    & 1.0 &26  & 13\\
Sr II &4215.52      & 0.000 & -0.16 (HFS)    & 1.0 &36  & 13\\
Y II  &4883.68      & 1.084 & 0.07  & 1.0   & 131&6\\ 
Y II  &5402.78      & 1.838  & -0.51 & 1.0   & 97&6\\     
Ba II &5853.69      & 0.604 & -1.01 (HFS)   & 1.0 &60  & 14\\
Ba II &6141.73      & 0.704 & -0.08 (HFS)   & 0.5 & 332&14\\
Ba II &6496.90      & 0.604 & -0.38 (HFS)   & 1.0  &340 & 14\\
Eu II &4205.05      & 0.000 & 0.21 (HFS)   & 1.0   &31& 2\\
Eu II &6645.21      & 1.379 & 0.12 (HFS)   & 1.0  &181 & 2\\

\hline                                   
\end{tabular}
\tablebib{(1)~\citet{Pehlivan2017}; (2)~\citet{Lawler1989}; (3)~\citet{Wood2013}; (4)~\citet{Martin1988b}; (5)~\citet{Sobeck2007}; (6)~\citet{Kurucz2011}; (7)~\citet{DenHartog2011}; (8)~\citet{Roederer2012}; (9)~\citet{Wood2014}; (10)~\citet{Kostyk1982}; (11)~\citet{Lennard1975}; (12)~\citet{Biemont1980}  (13)~\citet{Bergemann2012};  (14)~\citet{Gallagher2012};  } 
\end{table*}

\section{Surface gravities from parallaxes}
\label{sct:appendix_surf_gravs}
As a test, we also derived surface gravities from parallaxes and derived values that are closer to the literature for the most uncertain parallaxes. In the following, we demonstrate our findings with the example star $[$LPP2003$]$ 420. We determine a heliocentric radial velocity of $v_\mathrm{helio} = 232.45\, \rm km\, s^{-1}$ (compare to the systemic velocity of Sextans of $v_\mathrm{sys,helio}=226.3 \pm 0.6 \, \rm km \, s^{-1}$ and a global dispersion of $\sigma _\mathrm{helio}=8.8 \pm 0.4 \, \rm km \, s^{-1}$, \citealt{Battaglia2011}). This radial velocity together with the EW of a magnesium absorption line to discriminate between a dwarf and a giant star, let \citet{Battaglia2011} derive the conclusion that this star is likely a member of the Sextans dSph galaxy. The magnitude of this star is given by $V=18.768 \pm 0.010$ \citep{Lee2003} and $G=18.4973 \pm 0.0031$ \citep{Gaia2018}. The reddening is given by $A_c(V) = 0.115$ \footnote{https://irsa.ipac.caltech.edu/applications/DUST/ , accessed on 13. October 2019} and the parallax from \citet{Gaia2018} is $\pi = 1.6914 \, \rm mas$ with $\sigma( \pi )/\pi = 0.35$. This parallax corresponds to a distance of $d=1/\pi \approx 591 \pm 208 \, \rm pc$. The distance to the Sextans dSph galaxy is $d_{\mathrm{sex}}=86 \, \rm kpc$ (see Table~\ref{tab:distances_dwarf}). With the parallax, we obtain a much higher surface gravity of $5.98 \ge \log g \ge 5.77$ compared to other methods. A mismatch of this star could be possible either by SIMBAD, NED, ESO headers, or {\it{Gaia}}. Another explanation for this could be that the star is a foreground star. Table~\ref{tab:parallax_dist} lists $38$ stars with contradictory parallaxes. Given the error of the parallax $\sigma(\pi_\mathrm{Star})$ and assuming a Gaussian distribution (this is not entirely true, but sufficient for large distances, see, for example, \citealt{Bailer-jones2015,Luri2018}), we can calculate the probability $P$ that the star has a parallax in agreement with the distance to the galaxy. We calculate $P$ as the integral of the probability density function:
\begin{equation}
P(\pi _\mathrm{G}) = 1-\frac{1}{\sigma(\pi_\mathrm{Star})\cdot \sqrt{2\pi}}\int \limits _{-\infty} ^{\pi _\mathrm{G}} e^{-0.5\cdot \left(\frac{\pi _\mathrm{G}-\pi_\mathrm{Star}}{\sigma(\pi_\mathrm{Star})}\right)^2}\mathrm{d}\pi _\mathrm{G},
\end{equation}
where $\pi _\mathrm{G}$ is the parallax of the corresponding galaxy, $\pi_\mathrm{Star}$ the parallax of the star, and $\sigma(\pi_\mathrm{Star})$ the error of the parallax. We note that a complete agreement of parallax and distance to the galaxy would yield a probability of $50\%$, because the star may also be closer. An agreement of one $\sigma$ translates into $\mathrm{P}>15.9\%$, $2\cdot \sigma$ to $\mathrm{P}>2.2\%$. Table~\ref{tab:parallax_dist} also lists the highly r-process enriched Reticulum~II star \object{DES J033607.75-540235.6} as possible foreground star, where the membership is only given within $2\cdot \sigma$. We note that $1/\pi$ may be a bad estimator that often leads to an overestimation of the distance or even negative distances \citep{Bailer-jones2015,Luri2018,Bailer-Jones2018}. For the here shown sample (Table~\ref{tab:parallax_dist}) the derived distances are too small. The distances given by \citet{Bailer-Jones2018} are even smaller ($\sim 2 - 3\, \rm kpc$ for most stars in Table~\ref{tab:parallax_dist}, irrespective of the galaxy they reside in). These small values originate from a small assumed prior ($\sim 600\, \rm pc$, for most of the here shown stars) in the Bayesian distance estimate, which has been adopted for near-by MW stars by \citet{Bailer-Jones2018}. A larger prior may lead to larger distances for stars with very uncertain parallaxes. However, for stars with certain parallaxes ($\sigma (\pi_\mathrm{Star}) \lesssim 50\%$), the distance may still disagree with the ones from the dSph galaxies.

\begin{table*}[htbp]
\caption{Maximum and minimum distances calculated with the {\it Gaia} DR2 parallax for the brighter stars in our sample.}
\centering
\begin{tabular}{llcrrrrr}
\hline  
Galaxy & Object & $D(\pi )_\mathrm{min,max}\, \rm [kpc]$  & $D(\mathrm{Galaxy}) \, \rm [kpc]$ &$\mathrm{[Fe/H]}$  & $\pi \, \rm [mas]$ & $\sigma(\pi_\mathrm{Star})$ [\%]& P [\%]\\
\hline
Fnx       &  \object{$[$WMO2009$]$ For-0949}          &  2,    6 &  $146 \pm 8$  &  -0.42 & $0.30 \pm 0.14$& 45 &  1.56 \\
Fnx       &  \object{2MASS J02393412-3433096}         &  3,   19 &  $146 \pm 8$  &  -0.48 & $0.20 \pm 0.15$& 74 &  9.69 \\
Fnx       &  \object{2MASS J02394195-3430361}         &  3,   20 &  $146 \pm 8$  &  -0.54 & $0.19 \pm 0.14$& 73 &  9.47 \\
Fnx       &  \object{$[$LDH2014$]$ Fnx-mem0717}       &  2,    7 &  $146 \pm 8$  &  -0.59 & $0.29 \pm 0.13$& 47 &  1.91 \\
Fnx       &  \object{2MASS J02390853-3430556}         &  2,  106 &  $146 \pm 8$  &  -0.62 & $0.22 \pm 0.21$& 96 & 15.58 \\
Fnx       &  \object{$[$LDH2014$]$ Fnx-mem0638}       &  2,    5 &  $146 \pm 8$  &  -0.67 & $0.36 \pm 0.14$& 40 &  0.69 \\
Fnx       &  \object{$[$WMO2006$]$ F01-6}             &  4,   30 &  $146 \pm 8$  &  -0.73 & $0.16 \pm 0.12$& 79 & 11.19 \\
Fnx       &  \object{$[$LDH2014$]$ Fnx-mem0572}       &  3,   67 &  $146 \pm 8$  &  -0.77 & $0.15 \pm 0.14$& 90 & 14.48  \\
Fnx       &  \object{$[$LDH2014$]$ Fnx-mem0633}       &  3,   13 &  $146 \pm 8$  &  -0.81 & $0.21 \pm 0.14$& 65 &  6.72  \\
Fnx       &  \object{$[$LDH2014$]$ Fnx-mem0629}       &  3,   32 &  $146 \pm 8$  &  -0.88 & $0.19 \pm 0.16$& 84 & 12.47  \\
Fnx       &  \object{$[$WMO2009$]$ For-0361}          &  3,  118 &  $146 \pm 8$  &  -1.01 & $0.15 \pm 0.14$& 94 & 15.58  \\
Fnx       &  \object{2MASS J02392218-3419407}         &  3,    8 &  $146 \pm 8$  &  -1.45 & $0.27 \pm 0.13$& 50 &  2.65  \\
Fnx       &  \object{$[$LDH2014$]$ Fnx-mem0612}       &  3,   13 &  $146 \pm 8$  &  -1.54 & $0.22 \pm 0.14$& 64 &  6.63  \\
Fnx       &  \object{$[$LDH2014$]$ Fnx-mem0732}       &  3,  139 &  $146 \pm 8$  &  -1.81 & $0.15 \pm 0.14$& 95 & 15.81  \\
Fnx       &  \object{$[$LDH2014$]$ Fnx-mem0714}       &  4,   46 &  $146 \pm 8$  &  -1.84 & $0.14 \pm 0.12$& 84 & 12.96  \\
Sgr       &  \object{2MASS J18534878-3029391}         &  4,    7 &   $27 \pm 1$  &  -0.23 & $0.18 \pm 0.04$& 24 &  0.05  \\
Sgr       &  \object{$[$MBC98a$]$ Sgr1  432}          &  2,    8 &   $27 \pm 1$  &  -1.22 & $0.27 \pm 0.14$& 54 &  5.51  \\
Scl       &  \object{2MASS J00594490-3344350}         &  4,   27 &   $86 \pm 6$  &  -0.93 & $0.16 \pm 0.12$& 77 & 11.37  \\
Scl       &  \object{2MASS J00595827-3341087}         &  5,   41 &   $86 \pm 6$  &  -1.01 & $0.11 \pm 0.09$& 78 & 12.65  \\
Scl       &  \object{2MASS J00594258-3342182}         &  4,    9 &   $86 \pm 6$  &  -1.06 & $0.19 \pm 0.08$& 42 & 1.21   \\
Scl       &  \object{2MASS J00591514-3339438}         &  3,   12 &   $86 \pm 6$  &  -1.30 & $0.21 \pm 0.13$& 61 & 6.11   \\
Scl       &  \object{2MASS J00593811-3335080}         &  4,   57 &   $86 \pm 6$  &  -1.31 & $0.12 \pm 0.10$& 86 & 14.57  \\
Scl       &  \object{2MASS J01001051-3349366}         &  3,   10 &   $86 \pm 6$  &  -1.45 & $0.20 \pm 0.11$& 52 & 3.62   \\
Scl       &  \object{2MASS J00591134-3337281}         &  3,   13 &   $86 \pm 6$  &  -1.47 & $0.20 \pm 0.12$& 62 & 6.56   \\
Scl       &  \object{2MASS J01000758-3337039}         &  5,   56 &   $86 \pm 6$  &  -1.58 & $0.11 \pm 0.09$& 83 & 14.24  \\
Scl       &  \object{2MASS J01002463-3344287}         &  4,   24 &   $86 \pm 6$  &  -1.96 & $0.14 \pm 0.10$& 71 & 9.8    \\
Scl       &  \object{2MASS J01001736-3343595}         &  4,   36 &   $86 \pm 6$  &  -2.16 & $0.14 \pm 0.11$& 81 & 12.73  \\
Scl       &  \object{$[$SHT2013$]$ Scl024 01}         &  2,    9 &   $86 \pm 6$  &  -2.61 & $0.31 \pm 0.19$& 62 & 6.13   \\
Scl       &  \object{$[$WMO2009$]$ Scl-927}           &  2,   10 &   $86 \pm 6$  &  -3.13 & $0.26 \pm 0.16$& 60 & 5.65   \\
Car       &  \object{$[$MOP2000$]$ C130725}           &  3,   44 &  $106 \pm 6$  &  -1.32 & $0.19 \pm 0.17$& 88 & 14.01  \\
Car       &  \object{$[$MOP2000$]$ C120769}           &  2,    8 &  $106 \pm 6$  &  -1.41 & $0.32 \pm 0.20$& 61 & 5.66   \\
Car       &  \object{2MASS J06411632-5100185}         &  4,   22 &  $106 \pm 6$  &  -2.06 & $0.14 \pm 0.09$& 66 & 7.99   \\
Car       &  \object{$[$MOP2000$]$ C120721}           &  3,    8 &  $106 \pm 6$  &  -2.68 & $0.24 \pm 0.13$& 51 & 3.02   \\
Sex       &  \object{$[$LPP2003$]$   420}             &  0,    1 &   $86 \pm 6$  &  -1.63 & $1.69 \pm 0.60$& 35 &  0.24  \\
Sex       &  \object{$[$LPP2003$]$   570}             &  0,    1 &   $86 \pm 6$  &  -2.53 & $1.78 \pm 1.04$& 58 &  4.46  \\
Boo I     &  \object{$[$NGW2008$]$ 41}                &  3,   51 &   $66 \pm 2$  &  -2.16 & $0.17 \pm 0.15$& 88 & 15.14  \\
Ret II    &  \object{DES J033607.75-540235.6}         &  5,   13 &   $32 \pm 2$  &  -2.83 & $0.15 \pm 0.07$& 46 &  4.28  \\
Draco     &  \object{SDSS J172034.19+575331.9}        &  5,   14 &   $79 \pm 6$  &  -1.54 & $0.13 \pm 0.06$& 46 &  2.47  \\
\hline
\end{tabular}
\label{tab:parallax_dist}
\end{table*}

\section{Comparison of the barium abundances in Fornax with previous studies}
\label{sct:appendix_ba_fnx}
\begin{figure}
   \centering
   \includegraphics[width=\hsize]{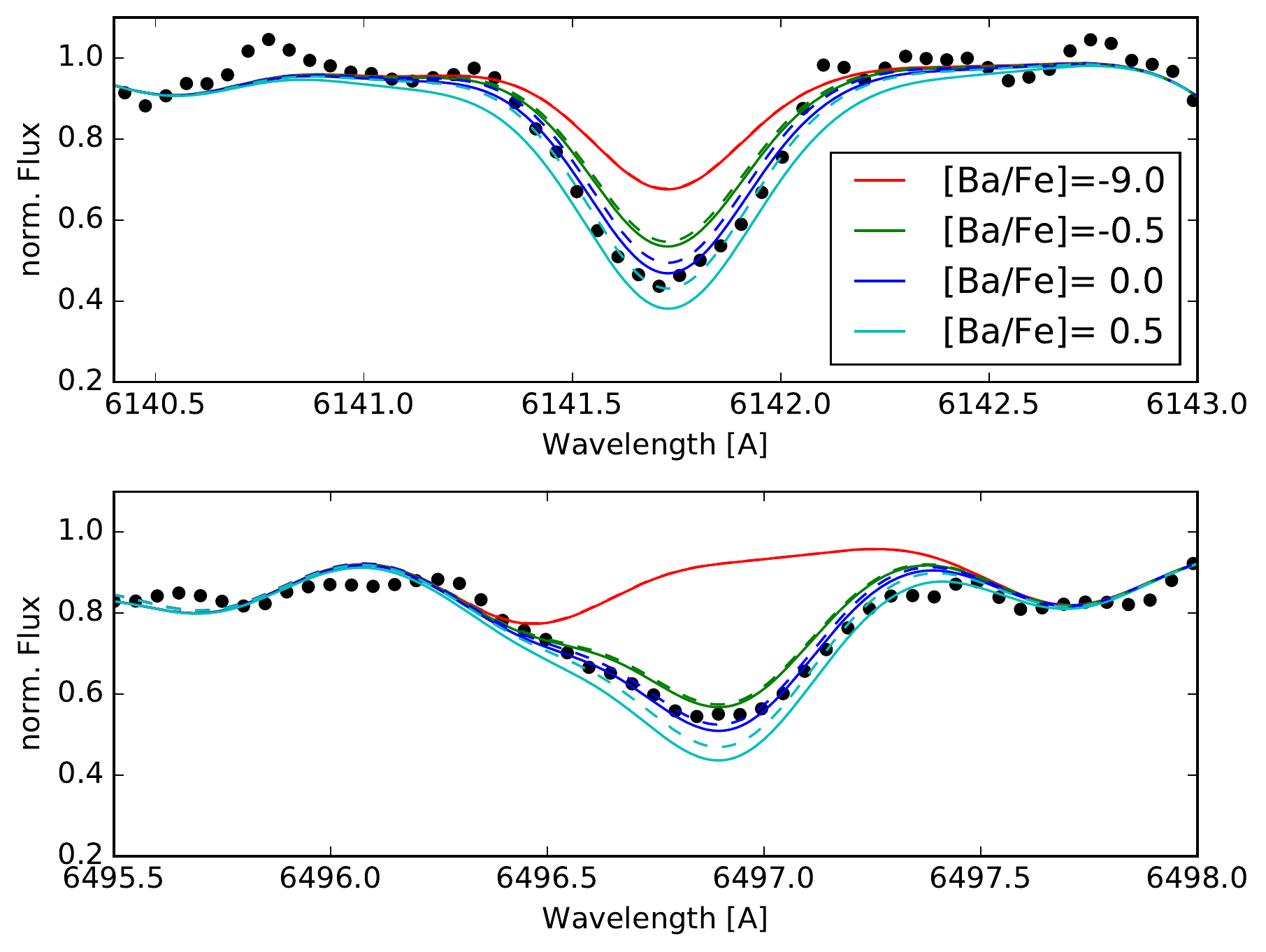}
      \caption{Spectrum synthesis for \object{2MASS J02390853-3430556}, a member of the Fornax dSph galaxy. Dashed lines indicate a damping according to the Unsöld approximation \citep{Unsold1927}, whereas straight lines indicate a damping according to \citet{Barklem2000}.}
         \label{fig:letarte_synth}
 \end{figure}

In comparison with previous studies of Fornax \citep{Letarte2010,Letarte2018,Lemasle2014}, we obtain lower $\mathrm{[Ba/Fe]}$ values \citep[cf. Fig.~\ref{fig:ba_eu} with Fig.~14. of ][]{Letarte2018}. The star \object{2MASS J02390853-3430556}, which is named BL125 in \citet{Letarte2018} can be used as an example. \citet{Letarte2018} derived a value of $\mathrm{[Ba/Fe]}=0.53 \pm 0.17\, \rm dex$, whereas we derive $\mathrm{[Ba/Fe]}=-0.08 \pm 0.15\, \rm dex$ for this star. There are multiple factors that let us derive a lower value. First, we use a different weighting of the lines (see Sect.~\ref{ssct:barium}). Second, we note that the Ba line at $\lambda = 6141.73 \, \AA$ is heavily blended by an \ion{Fe}{I} line (Fig.~\ref{fig:letarte_synth}) and it might be difficult to determine the abundance by measuring equivalent widths as done in previous studies by DAOSPEC (see also \citealt{Gallagher2020}, for a discussion of blends on barium lines). 
Measuring the EWs for these lines gives us almost the same values as given in \citet{Letarte2010}. However, converting this to abundances without taking the \ion{Fe}{I} blend into account, results in a still lower value of $\mathrm{[Ba/Fe]}=0.19$. By considering a different treatment of damping, we are able to reproduce the values given in \citet{Letarte2018}. The strong dependence on the damping states that these lines are already close to saturation. We use data of \citet{Barklem2000} and \citet{Barklem2005} to take damping into account. A comparison of the different treatments as well as the iron blend is shown in Fig.~\ref{fig:letarte_synth}. \\

\section{The evolution of Mg, Sc, and Ti}
\label{sct:construction_galaxies}
In order to create Fig.~\ref{fig:constr_gal} we used the following equation:
\begin{equation}
\label{eq:gal_reconstruction}
   \mathrm{[X/Fe]}= \begin{cases}  \left(a\cdot M_V - b\right)\cdot \mathrm{[Fe/H]} + b-a\cdot c & \mathrm{[Fe/H]}>\mathrm{[Fe/H]}_\mathrm{\alpha,knee}\\ c &\text{else} \end{cases}
\end{equation}
with the parameters as given in Table~\ref{tab:reconstruction_galaxies}.

\begin{table}
\caption{Fit parameter for the evolution of Mg, Sc, and Ti as used in Eq.~\ref{eq:gal_reconstruction}.}            
\label{tab:reconstruction_galaxies}     
\centering                         
\begin{tabular}{lrrr}
\hline\hline                
Parameter & Mg & Sc & Ti\\
\hline    
a & $-0.102$ & $-0.127$ & $-0.081$ \\
b & $-1.822$ & $-2.124$ & $-1.500$ \\
c & $0.326$  & $-0.056$ & $0.334$ \\
\hline                                   
\end{tabular}
\end{table}

\section{Abundance plots}
\begin{figure}
   \centering
   \includegraphics[width=\hsize]{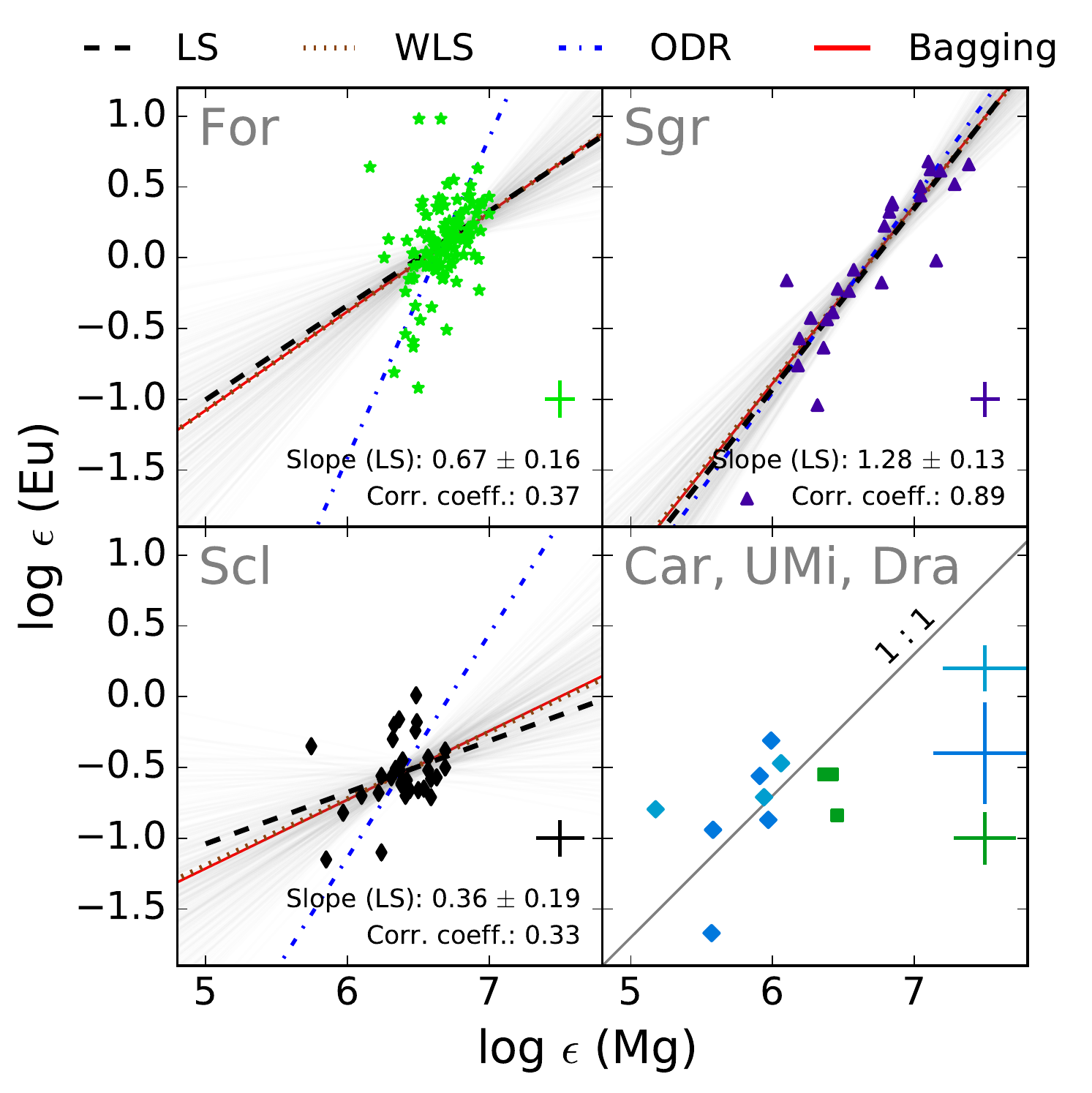}
      \caption{Same as Fig.~\ref{fig:bar_alpha_corr}, but for Eu.}
         \label{fig:Eu_alpha_corr}
 \end{figure}

In our sample, europium is the cleanest tracer of the r-process. Therefore, a correlation of europium with Mg would be the best indicator of the time delays of the individual production channels. However, the limited detections of europium cause a low correlation between europium and Mg. As a consequence, different fitting techniques result in different slopes (Fig.~\ref{fig:Eu_alpha_corr}).
   
\end{appendix}

\end{document}